\documentstyle[amssymb,amstex,12pt,a4wide]{article}
\newtheorem{theorem}{Theorem}[section]
\newtheorem{lemma}[theorem]{Lemma}

\newtheorem{corollary}[theorem]{Corollary}
\newtheorem{remark}[theorem]{Remark}
\newcommand{\lgth}{\operatorname{length}}
\newcommand{\dist}{\operatorname{dist}}
\newcommand{\An}{\operatorname{An}}
\newcommand{\innt}{\operatorname{Int}}
\newcommand{\diam}{\operatorname{diam}}

\newcommand{\Area}{\operatorname{Area}}

\newcommand{\card}{\operatorname{card}}
\newcommand{\Ver}{\operatorname{Vertices}}
\newcommand{\blk}{\operatorname{{\Bbb M}}}
\newcommand{\Var}{\operatorname{Var}}
\newcommand{\Cov}{\operatorname{Cov}}
\newcommand{\e}{\operatorname{e}}
\newcommand{\iin}{\operatorname{in}}
\newcommand{\oout}{\operatorname{out}}
\newcommand{\seen}{\operatorname{seen}}

\begin{document}
\setlength{\baselineskip}{1.3\baselineskip}

\title{Dobrushin-Koteck\'y-Shlosman theorem for polygonal Markov fields in the plane}
\author{Tomasz Schreiber\footnote{Research supported by the Foundation for
 Polish Science (FNP), by the Polish Minister of Scientific Research and
 Information Technology grant 1 P03A 018 28 (2005-2007) and from the
 EC 6th Framework Programme
 Priority 2 Information Society Technology Network of Excellence MUSCLE 
 (Multimedia Understanding through Semantics, Computation and Learning;
 FP6-507752). A part of this research was completed while staying
 at the Centrum voor Wiskunde en Informatica (CWI), Amsterdam, The
 Netherlands},\footnote{Mailing address: Tomasz Schreiber, Faculty of
 Mathematics \& Computer Science, Nicolaus Copernicus University,
 ul. Chopina 12 $\slash$ 18, 87-100 Toru\'n, Poland; tel.: (++48) (+56)
 6113442, fax: (++48) (+56) 6228979; e-mail: {\tt tomeks at mat.uni.torun.pl}}\\
        Faculty of Mathematics \& Computer Science,\\
        Nicolaus Copernicus University,\\
        Toru\'n, Poland.}
\date{}
\maketitle

\paragraph{Abstract:} 
 {\it We consider the so-called length-interacting Arak-Surgailis
      polygonal Markov fields with V-shaped nodes \--- a continuum 
      and isometry invariant process in the plane sharing a number of
      properties with the two-dimensional Ising model. For these polygonal
      fields we establish a low-temperature phase separation theorem in the
      spirit of the Dobrushin-Koteck\'y-Shlosman theory, with the corresponding 
      Wulff shape deteremined to be a disk due to  the rotation invariant
      nature of the considered model. As an important tool replacing the
      classical cluster expansion techniques and very well suited for our geometric
      setting we use a graphical construction built on contour birth and death
      process, following the ideas of F\'ernandez, Ferrari and Garcia.}

\section{Introduction and main results}
 An example of a planar polygonal Markov field, referred to as the Arak process
 throughout this paper, was first introduced by Arak \cite{A1}. Further developments are due
 to Arak \& Surgailis \cite{AS1},\cite{AS2}, Surgailis \cite{SU1},
 Arak, Clifford \& Surgailis
 \cite{ACS}. In this paper we focus our attention on polygonal Markov fields with V-shaped
 nodes, which are a particular class of ensembles of self-avoiding polygonal
 loops (contours) in the plane, interacting only by the requirement of disjointness.
 Not unexpectedly, these objects share a number of properties of the two-dimensional
 Ising model, including the presence of spontaneous magnetisation and absence of
 infinite contour nesting in low temperature region, see Nicholls \cite{N1} and
 Schreiber \cite{SC2}. An important property of the Arak process and its length-interacting
 Gibbsian modifications is their isometry invariance. One might be tempted to view these
 purely continuum polygonal fields as a kind of continuum version of the Ising
 model. For low temperatures this opinion seems to be well founded. There is
 a number of relevant differences though in the much less understood high
 temperature region. In sharp contrast to the Ising model it is not clear
 how to define the infinite temperature non-interacting field, since some
 spatial correlation is always present due to the imposed polygonal nature
 of the contours. Therefore as the reference field for length-interacting Gibbsian
 modifications we choose the original Arak process, enjoying a number of striking
 properties including consistency, exact solubility and admitting the so-called
 {\it dynamic representation} in terms of equilibrium evolution of one-dimensional
 particle systems tracing the polygonal boundaries of the process in two-dimensional
 space-time, see Arak \& Surgailis \cite{AS1} and the Appendix below for details.

 The purpose of this paper is to show that, in analogy with the Ising model,
 the phase separation phenomenon is present for length-interacting polygonal
 Markov fields and it is gouverned by the Wulff construction, see Bodineau, Ioffe \& Velenik 
 \cite{BIV} for an extensive reference. We establish our main Theorem \ref{GLOWNE} in the DKS
 set-up, as introduced by Dobrushin, Koteck\'y \& Shlosman in their seminal
 monograph \cite{DKS}, and we only work at low enough temperatures. As a crucial tool
 replacing cluster expansion techniques and very well suited for our geometric
 setting we use a graphical construction built on contour birth and death process,
 as introduced by F\'ernandez, Ferrari \& Garcia
 \cite{FFG1},\cite{FFG2},\cite{FFG3}, see Subsection 
 \ref{KONSGRAF}. We took advantage of the particular properties of polygonal
 fields in order to characterise the model-specific surface tension, defined
 in Subsection \ref{NAPPOWSE}, in terms of hitting probabilities of
 appropriate planar random walks in random environment provided by
 the graphical construction. Even though we were only able to establish relatively
 weak results for the quality of approximation of the surface tension by its finite
 volume versions, we used the isometry invariance of the model to circumvent
 this problem. A particular feature of our approach is that rather than
 imposing periodic [as in Dobrushin, Koteck\'y \& Shlosman \cite{DKS}] or fixed sign
 boundary conditions [as Ioffe \& Schonmann \cite{IS1}], we work directly in the
 thermodynamic limit conditioned on the event that no large contours hit the
 boundary of the region. This allows us to avoid technical difficulties
 which would arise if we had to control our surface tension estimates in
 close vicinity of the boundary. Finally, the micro-canonical constraint
 considered in this paper requires that the excess of total magnetisation
 be  {\it larger or equal} rather than just equal to a given positive
 threshold value - this avoids a number of technical complications
 which would otherwise arise due to the continuum nature of our setting
 and allows us to work with weaker versions of moderate deviation estimates
 and to rely upon general local central limit (LCL) results available in the
 literature rather than establishing an LCL theorem in its full strength
 specialised for our model. 

 In analogy to the original DKS approach, the crucial
 ingredients of the proof of our main Theorem \ref{GLOWNE} are 
 \begin{itemize}
  \item the coarse graining estimates of Section \ref{SZKIEE}, based on
        skeleton techniques slightly modified and specialised for our
        particular setting. The graphical construction of Subsection
        \ref{KONSGRAF} is used as a crucial tool replacing
        cluster expansion techniques, 
  \item moderate deviation estimates for cut-off ensembles, stated 
        in Section \ref{MDCE} and established by the classical method
        of studying the restricted phase modified by actions of finely-tuned
        external magnetic fields, see e.g. Section 2 of Ioffe \& Schonmann
        \cite{IS1}. The graphical construction of Subsection \ref{KONSGRAF}
        admits an extension for these modified fields.  
 \end{itemize}         

 We believe the techniques developed in the present paper should in
 principle be applicable to general continuum models exhibiting
 isometry invariance and admitting polymer representation.  

 The remaining part of the introductory section is organised as follows.
 Below, in Subsection \ref{LIPMF} we give a formal construction of the
 polygonal Markov fields. The next Subsection \ref{KONSGRAF} is devoted
 to the graphical construction. The surface tension specific for our
 model is defined in Subsection \ref{NAPPOWSE}. Finally, our main
 results are formulated in Subsection \ref{MARE}.  

 Throughout the paper we make extensive use of the '$O,\Omega,\Theta$' notation,
 with $O(X)$ and $\Omega(X)$ standing respectively for quantities bounded 
 in their absolute value above and below by a constant times $X,$
 and with $\Theta(X) = O(X) \cap \Omega(X).$ Moreover, we use $c, C, C_1, C_2$
 etc. to denote generic constants which can change their values from one
 statement to another.

\subsection{Length-interacting polygonal Markov fields}\label{LIPMF}
 The formal construction of the basic Arak process with empty boundary
 conditions in a bounded open set $D \subseteq {\Bbb R}^2$ goes as
 briefly discussed below [we refer the reader to \cite{AS1}
 and \cite{ACS} for further details]. In the sequel
 we assume that the boundary $\partial D$ is piecewise smooth. We define
 the family $\Gamma_D$ of admissible polygonal configurations on $D$ by
 taking all the planar graphs $\gamma$ in $D$ such that
 \begin{description}
  \item{\bf (P1)} $\gamma \cap \partial D = \emptyset,$
  \item{\bf (P2)} all the vertices of $\gamma$ are of degree $2,$
  \item{\bf (P3)} the edges of $\gamma$ do not intersect,
  \item{\bf (P4)} no two edges of $\gamma$ are co-linear.
 \end{description}
 In other words, $\gamma$ consists of a finite number of disjoint polygons fully
 contained in $D$ and possibly nested. Further, for a finite collection $(l) = (l_i)_{i=1}^n$
 of straight lines intersecting $D$ we denote by $\Gamma_D(l)$ the family of admissible
 configurations $\gamma$ with the additional properties that $\gamma \subseteq
 \bigcup_{i=1}^n l_i$ and $\gamma \cap l_i$ is a single interval of a strictly
 positive length for each $l_i, i=1,...,n,$ possibly with some isolated
 points added. Let $\Lambda_D$ be the restriction
 to $D$ of a homogeneous Poisson line process $\Lambda$ with intensity measure
 given by the standard isometry-invariant Lebesgue measure $\mu$ on the space of
 straight lines in ${\Bbb R}^2.$ One possible construction of $\mu$ goes by
 identifying a straight line $l$ with the pair $(\phi,\rho) \in [0,\pi) \times {\Bbb R},$
 where $(\rho \sin(\phi), \rho \cos(\phi))$ is the vector orthogonal to $l$ and joining
 it to the origin, and then by endowing the parameter space
 $[0,\pi) \times {\Bbb R}$ with the usual Lebesgue measure.  With the above notation,
 the basic polygonal Arak process ${\cal A}_D$ on $D$ arises as the Gibbsian
 modification of the process induced on $\Gamma_D$ by $\Lambda_D,$ with the Hamiltonian
 given  by the double total edge length, that is to say
 %\begin{equation}\label{GREPR1}
 %   d [{\cal L}({\cal A}_D)](\gamma) = \frac{{\bf 1}_{\Gamma_D(l)}(\gamma) \exp(-2\lgth(\gamma))
 %   d [{\cal L}(\Lambda_D)](l)}{{\Bbb E}\sum_{\delta \in \Gamma_D(\Lambda_D)}
 %                                     \exp(-2 \lgth(\delta))}
 %\end{equation}
 %with ${\cal L}(\cdot)$ standing for the distribution of the argument random object.
 \begin{equation}\label{GREPR1}
  {\Bbb P}\left({\cal A}_D \in G\right)
  = \frac{{\Bbb E}\sum_{\gamma \in \Gamma_D(\Lambda_D) \cap G}
                  \exp(-2\lgth(\gamma))}
         {{\Bbb E}\sum_{\gamma \in \Gamma_D(\Lambda_D)} \exp(-2\lgth(\gamma))}
 \end{equation}
 for all $G \subseteq \Gamma_D$ Borel measurable, say with respect to the
 usual Hausdorff distance topology, and where $\Gamma_D(\Lambda_D)$ 
 denotes $\Gamma_D(l)$ as defined above with $l$ set to be the 
 collection of all straight lines of $\Lambda_D.$ Note that by the
 {\it total edge length} $\lgth(\gamma)$ of a polygonal configuration 
 $\gamma$ we mean here and below the sum of lengths of all constituent
 polygons. The expectations in (\ref{GREPR1}) are taken with respect
 to the randomness of $\Lambda_D.$  
 It should be mentioned at this point that in the literature on consistent
 polygonal fields one
 usually considers free rather than empty boundary conditions, see
 \cite{AS1} and the Appendix below, yet the empty
 boundary object is better
 suited for the graphical construction below and for our further
 purposes.

 For a positive inverse temperature $\beta > 0$ we consider the length-interacting
 Arak process ${\cal A}^{[\beta]}_D$ in $D$ determined in distribution by
 \begin{equation}\label{GREPRBETA}
  \frac{d {\cal L}({\cal A}_D^{[\beta]})}{d{\cal L}({\cal A}_D)}[\gamma] :=
  \frac{\exp(-\beta \lgth(\gamma))}{{\Bbb E} \exp\left( -\beta \lgth\left( {\cal A}_D \right)
  \right)},
 \end{equation}
 with ${\cal L}(\cdot)$ standing for the law of the argument random object.
 As shown in Theorem 3 and Corollary 4 of \cite{SC2} for $\beta$ large enough
 the polygonal fields ${\cal A}^{[\beta]}_D,\; D \subseteq {\Bbb R}^2,$ admit
 a unique whole plane thermodynamic
 limit without infinite contours, denoted in the sequel by ${\cal A}^{[\beta]},$
 see also below for its construction. The field ${\cal A}^{[\beta]}$ is isometry
 invariant. The thermodynamic limit ${\cal A}^{[0]}$ can also be shown to exist
 for $\beta = 0,$ in the sequel it is denoted by ${\cal A}$ and its construction
 is given in the Appendix. 

 It is known that for the inverse temperature $\beta$ sufficiently large (in particular,
 for all $\beta$ within the validity region of the graphical construction below)
 the thermodynamic limit ${\cal A}^{[\beta]}$ exhibits only finite contour nesting,
 see Nicholls \cite{N1} and the discussion following Corollary 4 in Schreiber
 \cite{SC2}.
 Whence, the contour ensemble ${\cal A}^{[\beta]}$ partitions the plane into a unique
 infinite connected component ({\it the ocean}) and a countable number of finitely nested
 bounded regions (islands). We colour black and white the polygonal regions of this partition
 by declaring the infinite ocean white and by requiring that the collection of interfaces
 between black and white regions coincide with the collection of contours ${\cal A}^{[\beta]},$
 which uniquely determines the colouring. Denote the resulting union of black regions by
 ${\rm black}[{\cal A}^{[\beta]}]$ and the union of white regions by ${\rm white}[{\cal A}^{[\beta]}].$
 For a bounded region $U \subseteq {\Bbb R}^2$ let $\blk_U\left({\cal A}^{[\beta]}\right)$
 be the {\it magnetisation} in $U$ determined by the coloured contour ensemble ${\cal A}^{[\beta]}$
 under the assignment ${\rm black} \mapsto +,\; {\rm white} \mapsto -.$ In other words,
 $\blk_U\left({\cal A}^{[\beta]}\right)$ is the total area of the black-coloured regions
 in $U$ minus the total area of white-coloured regions in $U:$
 $$ \blk_U\left({\cal A}^{[\beta]}\right) := \Area\left({\rm black}[{\cal A}^{[\beta]}]
    \cap U \right) - \Area\left({\rm white}[{\cal A}^{[\beta]}] \cap U \right). $$
 For $L > 0$ we shall abbreviate $\blk_{{\Bbb B}_2(L)}\left({\cal A}^{[\beta]}\right)$
 to $\blk_L[\beta],$ where ${\Bbb B}_2(L)$ stands for the disk of radius $L$ centred at
 $0.$ The isometry invariance of the infinite-volume field ${\cal A}^{[\beta]}$ 
 implies that 
 \begin{equation}\label{MagnetyzacjaWlasciwa}
  {\Bbb E}\blk_U\left( {\cal A}^{[\beta]} \right) = \Area(U) \blk[\beta],\;
    \;\;\blk[\beta] \in (-1,0), 
 \end{equation}
 where $|\blk[\beta]|$ is further referred to as the specific spontaneous
 magnetisation at inverse temperature $\beta.$ 
 The '${\rm black}[\cdot]$' and '$\blk_{\cdot}(\cdot)$' notation will be
 also used for ${\cal A}^{[\beta]}$ replaced by a number of other polygonal
 fields enjoying the property that the corresponding contour ensemble
 determines a unique unbounded region, to be coloured white.

\subsection{Graphical construction}\label{KONSGRAF}
\subsubsection{Basic graphical construction}
 As argued in Schreiber \cite{SC2}, the polygonal field ${\cal A}^{[\beta]}_D$ admits
 a  natural representation in terms of a graphical construction in the spirit of
 Fern\'andez, Ferrari \& Garcia \cite{FFG1},\cite{FFG2},\cite{FFG3}, which will be a crucial tool in
 our argument in the sequel, as replacing cluster expansion techniques.
 Below, we provide a description of this construction borrowed
 from \cite{SC2}. Consider the space ${\cal C}_D$ consisting of
 all closed polygonal contours in $D$
 which do not touch the boundary $\partial D.$ For a given finite configuration
 $(l) := ( l_1,...,l_n )$ of straight lines intersecting $D$ denote by ${\cal C}_D(l)$
 the family of those polygonal contours in ${\cal C}_D$ which belong to $\Gamma_D(l).$
 We define the so-called free contour measure $\Theta_D$ on ${\cal C}_D$ by putting
 for $C \subseteq {\cal C}_D$ measurable, say with respect to the Borel $\sigma$-field
 generated by the Hausdorff distance topology,
 \begin{equation}\label{WOLNEKONTURY}
   \Theta_D(C) = \int_{{\rm Fin}(L[D])} \sum_{\theta \in C \cap {\cal C}_D(l)}
    \exp(-2\lgth(\theta)) d\mu^*((l))
 \end{equation}
 with ${\rm Fin}(L[D])$ standing for the for the family of finite line
 configurations intersecting $D$ and where $\mu^*$ is the measure
 on ${\rm Fin}(L[D])$ given by $d\mu^*((l_1,...,l_n)) := \prod_{i=1}^n d\mu(l_i)$
 with $\mu$ defined in the discussion preceding (\ref{GREPR1}).

 For $\beta > 0$ we consider the exponential modification $\Theta^{[\beta]}_D$
 of the free measure $\Theta_D,$
 \begin{equation}\label{THETAB}
  \Theta^{[\beta]}_D(d\theta) := \exp(-\beta \lgth(\theta)) \Theta_D(d\theta).
 \end{equation}
 Observe that for all bounded open sets $D$ with piecewise smooth boundary
 the free contour measures $\Theta_D$ as defined in (\ref{WOLNEKONTURY})
 arise as the respective restrictions to ${\cal C}_D$ of the same measure
 $\Theta$ on ${\cal C} := \bigcup_{n=1}^{\infty}
  {\cal C}_{(-n,n)^2},$ in the sequel referred to as the infinite volume
 free contour measure. Indeed, this follows easily by the observation
 that $\Theta_{D_1}$ restricted to ${\cal C}_{D_2}$ coincides with
 $\Theta_{D_2}$ for $D_2 \subseteq D_1.$ In the same way we construct
 the infinite-volume Gibbs-modified measures $\Theta^{[\beta]}.$
 The following  result, which is Lemma 1 of \cite{SC2} (note that
 the first result in this spirit is due to Nicholls \cite{N1}, see
 Lemma in the Appendix ibidem) will be crucial for our further purposes.
 \begin{lemma}\label{NICHOLLS}
  For $\beta \geq 2$ we have
  $$
   \Theta^{[\beta]}(\{ \theta\;|\; dx \cap \Ver(\theta) \neq \emptyset,\; \lgth(\theta) > R \})
   \leq 
   8 \pi \exp(-[\beta-2] R) dx,
  $$
  where the event $\{ dx \cap \Ver(\theta) \neq \emptyset \}$ is to be understood
  that a vertex of $\theta$ falls into $dx.$ 
  Moreover, there exists a constant $\varepsilon > 0$ such that, for $\beta \geq 2,$
  $$
   \Theta^{[\beta]}(\{ \theta\;|\; {\bf 0} \in \innt \theta,\;
   \lgth(\theta) > R \}) \leq \exp(-[\beta-2+\varepsilon] R + o(R)),
  $$
  with $\innt \theta$ standing for the region enclosed by $\theta$
  (recall that $\theta \in {\cal C}$ is always a single bounded contour).  
 \end{lemma}
 Let ${\cal P}_{\Theta_D^{[\beta]}}$ be the Poisson point process on ${\cal C}_D$
 with intensity measure $\Theta^{[\beta]}_D.$ It follows then directly by
 (\ref{WOLNEKONTURY}) and by (\ref{GREPR1}) that ${\cal A}^{[\beta]}_D$ coincides
 in distribution with the union of contours in ${\cal P}_{\Theta_D^{[\beta]}}$
 conditioned on the event that they are disjoint so that
 \begin{equation}\label{WARUNKOWKT}
  {\cal L}\left( {\cal A}^{[\beta]}_{D} \right) =
  {\cal L}\left( \bigcup_{\theta \in {\cal P}_{\Theta_D^{[\beta]}}} \theta \; \left|
  \; \forall_{\theta, \theta' \in {\cal P}_{\Theta_D^{[\beta]}}} \theta \neq \theta'
  \Rightarrow \theta \cap \theta' = \emptyset \right. \right),
 \end{equation}
 where the conditioning makes sense because $\Theta_D^{[\beta]}({\cal C}_D)$ is
 finite as shown in Subsection 2.2 of \cite{SC2}. In particular, as argued
 in Subsection 2.2 and Theorem 2 ibidem,
 the law of ${\cal A}^{[\beta]}_{D}$ is invariant and reversible with respect
 to the following contour birth and death dynamics $(\gamma^D_s)$ on $\Gamma_D.$
 \begin{description}
  \item{${\bf (C:birth[\beta])}$} With intensity $\Theta^{[\beta]}_D(d\theta) ds$ do
   \begin{itemize}
    \item Choose a new contour $\theta,$
    \item If $\theta \cap \gamma^D_s = \emptyset,$ accept $\theta$ and
          set $\gamma^D_{s+ds} := \gamma^D_s \cup \theta,$
    \item Otherwise reject $\theta$ and keep $\gamma^D_{s+ds} := \gamma^D_s,$
   \end{itemize}
  \item{${\bf (C:death[\beta])}$}
      With intensity $1 \cdot ds$ for each contour $\theta \in \gamma^D_s$
      remove $\theta$ from $\gamma^D_s$ setting $\gamma^D_s := \gamma^D_s \setminus \theta.$
 \end{description}
 Moreover, ${\cal L}({\cal A}^{[\beta]}_D)$ is the unique invariant distribution
 of the above dynamics, see Theorem 2 in \cite{SC2}. These observations
 place us within the framework of the general contour birth and death graphical
 construction as developed by Fern\'andez, Ferrari \& Garcia
 \cite{FFG1},\cite{FFG2},\cite{FFG3} and as briefly
 sketched below, see ibidem and Schreiber \cite{SC2} for further details. Choose $\beta$
 large enough, to be specified below.  Define ${\cal F}({\cal C})$ to be the space of
 countable and locally finite collections of contours from ${\cal C},$ with
 the local finiteness requirement meaning that at most a finite number of
 contours can hit a bounded subset of ${\Bbb R}^2.$ On the time-space
 ${\Bbb R} \times {\cal F}({\cal C})$ we construct the stationary free contour
 birth and death process $(\varrho_s)_{s \in {\Bbb R}}$ with the birth intensity
 measure given by $\Theta^{[\beta]}$ and with the death intensity $1.$
 Note that {\it free} means here that every new-born contour is accepted
 regardless of whether it hits the union of already existing contours or not,
 moreover we admit negative time here, letting $s$ range through ${\Bbb R}$
 rather than just ${\Bbb R}_+.$
 Observe also that we need the birth measure $\Theta^{[\beta]}$ to be finite on
 the sets $\{ \theta \in {\cal C} \;|\; \theta \cap A \neq \emptyset \}$
 for all bounded Borel $A \subseteq {\Bbb R}^2$ in order to have the process
 $(\varrho_s)_{s \in {\Bbb R}}$ well defined on ${\Bbb R} \times {\cal F}({\cal C}).$
 By Lemma \ref{NICHOLLS} this is ensured whenever $\beta \geq 2.$
 To proceed, for the free process $(\varrho_s)_{s \in {\Bbb R}}$ we perform the
 following {\it trimming} procedure. We place a directed connection from each
 time-space instance of a contour showing up in $(\varrho_s)_{s \in {\Bbb R}}$
 and denoted by $\theta \times [s_0,s_1),$ with $\theta$ standing for the contour
 and $[s_0,s_1)$ for its lifespan, to all time-space contour instances
 $\theta' \times [s'_0,s'_1)$ with $\theta' \cap \theta \neq \emptyset,
 s'_0 \leq s_0$ and $s'_1 > s_0.$ In other words, we connect $\theta \times
 [s_0,s_1)$ to those contour instances which may have affected the
 acceptance status of $\theta \times [s_0,s_1)$ in the {\it constrained}
 contour birth and death dynamics {\bf (C)} as discussed above.
 These directed connections give rise to directed ancestor chains of time-space
 contour instances, following \cite{FFG3} the union
 of all ancestor chains stemming from a given contour instance $\theta^* =
 \theta \times [s_0,s_1),$ including the instance itself,
 is referred to as its {\it clan of ancestors}
 and is denoted by $\An(\theta^*).$ More generally, for a bounded region
 $U$ in the plane we write $\An_s(U)$ for the union of ancestor clans of
 all contour instances $\theta \times [s_0,s_1)$ with $\theta \cap U \neq
 \emptyset$ and $s \in [s_0,s_1).$ Lemma \ref{NICHOLLS} allows us to apply
 the technique of domination by sub-critical branching processes, developed
 in \cite{FFG1},\cite{FFG2},\cite{FFG3}, in order to conclude
 that there exists $\beta_g$ such that for each $\beta > \beta_g$ there
 exists $c := c(\beta) > 0$ such that
 \begin{equation}\label{ZANIKGR}
  {\Bbb P}(\diam \An_s({\Bbb B}_2(x,1)) > R) \leq \exp(- c R),\; s \in {\Bbb R},\; x \in {\Bbb R}^2,
 \end{equation}
 with ${\Bbb B}_2(x,1)$ standing for the radius $1$ ball in ${\Bbb R}^2$
 centred at $x.$ In the sequel we shall always assume that $\beta > \beta_g,$
 that is to say that $\beta$ is in the {\it validity} region
 of the graphical construction. We see that for $\beta > \beta_g$ all the
 ancestor clans are a.s. finite and we can uniquely determine the acceptance
 status of all their members: contour instances with no ancestors are a.s.
 accepted, which automatically and uniquely determines the acceptance status
 of all the remaining members of the clan by recursive application of
 the inter-contour exclusion rule.
 In this case, discarding the unaccepted contour instances leaves us with a
 time-space representation of a stationary evolution $(\gamma_s)_{s \in {\Bbb R}}$
 on ${\cal F}({\cal C}),$ which is easily checked to evolve according to the
 whole-plane version of the dynamics {\bf (C)} above. In Section 4 and Theorem 4
 of \cite{SC2} we argue that for all $s \in {\Bbb R}$ the polygonal field
 $\gamma_s$ coincides in distribution with the thermodynamic limit (see Section
 3 ibidem) for ${\cal A}^{[\beta]}$ without infinite contours, which is unique
 (see Corollary 4 ibidem). It should be observed that for each $s \in {\Bbb R}$
 the free field $\varrho_s$ coincides in distribution with the Poisson contour
 process ${\cal P}_{\Theta^{[\beta]}}.$ Since almost surely we have
 $\gamma_s \subseteq \varrho_s,$ we get the stochastic domination of the
 contour ensemble ${\cal A}^{[\beta]}$ by ${\cal P}_{\Theta^{[\beta]}}.$

 We also consider finite-volume versions of the above
 graphical construction, replacing the infinite-volume birth intensity measure
 $\Theta^{[\beta]}$ with its finite-volume counterparts $\Theta^{[\beta]}_D$
 for bounded and open $D$
 with piecewise smooth boundary. Clearly, the graphical construction yields
 then a version of the finite-volume contour birth and death evolution ${\bf (C)}.$
 For each $D$ denote by $(\gamma_s^D)_{s \in {\Bbb R}}$ the resulting finite-volume
 stationary process on the space ${\cal F}({\cal C}_D)$ of finite contour configurations
 in $D$ and write $(\varrho^D_s)_{s \in {\Bbb R}}$ for the corresponding free process.
 It follows by Theorem 2 in \cite{SC2} that $\gamma_s^D$
 coincides in distribution with ${\cal A}^{[\beta]}_D$ for each $s
 \in {\Bbb R}.$ Likewise, $\varrho^D_s$ coincides in distribution with
 ${\cal P}_{\Theta_D^{[\beta]}}.$

 By representing the measures $\Theta^{[\beta]}_D$ as the corresponding restrictions
 of $\Theta^{[\beta]}$ we obtain a natural coupling of all the processes
 $\gamma^D_s, \varrho^D_s, \gamma_s$ and $\varrho_s$ on a common probability space.
 We shall also consider ${\cal A}_D^{[\beta]}$ coupled on the same probability
 space by putting ${\cal A}_D^{[\beta]} = \gamma^D_0.$ Likewise, we put
 ${\cal A}^{[\beta]} = \gamma_0,\; {\cal P}_{\Theta^{[\beta]}}
 = \varrho_0,\; {\cal P}_{\Theta^{[\beta]}_D} = \varrho_0^D.$ This coupling,
 referred to as the canonical coupling in the sequel, will be assumed without
 a further mention throughout this paper.

 A simple yet useful application of this coupling is that 
 $$ 
  \left| \Area(D) \blk[\beta] - {\Bbb E}\blk_D({\cal A}^{[\beta]}_D) \right|
  = O(\Area(\partial D \oplus {\Bbb B}_2(1)))
 $$ 
 with $\oplus$ standing for the usual Minkowski addition [i.e.
 $X \oplus Y := \{ x + y \;|\; x \in X,\; y \in Y \}$]. Indeed, this is
 immediately seen by observing that, by Lemma \ref{NICHOLLS} and in view
 of (\ref{ZANIKGR}) stating
 the exponential tail decay for ancestor clan diameters, under the
 canonical coupling of ${\cal A}^{[\beta]}_D$ and ${\cal A}^{[\beta]},$ 
 the probability that a given point $x \in D$ is assigned different colours
 by these ensembles decays as $\exp(-\Omega(\dist(x,\partial D))).$
 Integrating over $D$ and using (\ref{MagnetyzacjaWlasciwa}) we obtain
 the required relation.

% Dotad 7 wrzesnia rano.

\subsubsection{Modifications of the basic graphical construction}\label{MODGK}

 Below we discuss a number of modifications of the graphical construction,
 which will be of use for our further purposes. Apart from the area-interacting
 modifications all the remaining ones can be defined on the probability space
 of the basic construction, thus extending the canonical coupling.

\paragraph{Imposing forbidden regions}
 A particular property of the graphical construction which will be crucial for our
 further purposes is that it admits, on the same probability space, conditional versions
 on the events of the type {\it no contour of the polygonal field hits [intersects] a given
 region $U$}. Indeed, let $U$ be a bounded subset of the plane ${\Bbb R}^2.$ Then,
 adding the rule that all new-born contours hitting $U$ [intersections of $U$ only
 with the interior of a contour are not taken into account] be immediately discarded, to
 the trimming procedure constructing $(\gamma_s)$ out of $(\varrho_s)$ or, equivalently,
 to the dynamics ${\bf (C)},$ we obtain a stationary and reversible process
 $(\gamma_{s:U})$ easily seen to enjoy the property that the distribution of
 $\gamma_{s:U}$ for each fixed $s$
 coincides with the law of ${\cal A}^{[\beta]}$ conditioned on the event that
 ${\cal A}^{[\beta]} \cap U = \emptyset.$ Put ${\cal A}^{[\beta]}_{{\Bbb R}^2:U} :=
 \gamma_{0:U}.$  Likewise, we define the conditioned
 version $(\gamma^D_{s:U})$ of the finite volume process $(\gamma^D_s)$ for
 which the distribution of $\gamma^D_{s:U}$ coincides for each $s \in {\Bbb R}$
 with the law of ${\cal A}^{[\beta]}_D$ conditioned on $\{ {\cal A}^{[\beta]}_D
 \cap U = \emptyset \}.$ We put ${\cal A}^{[\beta]}_{D:U} := \gamma^D_{0:U}.$
 In full analogy with the similar discussion above, the conditioned field ${\cal A}^{[\beta]}_{{\Bbb
 R}^2:U}$ is stochastically bounded by the Poisson contour process
 ${\cal P}_{\Theta^{[\beta]}:U} := \{ \theta \in {\cal P}_{\Theta^{[\beta]}}
 \;|\; \theta \cap U = \emptyset \}.$ Likewise, ${\cal A}^{[\beta]}_{D:U}$ is stochastically
 bounded by ${\cal P}_{\Theta^{[\beta]}_D :U} := \{ \theta \in {\cal P}_{\Theta^{[\beta]}_D}\;|\;
 \theta \cap U = \emptyset \}.$

\paragraph{Cut-off ensembles}
 An important family of processes we embed into the original graphical
 construction are the cut-off ensembles for ${\cal A}^{[\beta]}.$ They are
 defined as follows. For a positive {\it cut-off threshold} $\alpha$ and a
 bounded region $V \subseteq {\Bbb R}^2$ we consider the measure
 $\Theta^{[\beta];\alpha,V}$ which is the restriction of
 $\Theta^{[\beta]}$ to the family of polygonal contours which
 either do not hit $V,$ or if they do hit $V$ then their diameter does not
 exceed $\alpha.$ In this context, it is convenient to say that a contour
 $\gamma$ is $\alpha$-{\it large} iff $\diam(\gamma) > \alpha$ and
 that it is $\alpha$-{\it small} otherwise.  
 Using $\Theta^{[\beta];\alpha,V}$ instead of
 $\Theta^{[\beta]}$ for the contour birth intensity in the graphical
 construction we obtain $\alpha$-cut-off version $(\gamma^{{\Bbb R}^2;\alpha,V}_s)_{s \in {\Bbb R}}$
 of the process $(\gamma_s)_{s \in {\Bbb R}}$ (equivalently, we can simply reject all
 $\alpha$-large contours hitting $V$ upon their birth in the
 original graphical construction, which naturally extends the canonical coupling). Put
 ${\cal A}^{[\beta];\alpha,V} :=
 \gamma^{{\Bbb R}^2;\alpha,V}_0.$ It is easily seen that the $\alpha$-cut-off
 polygonal field ${\cal A}^{[\beta];\alpha,V} := \gamma^{{\Bbb R}^2;\alpha,V}_0$
 coincides in law with ${\cal A}^{[\beta]}$ conditioned
 on the event that no contour hitting $V$ has its diameter larger than $\alpha.$
 Likewise, we consider with obvious definition
 the finite volume cut-off processes $(\gamma^{D;\alpha,V}_s)_{s \in {\Bbb R}}$
 for open and bounded $D$ with piecewise smooth boundary. Clearly, the finite
 volume $\alpha$-cut-off polygonal field ${\cal A}^{[\beta];\alpha,V}_D
 := \gamma^{D;\alpha,V}_0$ arises as ${\cal A}^{[\beta]}_D$ conditioned
 on the event that no contour hitting $V$ is $\alpha$-large.
 In analogy with the similar discussion above, the cut-off field
 ${\cal A}^{[\beta];\alpha,V}$ is stochastically dominated by
 ${\cal P}_{\Theta^{[\beta];\alpha,V}}$ and
 ${\cal A}^{[\beta];\alpha,V}_D$ is stochastically dominated by
 ${\cal P}_{\Theta_D^{[\beta];\alpha,V}}.$

 Clearly, we can combine the cut-off operation with imposing a forbidden region
 which leads to processes $\gamma^{{\Bbb R}^2;\alpha,V}_{s:U}, \gamma^{D;\alpha,V}_{s:U},
 {\cal A}^{[\beta];\alpha,V}_{{\Bbb R}^2:U}$ and ${\cal A}^{[\beta];\alpha,V}_{D:U}$
 with obvious definitions, stochastically dominated by
 ${\cal P}_{\Theta^{[\beta];\alpha,V}:U}$
 and ${\cal P}_{\Theta^{[\beta];\alpha,V}_{D}:U}$ respectively.
 The canonical coupling is extended in the obvious way. 

% Dotad 7 wrzesnia wieczorem 

\paragraph{Area-interacting fields}
 The final modification considered involves introducing an area-order term
 to the Hamiltonian of (\ref{GREPRBETA}). To this end, for a bounded region
 $W \subseteq {\Bbb R}^2$ and $h \in {\Bbb R}$ we consider the polygonal
 field ${\cal A}^{[\beta,h]}_W$ on $W$, given in distribution by
 $$ \frac{d{\cal L}({\cal A}^{[\beta,h]}_W)}{d{\cal L}({\cal A}^{[\beta]} \cap W)}[\gamma]
 = \frac{\exp(h \blk_W(\gamma))}{{\Bbb E} \exp(h \blk_W({\cal A}^{[\beta]}))}.
 $$
 Note that, unlike ${\cal A}^{[\beta]}_D,$ the field ${\cal A}^{[\beta,h]}_W$
 is defined as a Gibbsian modification of the thermodynamic limit ${\cal A}^{[\beta]}$
 restricted to $W$ rather than as a Gibbsian modification of the finite volume field
 ${\cal A}_W.$ In particular, the laws of the fields ${\cal A}^{[\beta,0]}_W$
 and ${\cal A}^{[\beta]}_W$ do not coincide; in fact ${\cal A}^{[\beta,0]}_W$
 coincides in distribution with ${\cal A}^{[\beta]} \cap W.$ 
 We will mainly use the area-interacting modification combined
 with the cut-off operation. The field ${\cal A}^{[\beta,h];\alpha,V}_W$ is
 given in law by
 \begin{equation}\label{ODDZPOL}
  \frac{d{\cal L}({\cal A}^{[\beta,h];\alpha,V}_W)}{d{\cal L}({\cal A}^{[\beta];\alpha,V} \cap W)}
  [\gamma] = \frac{\exp(h \blk_W(\gamma))}{{\Bbb E}\exp(h \blk_W({\cal A}^{[\beta];\alpha,V}))}.
 \end{equation}
 To proceed with the graphical construction we assume that
 \begin{equation}\label{KGDLAPOL}
  |h| \leq \frac{\beta}{\pi \alpha}
 \end{equation}
 and observe that adding a single $\alpha$-small contour $\theta$ to
 a contour configuration $\gamma,\; \gamma \cap \theta = \emptyset,$
 can change the {\it magnetisation} $\blk_W(\gamma)$ by at most
 $\pi \lgth(\theta)^2 \slash 2$ whence the value of $h \blk_W(\gamma)$
 can change by at most $\beta \lgth(\theta) \slash 2.$ With $\gamma$ standing for
 the current contour configuration, we modify the
 original graphical construction by
 \begin{itemize}
  \item constructing the free birth and death process $\hat{\varrho}_s,\; s \in {\Bbb R},$
        with birth intensity measure
        $\Theta^{[\beta \slash 2];\alpha,V}$ and death
        intensity $1,$
  \item at the {\it trimming stage}, by accepting a time-space contour instance $\theta
        \times [s_0,s_1)$
        \begin{itemize}
         \item with probability $0$ if $\theta$ hits $\theta'$ for some previously
               accepted contour instance $\theta' \times [s_0',s_1')$ alive at time
               $s_0,$
         \item with probability $\exp\left(-\frac{\beta}{2} \lgth(\theta) +
                                 h[\blk_W(\gamma \cup \theta)- \blk_W(\gamma)]\right)$
               otherwise.
        \end{itemize}
 \end{itemize}
 Observe that the last probability falls into $(0,1]$ because of (\ref{KGDLAPOL}).
 Denote the resulting {\it trimmed} process by $\hat{\gamma}_s.$ The validity of
 this construction requires a justification. In fact, we have to redefine here
 the notion of an ancestor clan. We set a directed connection from a contour
 instance $\theta^* = \theta \times [s_0,s_1)$ to all contour instances
 ${\theta'}^* = \theta' \times [s_0',s_1')$
 such that $\innt \theta \cap \innt \theta' \neq \emptyset$
 (which is weaker than the condition $\theta \cap \theta' \neq \emptyset$
  of the original definition) and $s_0' \leq s_0, s_1' > s_0.$ Clearly, these are all
 contour instances which may affect the acceptance status of $\theta^*.$
 The union of all the directed chains stemming from $\theta^*$ is called
 the ancestor clan of $\theta^*$ and denoted by $\hat{\An}(\theta^*).$
 Likewise, for $s \in {\Bbb R}$ and $W \subseteq {\Bbb R}^2$ we write
 $\hat{\An}_s(U)$ for the union of all the ancestor chains of contour
 instances $\theta \times [s_0,s_1)$ alive at time $s$ (i.e. $s_0 \leq s < s_1$)
 and such that $\innt \theta \cap U \neq \emptyset.$ In full analogy with (\ref{ZANIKGR}),
 Lemma \ref{NICHOLLS} guarantees that for $\beta$ large enough (larger than some
 $\hat{\beta}_g$) we have
 \begin{equation}\label{ZANIKGRPOL}
  {\Bbb P}(\diam \hat{\An}_s({\Bbb B}_2(x,1)) > R) \leq \exp(- c R),\; s \in {\Bbb R},\; x \in {\Bbb R}^2,
 \end{equation}
 for some $c = c(\beta) > 0.$ Clearly, this implies that the ancestor clans are a.s.
 finite, thus ensuring the validity of the construction. In the sequel we shall
 always assume that $\beta > \hat{\beta}_g$ so that (\ref{ZANIKGRPOL}) holds.
 It follows by the general theory developed by F\'ernandez, Ferrari \& Garcia
 \cite{FFG1},\cite{FFG2},\cite{FFG3},
 that so constructed $\hat{\gamma}_s$ for each fixed $s$ coincides in law with
 ${\cal A}^{[\beta,h];\alpha,V}_W.$ Moreover, it is easily seen that, for each
 $s \in {\Bbb R},$ $\hat{\varrho}_s$ coincides in law with the Poisson contour
 process ${\cal P}_{\Theta^{[\beta \slash 2];\alpha,V}} \cap W.$ Consequently, the
 almost sure inclusion $\hat{\gamma}_s \subseteq \hat{\varrho}_s$ yields the
 stochastic domination of ${\cal A}^{[\beta,h];\alpha,V}_W$ by
 ${\cal P}_{\Theta^{[\beta \slash 2];\alpha,V}} \cap W.$

 Clearly, the above construction can be easily extended to take into account
 forbidden regions. For bounded measurable $U \subseteq {\Bbb R}^2$ denote by
 ${\cal A}^{[\beta,h];\alpha,V}_{W:U}$ the polygonal fields arising by
 conditioning ${\cal A}^{[\beta,h];\alpha,V}_{W}$ on none of its contours
 hitting $U.$ It is easily seen that ${\cal A}^{[\beta,h];\alpha,V}_{W:U}$
 can be represented by the graphical construction of this paragraph, with
 the additional rule that all contours hitting $U$ be immediately discarded.
 In analogy with a similar observation made above for ${\cal A}^{[\beta];\alpha,V}_{W},$
 also here it should be noted that ${\cal A}^{[\beta,0];\alpha,V}_{W:U}$
 coincides in law with ${\cal A}^{[\beta];\alpha,V}_{{\Bbb R}^2:U} \cap W$
 rather than with ${\cal A}^{[\beta];\alpha,V}_{W:U}.$

 Moreover, in full analogy
 with the argument above, we see that ${\cal A}^{[\beta];\alpha,V}_{W:U}$ is
 stochastically dominated in the sense of inclusion by the Poisson contour
 process ${\cal P}_{\Theta^{[\beta \slash 2];\alpha,V}:U} \cap W.$

 Note that the above construction provides a natural coupling for 
 area-interacting fields with cut-off and (possibly) forbidden 
 regions imposed, under the constraint (\ref{KGDLAPOL}). To
 distinguish it from the canonical coupling available for fields
 with no area interaction as discussed above, we shall call this
 coupling the {\it canonical coupling for area-interacting fields}.  

% Dotad 9 wrzesnia rano

\subsection{Surface tension}\label{NAPPOWSE}
 The purpose of this section is to define the surface tension functional specific for
 our model.
 To this end, for a given bounded and convex domain $D \subseteq {\Bbb R}^2$
 and $\delta > 0$ we consider the family ${\cal C}_D^{x \leftrightarrow y; \delta}$
 of self-avoiding polygonal paths in $D$ connecting the balls
 ${\Bbb B}_2(x,\delta) \subseteq D$ and ${\Bbb B}_2(y,\delta) \subseteq D,$
 with the additional property that the first and last segments of the paths
 do not intersect the interiors of the balls ${\Bbb B}_2(x,\delta)$ and
 ${\Bbb B}_2(y,\delta)$ respectively, but they do touch their respective
 boundaries and the intersection points coincide with the initial and
 final point of the path. In other words, moving along a path in 
 ${\cal C}_D^{x \leftrightarrow y; \delta}$ we travel from
 $\partial {\Bbb B}_2(x,\delta)$ to $\partial {\Bbb B}_2(y,\delta),$
 with the initial segment falling outside ${\Bbb B}_2(x,\delta)$ and
 with the final segment outside ${\Bbb B}_2(y,\delta),$ which does
 not prevent us though from passing through ${\Bbb B}_2(x,\delta)$
 and ${\Bbb B}_2(y,\delta)$ along the remaining segments. Next,
 we introduce on ${\cal C}_D^{x \leftrightarrow y; \delta}$ the free
 measure $\Theta_{D}^{x \leftrightarrow y;\delta},$ constructed in
 full analogy with the definition of the free contour measure as
 given in (\ref{WOLNEKONTURY}).
 For a finite configuration $(l)$ of straight lines
 crossing $D$ write ${\cal C}_D^{x \leftrightarrow y;\delta}(l)$ for the collection
 of those paths in ${\cal C}_D^{x \leftrightarrow y;\delta}$ which only contain
 segments of the lines in $(l)$ and exactly one non-zero length segment on
 each line. For measurable $C \subseteq {\cal C}_D^{x\leftrightarrow y;\delta}$
 we put
 \begin{equation}\label{WOLNESCIEZKI}
   \Theta^{x\leftrightarrow y;\delta}_D(C) =
   \int_{{\rm Fin}(L[D])}
    \sum_{\theta \in C \cap {\cal C}^{x\leftrightarrow y; \delta}_D(l)}
    \exp(-2\lgth(\theta)) d\mu^*((l))
 \end{equation}
 with ${\rm Fin}(L[D])$ and $\mu^*$ as in (\ref{WOLNEKONTURY}).
 Note that the initial point and the endpoint of the path $\theta$
 in the above definition are uniquely determined, respectively as
 the intersection of the first and last segment of the path with
 $\partial {\Bbb B}_2(x,\delta)$ and $\partial {\Bbb B}_2(y,\delta).$
 Likewise, we define the $\beta$-tilted measures
 $[\Theta_D^{x\leftrightarrow y;\delta}]^{[\beta]}$ by
 \begin{equation}\label{WOLNESCIEZKIB}
  [\Theta_D^{x\leftrightarrow y;\delta}]^{[\beta]}(d\theta)
  := \exp(-\beta \lgth(\theta))\Theta_D^{x\leftrightarrow y;\delta}(d\theta).
 \end{equation}
 As observed above for the free contour measures, also the path measures
 are consistent in that $\Theta^{x\leftrightarrow y;\delta}_D
 = [\Theta^{x\leftrightarrow y;\delta}_{D'}]_{|D}$ for $D \subseteq D'$
 and, consequently, we can construct the whole plane free measure
 $\Theta^{x \leftrightarrow y;\delta}$ and its tilted version
 $[\Theta^{x \leftrightarrow y;\delta}]^{[\beta]},$ both defined
 on ${\cal C}^{x \leftrightarrow y;\delta} =
 {\cal C}^{x \leftrightarrow y;\delta}_{{\Bbb R}^2}
 := \bigcup_{n=1}^{\infty} {\cal C}^{x \leftrightarrow y;\delta}_{(-n,n)^2}.$

 To proceed, write
 \begin{equation}\label{T1}
  T^{[\beta]}_{(\delta)}[x \leftrightarrow y] :=
  \int_{{\cal C}^{x \leftrightarrow y;\delta}} {\Bbb P}({\cal A}^{[\beta]} \cap \theta = \emptyset)
  [\Theta^{x \leftrightarrow y;\delta}]^{[\beta]}(d\theta).
 \end{equation}
 Put ${\bf e}_x := (1,0),$ fix some small $\delta > 0$ and let
 \begin{equation}\label{PRZYBLNAP}
   \tau^{[\beta]}_{\lambda} := - \frac{1}{\lambda} \log T^{[\beta]}_{(\delta)}[0 \leftrightarrow \lambda
   {\bf e}_x].
 \end{equation}
 The surface tension is defined as the limit
 \begin{equation}\label{NAPPOW}
  \tau^{[\beta]} := \lim_{\lambda\to\infty} \tau^{[\beta]}_{\lambda} =
     - \lim_{\lambda\to\infty} \frac{1}{\lambda} \log T^{[\beta]}_{(\delta)}[0 \leftrightarrow \lambda
    {\bf e}_x].
 \end{equation}
 It is clear that the asymptotic behaviour of $\tau^{[\beta]}_{\lambda}$ as $\lambda \to \infty$
 does not depend on the choice of $\delta$ above \--- indeed, changing $\delta$ is easily
 seen to result only in a bounded and uniformly non-zero prefactor before
 $T^{[\beta]}_{(\delta)}[0 \leftrightarrow \lambda {\bf e}_x],$ which is negligible in
 the logarithmic large $\lambda$ asymptotics.
 This is why our notation does not take into account the dependency of
 $\tau^{[\beta]}_{\lambda}$ on $\delta.$ The existence, finiteness and 
 strict positivity of the limit in (\ref{NAPPOW}) for $\beta$ large enough
 and other properties of the surface tension are discussed in 
 Section \ref{SEKCJANAPPOW} below, see Lemma \ref{NPISTN} there. 

\subsection{Main results}\label{MARE}
 % Dla bialej granicy termodynamicznej !
 % Tutaj o malych i duzych konturach, o pojeciu fazy ! 
 
 Our main result below states that, at low enough temperatures, conditioning
 the white-dominated phase of polygonal Markov field to contain black-coloured 
 regions of total area exceeding its expectation by an area-order quantity
 results in aggregation of the excess black area and in formation of a
 macroscopic-size disk-shaped region (Wulff crystal) of black-dominated
 phase, separated from the outside white phase by a single large contour. 
 Moreover, the probability of such area-order black exceedances exhibits
 perimeter-order exponential decay. 
 
 As shown in Section \ref{EXPTI} below, 
 for $\alpha \gg \log L$ with overwhelming probability there are 
 no $\alpha$-large contours of ${\cal A}^{[\beta]}$ in ${\Bbb B}_2(L).$
 Thus, it is natural to consider the regions separated by
 $\Omega(\log L)$-large contours of ${\cal A}^{[\beta]}$ and to assign
 them, in the obvious way, black or white {\it phase labels}.
 In this language, we show in this paper that the single large contour
 determining the Wulff shape encloses a disk-shaped portion of
 black-labeled phase region surrounded by {\it ocean} of white-labeled
 phase. 

 As already discussed in the introductory section, since our main results are
 formulated directly under boundary conditions induced by the thermodynamic limit
 ${\cal A}^{[\beta]}$ rather than with periodic or monochromatic boundary conditions, we have
 to explicitly rule out the situation where the phase separating curves cross or go along
 the boundary of the considered finite volume region ${\Bbb B}_2(L).$ To this end,
 for $\alpha > 0$ shall write ${\cal N}[\alpha,L]$ for the event
 that no $\alpha$-large contour of ${\cal A}^{[\beta]}$ gets
 closer than at the distance $6\alpha$ to the circle ${\Bbb S}_1(L) :=
 \partial {\Bbb B}_2(L).$ In what follows we shall write
 \begin{equation}\label{ALFAZWYKLE}
  \alpha[L] := \sqrt{L} \log L.
 \end{equation}

 Our main result is the following theorem.
 \begin{theorem}\label{GLOWNE}
  For $0 < a < 2\pi |\blk[\beta]|$ we have
  \begin{equation}\label{PSTWAA}
   {\Bbb P}\left( \blk_L\left( {\cal A}^{[\beta]} \right) \geq \blk[\beta] \pi L^2 + a L^2,\;
                  {\cal N}[\alpha[L],L]\; {\rm holds} \right)
     =
    \exp\left( -  \sqrt{\frac{2\pi a}{|\blk[\beta]|}} L
                     \tau^{[\beta]}_{\alpha[L]} + O(\alpha[L]) \right)
  \end{equation}
  $$ = \exp\left( - \sqrt{\frac{2\pi a}{|\blk[\beta]|}} L \tau^{[\beta]} + o(L) \right). $$
  Moreover, there exists a constant $C_{\rm large}$ such that on the event
  $$\left\{ \blk_L \left( {\cal A}^{[\beta]} \right) \geq \blk[\beta] \pi L^2 + a L^2,\;
  {\cal N}[\alpha[L],L] \; {\rm holds} \right\},$$
  for sufficiently large $L$ we have with probability arbitrarily close to $1$
  \begin{itemize}
   \item There is exactly one $C_{\rm large} \log L$-large contour
         $\theta_{\rm large},$
   \item This $\theta_{\rm large}$ satisfies
         $$ \min_x \rho_H\left( \theta_{\rm large},
            {\Bbb S}_1\left(x,L\sqrt{\frac{a}{2\pi |\blk[\beta]|}}\right)
            \right) = O\left(L^{3\slash 4} \sqrt{\log L}\right), $$
         with $\rho_H$ standing for the usual Hausdorff distance.
 \end{itemize}
\end{theorem}
 Note that in the sequel we shall refer to the condition 
 $\blk_L \left( {\cal A}^{[\beta]} \right) \geq \blk[\beta] \pi L^2 + a L^2$
 as to the {\it micro-canonical constraint}. 

 The remaining part of the paper is the proof of Theorem \ref{GLOWNE}
 and is organised as follows. In Section \ref{EXPTI} below
 we establish upper bounds on occurrence probabilities of large contours. Next,
 in Section \ref{MDCE} we study moderate deviation probabilities for cut-off 
 contour ensembles of polygonal fields. Section \ref{DECRE} provides a simple
 yet important lemma allowing us to factorise the avoidance probabilities of
 ${\cal A}^{[\beta]}$ over disjoint regions. This is followed by Section 
 \ref{SEKCJANAPPOW} dealing with properties of the surface tension, and 
 then by Section \ref{SZKIEE}, where we establish coarse-graining 
 skeleton estimates. The complementary lower bounds for occurrence
 probabilities of large contours are stated in Section \ref{LOBO}. 
 Finally, in Section \ref{GLOTWI} we complete the proof of Theorem
 \ref{GLOWNE} by putting together the results of previous sections.   

\section{Exponential tightness bounds}\label{EXPTI}
 In this section we show that although the total length of the contour ensemble
 ${\cal A}^{[\beta]} \cap {\Bbb B}_2(L)$ is clearly of the area order $\Theta(L^2),$
 this is due to the contributions of $O(\log L)$-small contours, while the
 contribution of $\Omega(\log L)$-large contours is of order $O(1)$ with
 the corresponding large deviation probabilities exhibiting exponential decay.
 To put it in formal terms, with $\alpha > 0$ not necessarily given by 
 (\ref{ALFAZWYKLE}),  write ${\Bbb L}_{\alpha,L} :=
 {\Bbb L}_{\alpha,L}\left[{\cal A}^{[\beta]}\right]$ for the family of
 $\alpha$-large contours of ${\cal A}^{[\beta]}$ hitting ${\Bbb B}_2(L)$
 and, in general, let ${\Bbb L}_{\alpha,L}[\gamma]$ stand for the 
 family of $\alpha$-large contours of a contour collection $\gamma$
 which hit ${\Bbb B}_2(L).$  
 We claim that
 \begin{lemma}\label{DLG}
  For each $\kappa < \beta \slash 2 - 2$ there exist $M,
  C = C(\beta,\kappa) < \infty$ such
  that for all $\alpha > C \log L$ and $\lambda > 0$ we have
  $$ {\Bbb P}\left(\lgth({\Bbb L}_{\alpha,L}) > \lambda\right) \leq M \exp(-\kappa \lambda) $$
  and the same applies for ${\Bbb L}_{\alpha,L}$ replaced with
  ${\Bbb L}_{\alpha,L}[{\cal A}^{[\beta];\alpha,\cdot}_{\cdot}], 
   {\Bbb L}_{\alpha,L}[{\cal A}^{[\beta];\alpha,(\cdot)}_{(\cdot):(\cdot)}]$
  and 
  ${\Bbb L}_{\alpha,L}[{\cal A}^{[\beta,h];\alpha,(\cdot)}_{(\cdot)}],$ $
   {\Bbb L}_{\alpha,L}[{\cal A}^{[\beta,h];\alpha,(\cdot)}_{(\cdot):(\cdot)}]$
  for $h$ within the validity range of (\ref{KGDLAPOL}). 
 \end{lemma}
  Note that it is natural to regard this lemma as an exponential tightness
  statement for $\lgth({\Bbb L}_{\alpha,L}),$ whence the title of the section.
 % Even though Lemma \ref{DLG} is formulated for the polygonal field
 % ${\cal A}^{[\beta]},$ it is clear from the proof below that it is valid
 % as well for all the modified fields obtained from the variants of the
 % graphical construction discussed in Subsubsection \ref{MODGK} and
 % admitting stochatically dominating Poisson contour processes.

 \paragraph{Proof}
  We provide the proof for the polygonal field ${\cal A}^{[\beta]}$ only,
  since the argument goes exactly along the same lines for all the modified
  fields obtained from the variants of the graphical construction
  discussed in Subsubsection \ref{MODGK} and admitting stochastically
  dominating Poisson contour processes. Note that the assumption
  $\kappa < \beta \slash 2-2$ was imposed for the purpose of dealing 
  with area-interacting processes with the external field $h$ within the
  validity range of (\ref{KGDLAPOL}), which admit stochastic domination
  by the Poisson contour process ${\cal P}_{\Theta^{[\beta \slash 2]}}.$
  For the remaining polygonal fields considered in the statement of
  the lemma, with no area interaction, a stronger stochastic domination 
  by ${\cal P}_{\Theta^{[\beta]}}$ is available and the assertion
  of the lemma still holds if we choose $\kappa < \beta - 2$ rather
  than $\kappa < \beta \slash 2-2.$ 

  To proceed, use the graphical construction to conclude that the total length of contours
  in ${\Bbb L}_{\alpha,L}$ is stochastically bounded by the total length of
  $\alpha$-large contours of ${\cal P}_{\Theta^{[\beta \slash 2]}}$ hitting ${\Bbb L}_{\alpha,L}.$
  Thus, by the definition of a Poisson point process, 
  $$ {\Bbb E}\exp\left(\kappa \lgth({\Bbb L}_{\alpha,L}) \right) \leq
     \exp \left[ \int_{\{ \theta \in {\cal C}\;|\; \theta \cap {\Bbb B}_2(L) \neq \emptyset,\;
     \theta \mbox{ is $\alpha$-large } \}} (\e^{\kappa \lgth(\theta)} - 1) d \Theta^{[\beta \slash 2]} 
     (\theta) \right] \leq $$
  $$ \e^{\kappa \alpha} \zeta(\alpha) + \kappa \int_{\alpha}^{\infty} \e^{\kappa \lambda} \zeta(\lambda)
     d\lambda, $$
  where the last inequality follows by integration by parts with 
  $$ \zeta(\lambda) := \Theta^{[\beta \slash 2]}(\{ \theta \in {\cal C}\;|\; \theta \cap {\Bbb B}_2(L)
     \neq \emptyset,\; \lgth(\theta) > \lambda \}). $$  
  In view of Lemma \ref{NICHOLLS} this means that
  $$ {\Bbb E}\exp\left(\kappa \lgth({\Bbb L}_{\alpha,L}) \right)
     \leq \exp\left[C_1(\beta,\kappa) L^2 \exp([\kappa+2-\frac{\beta}{2}] \alpha)\right] $$
  with some constant $C_1(\beta,\kappa).$ Thus, using Markov inequality we get
  $$ {\Bbb P}(\lgth({\Bbb L}_{\alpha,L}) > \lambda) \leq
     \frac{{\Bbb E}\exp\left(\kappa \lgth({\Bbb L}_{\alpha,L})\right)}{\exp(\kappa \lambda)}
     \leq \exp\left[C_1(\beta,\kappa) L^2 \exp([\kappa+2-\frac{\beta}{2}] \alpha)\right] \e^{-\kappa \lambda} $$
  which completes the proof of the lemma for $\alpha > C \log L$ with large enough
  $C = C(\beta,\kappa).$ $\Box$

\section{Moderate deviations for cut-off ensembles}\label{MDCE}
 The current section deals with the properties of the cut-off ensembles ${\cal A}^{[\beta];\alpha,V}$
 arising by conditioning the original field ${\cal A}^{[\beta]}$ on containing no $\alpha$-large
 contours hitting $V \subseteq {\Bbb R}^2.$ Recall that we assume here that $\beta > \beta_g$
 and $\beta > \hat{\beta}_g$ so that $\beta$ falls into the validity regions of the graphical
 construction discussed in Subsection \ref{KONSGRAF} as well as of its area-interacting
 modification discussed in Subsubsection \ref{MODGK}. We consider $\alpha$ not
 necessarily given by (\ref{ALFAZWYKLE}). Our first observation is that imposing
 a cut-off does not change the expected magnetisation by too much
 \begin{equation}\label{PRAWIEROWNOSC}
  \left| \pi L^2 \blk[\beta] - {\Bbb E}\blk_L\left( {\cal A}^{[\beta];\alpha,
   {\Bbb B}_2(L)} \right) \right|
  = O(L^4 \exp(-c \alpha))
 \end{equation}
 with some $c > 0.$ Indeed, this follows by the fact that, in view  the
 stochastic domination of ${\cal A}^{[\beta]}$ by ${\cal P}_{\Theta^{[\beta]}}$
 and in view of Lemma \ref{NICHOLLS}, an
 $\alpha$-large contour shows up in ${\cal A}^{[\beta]} \cap {\Bbb B}_2(L)$ with
 probability $O(L^2 \exp(-c \alpha)),$ whence conditioning on the absence of this
 event can change the probabilities of other events by at most $O(L^2 \exp(-c \alpha)),$ 
 consequently the variational distance between the laws
 ${\cal L}({\cal A}^{[\beta]})$ and
 ${\cal L}({\cal A}^{[\beta];\alpha,{\Bbb B}_2(L)})$
 is of the same order $O(L^2 \exp(-c\alpha)).$ To get (\ref{PRAWIEROWNOSC}) it suffices
 now to observe that the magnetisation over ${\Bbb B}_2(L)$ is a.s. bounded
 in absolute value by $\pi L^2.$

 Another useful observation is that the impact of imposing a forbidden region for
 cut-off ensembles can also be very well controlled. In formal terms, we claim
 that for a collection $\gamma$ of $\alpha$-large contours, $\alpha > 1,$ in
 ${\Bbb B}_2(L)$ we have
 \begin{equation}\label{PRZYBLOBC}
  \left| {\Bbb E}\blk_L\left({\cal A}^{[\beta];\alpha,{\Bbb B}_2(L)}\right) -
         {\Bbb E}\blk_L\left({\cal A}^{[\beta];\alpha,{\Bbb B}_2(L)}_{{\Bbb R}^2:\gamma}\right) \right|
   = O\left( \Area(\gamma \oplus {\Bbb B}_2(1))\right) = O\left(\lgth(\gamma)\right).
 \end{equation}
 This is an immediate consequence of the fact that, by (\ref{ZANIKGR}), under the canonical coupling of
 ${\cal A}^{[\beta];\alpha,{\Bbb B}_2(L)}$ and
 ${\cal A}^{[\beta];\alpha,{\Bbb B}_2(L)}_{{\Bbb R}^2:\gamma}$
 the probability that the colours assigned to a given point $x$ by these ensembles differ, is of
 order $O(\exp(- c \dist(x,\gamma))),\; c > 0.$

 The argument leading to (\ref{PRAWIEROWNOSC}) and (\ref{PRZYBLOBC}) above can
 be easily modified to yield the following combination of these relations. 
 Let $\gamma$ be a collection of $\alpha$-large contours in ${\Bbb B}_2(L).$
 Then 
 \begin{equation}\label{RATUJACE}
  \left| {\Bbb E}\blk_L\left({\cal A}^{[\beta];\alpha,{\Bbb B}_2(L)}_{{\Bbb R}^2;\gamma} \cup \gamma \right)
         - |\blk[\beta]| \blk_L(\gamma) \right| = O(\lgth(\gamma))
 \end{equation}
 provided $\alpha \geq C \log L$ for sufficiently large $C.$ 

 The main result of this section is the following moderate deviation bound
\begin{theorem}\label{SREDNIEO}
 For each $\beta$ large enough there exists a positive constant $C_1 = C_1(\beta)$
 such that, uniformly in $L, \alpha \geq  C_1 \log L$ and in finite
 collections $\gamma$ of polygonal contours in ${\Bbb R}^2$ we have for all
 $0 < A \leq C_1^{-1} L^2 \slash \log L$
 $$ {\Bbb P}\left( \left| 
    \blk_L\left({\cal A}^{[\beta];\alpha,{\Bbb B}_2(L)}_{{\Bbb R}^2:\gamma} \cup \gamma \right) -
    {\Bbb E}\blk_L\left({\cal A}^{[\beta];\alpha,{\Bbb B}_2(L)}_{{\Bbb R}^2:\gamma} \cup \gamma
    \right) \right| >  A \right)
    \leq \exp \left( - c \left[ \frac{A^2}{L^2} \wedge \frac{A}{\alpha} \right]\right) $$
 with some constant $c > 0.$
\end{theorem}

\paragraph{Proof}
 Write
 $$ \mu_{L,\gamma}^{\alpha} :=
    \blk_L\left({\cal A}^{[\beta];\alpha,{\Bbb B}_2(L)}_{{\Bbb R}^2:\gamma}
    \cup \gamma \right) $$
 and let
 $$ \mu_L^{\alpha} = \mu_{L,\emptyset}^{\alpha} = 
    \blk_L\left({\cal A}^{[\beta];\alpha,{\Bbb B}_2(L)}\right),\;
    \mu_L = \mu_L^{\infty} = \blk_L\left({\cal A}^{[\beta]}\right). $$
 For $h \in {\Bbb R}$ consider the partition function
 $$ Z[h] := {\Bbb E}\exp\left(h \mu_{L,\gamma}^{\alpha} \right) $$
 The following estimate, valid for all $h$ satisfying (\ref{KGDLAPOL}), is the crucial
 ingredient of our proof:
 \begin{equation}\label{OGRTRLA}
  \log Z[h] \leq h {\Bbb E} \mu_{L,\gamma}^{\alpha} + h^2 L^2 \sigma^2 \slash 2
 \end{equation}
 for some $\sigma > 0,$ uniformly in $L,\gamma,\alpha$ and $h$ within the validity
 region of (\ref{KGDLAPOL}). To see that (\ref{OGRTRLA}) suffices to complete the proof
 of the theorem, take first $0 < A \leq \frac{\beta L^2 \sigma^2}{\pi^2 \alpha},$ put
 $h := \frac{A}{\sigma^2 L^2}$ which clearly satisfies (\ref{KGDLAPOL}), and
 then use Markov's inequality to conclude that
 \begin{equation}\label{A1}
   {\Bbb P}\left( \mu^{\alpha}_{L,\gamma} > {\Bbb E}\mu^{\alpha}_{L,\gamma} +
    A \right)
    \leq \frac{Z[h]}{\exp(h({\Bbb E}\mu^{\alpha}_{L,\gamma}+A))} \leq
    \exp(h^2 L^2\sigma^2 \slash 2 - Ah) = \exp\left( - \frac{A^2}{2\sigma^2 L^2} \right).
 \end{equation}
 Next, for $A > \frac{\beta L^2 \sigma^2}{\pi^2 \alpha}$ choose
 $\kappa < \beta \slash 2 - 2$ and $C(\beta,\kappa)$ as in Lemma \ref{DLG}
 and assume that $C_1$ in the present lemma is chosen so that 
 $\tilde{\alpha} := 2 C(\beta,\kappa)\log L < \alpha.$ Then, on the event
 $\{ \mu^{\alpha}_{L,\gamma} > {\Bbb E}\mu^{\alpha}_{L,\gamma} + A \}$
 there are two possible scenarios:
 \begin{itemize}
 \item The total length of $\tilde{\alpha}$-large contours in 
  ${\Bbb L}_{\tilde{\alpha},L}\left({\cal A}^{[\beta];\alpha,{\Bbb B}_2(L)}_{{\Bbb R}^2:\gamma}\right)$
  exceeds $\frac{2A}{\pi A},$ which can happen with probability at most 
  $M \exp(-\frac{2\kappa A}{\pi \alpha})$ by the exponential tightness Lemma \ref{DLG},
 \item The total length of $\tilde{\alpha}$-large contours in    
 ${\Bbb L}_{\tilde{\alpha},L}\left({\cal A}^{[\beta];\alpha,{\Bbb B}_2(L)}_{{\Bbb R}^2:\gamma}\right)$
 does not exceed $\frac{2A}{\pi \alpha}.$ Since $\frac{8}{\pi \alpha}$ is the lower bound for the 
 length-to-enclosed-area ratio for an $\alpha$-small contour, this means in particular that 
 the total area enclosed by $\tilde{\alpha}$-large contours of 
 ${\cal A}^{[\beta];\alpha,{\Bbb B}_2(L)}_{{\Bbb R}^2:\gamma}$ falls below $A/4.$
 Denoting by $\tilde{\gamma}$ the family of contours 
 ${\Bbb L}_{\tilde{\alpha},L}\left({\cal A}^{[\beta];\alpha,{\Bbb B}_2(L)}_{{\Bbb R}^2:\gamma}\right)
 \cup \gamma$ and taking into account that the change of magnetisation induced by adding a contour
 is bounded in absolute value by twice the area it encloses, conditionally on given $\tilde{\gamma},$
 we are led to 
 $$ {\Bbb E}\mu^{\tilde{\alpha}}_{L,\tilde{\gamma}} \leq 
    {\Bbb E} {\Bbb M}_L
    \left({\cal A}^{[\beta];\tilde{\alpha},{\Bbb B}_2(L)}_{{\Bbb R}^2:\tilde{\gamma}} \cup \gamma \right)
    + A/2. $$
 Now, in full analogy with (\ref{PRZYBLOBC}), on the considered event we get
 $$ \left| {\Bbb E} {\Bbb M}_L
           \left({\cal A}^{[\beta];\tilde{\alpha},{\Bbb B}_2(L)}_{{\Bbb R}^2:\tilde{\gamma}} \cup 
           \gamma \right)
          - {\Bbb E}\mu^{\tilde{\alpha}}_{L,\gamma} \right| = O\left(\frac{2A}{\pi \alpha}\right). $$
 Next, in full analogy with (\ref{PRAWIEROWNOSC}), we have
 $$ \left| {\Bbb E}\mu^{\tilde{\alpha}}_{L,\gamma} - \mu^{\alpha}_{L,\gamma} \right| =
    O(L^4 \exp(-c \tilde{\alpha})) $$
 which goes to $0$ faster than the inverse of any polynomial under appropriate choice of $C(\beta,\kappa)$
 in Lemma \ref{DLG}. Putting the above relations together we conclude that 
 $$ {\Bbb E}\mu^{\tilde{\alpha}}_{L,\tilde{\gamma}} \leq
    {\Bbb E}\mu^{\alpha}_{L,\gamma} + A/2 (1+o(1)). $$
 Recalling that $\mu^{\alpha}_{L,\gamma}$ given $\tilde{\gamma}$ coincides in law with 
 $\mu^{\tilde{\alpha}}_{L,\tilde{\gamma}},$ we are led to
 $$ {\Bbb P}\left(\mu^{\alpha}_{L,\gamma} > {\Bbb E}^{\alpha}_{L,\gamma} | \tilde{\gamma} \right)
    \leq {\Bbb P}\left( \mu^{\tilde{\alpha}}_{L,\tilde{\gamma}} > 
    {\Bbb E}^{\tilde{\alpha}}_{L,\tilde{\gamma}} + A/2 (1-o(1)) \right). $$
 Now, choosing $C_1$ so that $A \leq \frac{\beta L^2 \sigma^2}{\pi^2 \tilde{\alpha}},$ we can bound 
 above the last probability by $\exp(-\frac{A^2}{8 \sigma^2 L^2})$ applying (\ref{A1}) with 
 $\alpha$ and $\gamma$ replaced there by $\tilde{\alpha}$ and $\tilde{\gamma}$ respectively.
\end{itemize}
 Combining the above two points with (\ref{A1}) and noting that 
 the probability of $\{ \mu^{\alpha}_{L,\gamma} < {\Bbb E}
 \mu^{\alpha}_{L,\gamma} - A \}$ can be dealt with in a
 completely analogous way, we obtain the assertion of the theorem. 

 It remains to verify (\ref{OGRTRLA}) for $h$ satisfying (\ref{KGDLAPOL}).
 We extend the notation by putting
 $$ \mu^{\alpha,h}_{L,\gamma} :=
    \blk_L\left({\cal A}^{[\beta,h];\alpha,{\Bbb B}_2(L)}_{
    {\Bbb B}_2(L):\gamma} \cup \gamma \right). $$
 Noting that
 \begin{equation}\label{PRZEDTAYLOR}
  \frac{\partial}{\partial h} \log Z[h] = {\Bbb E}\mu^{\alpha,h}_{L,\gamma},\;
    \frac{\partial^2}{\partial h^2} \log Z[h] =
    \Var[\mu^{\alpha,h}_{L,\gamma}]
 \end{equation}
 and 
 Taylor expanding the logarithm of the partition function up to the second order term yields
 \begin{equation}\label{TAYLOR}
  \log Z[h] = h {\Bbb E}\mu^{\alpha}_{L,\gamma} + h^2 \frac{\Var[\mu^{\alpha,h^*}_{L,\gamma}]}{2}
  \end{equation}
 for some $h^*$ between $0$ and $h.$ We claim that, uniformly in $L,\alpha,\gamma$ and $h$
 satisfying (\ref{KGDLAPOL}), the variance $\Var[\mu^{\alpha,h}_{L,\gamma}]$
 is of the area order $O(L^2)$
 \begin{equation}\label{KUM2}
  \Var[\mu^{\alpha,h}_{L,\gamma}] = O(L^2),
 \end{equation}
 which, once established, will immediately yield the required relation (\ref{OGRTRLA})
 as a conclusion of (\ref{TAYLOR}). To prove (\ref{KUM2}) we show that, for $U_1,U_2
 \subseteq {\Bbb B}_2(L),$ uniformly in $\gamma$ and in $h$ satisfying (\ref{KGDLAPOL})
 $$ 
  \Cov\left[\blk_{U_1}({\cal A}^{[\beta,h];\alpha,{\Bbb B}_2(L)}_{{\Bbb B}_2(L):\gamma} \cup \gamma);
       \blk_{U_2}({\cal A}^{[\beta,h];\alpha,{\Bbb B}_2(L)}_{{\Bbb B}_2(L):\gamma} \cup \gamma)\right]
  = $$
 \begin{equation}\label{DOKUM2}
   O\left(\Area(U_1) \Area(U_2)[\Area(U_1 \oplus {\Bbb B}_2(1)) + \Area(U_2 \oplus {\Bbb B}_2(1))]
   \e^{-c \dist(U_1,U_2)}\right)
 \end{equation}
 for a positive constant $c,$ with $\oplus$ standing for the usual Minkowski 
 addition. Indeed, with the representation provided by the
 graphical construction for area-interacting fields in Subsubsection \ref{MODGK},
 conditionally on the event $\{ \hat{\An}_0(U_1) \subseteq U_1 \oplus {\Bbb
 B}_2(\frac{\dist(U_1,U_2)}{2}),\; \hat{\An}_0(U_2) \subseteq U_2 \oplus
 {\Bbb B}_2(\frac{\dist(U_1,U_2)}{2}) \}$ the random variables
 $\blk_{U_1}({\cal A}^{[\beta,h];\alpha,{\Bbb B}_2(L)}_{{\Bbb B}_2(L):\gamma} \cup \gamma)$ and
 $\blk_{U_2}({\cal A}^{[\beta,h];\alpha,{\Bbb B}_2(L)}_{{\Bbb B}_2(L):\gamma} \cup \gamma)$ are
 independent. But in view of (\ref{ZANIKGRPOL}) the probability of this
 event does not fall below $1-O([\Area(U_1 \oplus {\Bbb B}_2(1)) +
  \Area(U_2 \oplus {\Bbb B}_2(1))]\exp(-c \dist(U_1,U_2))).$ This observation
 combined with the fact that
 $|\blk_{U_i}({\cal A}^{[\beta,h];\alpha,{\Bbb B}_2(L)}_{{\Bbb B}_2(L):\gamma})| \leq
  \Area(U_i),\; i=1,2$ implies (\ref{DOKUM2}). The required relation (\ref{KUM2})
 follows now from (\ref{DOKUM2}) by usual argument based on splitting ${\Bbb B}_2(L)$
 into $\Theta(L^2)$ disjoint regions of diameter and area $\Theta(1)$ and then noting
 that, with the magnetisation contributions coming from distant regions exhibiting
 exponentially decaying covariances, the asymptotic order of the total magnetisation
 variance $\Var[\mu^{\alpha,h}_{L,\gamma}]$ is determined by the sum of covariances
 between regions within distance $\Theta(1)$ from each other, which yields the
 desired order $O(L^2).$ The proof is complete. $\Box$\\

 \begin{remark}\label{GAUSS}
  We note that the bounds in Theorem \ref{SREDNIEO} are of optimal
  order only for the probabilities of positive deviations $\{ \mu_{L,\gamma}^{\alpha}
  > {\Bbb E} \mu_{L,\gamma}^{\alpha} + A \},\; A>0.$ We believe that the probabilities
  of negative moderate deviations $\{ \mu_{L,\gamma}^{\alpha} < {\Bbb E}\mu^{\alpha}_{L,\gamma} - A \}$
  as well as $\{ \mu_{L} < {\Bbb E}\mu_L - A \},\; A \ll L^2,$ exhibit Gaussian-type decay
  $\exp(-\Omega(A^2 \slash L^2))$ as in classical moderate deviation regime,
  in full analogy with similar phenomenon for the Ising model, see \cite{DS1},
  (1.1.2), (2.3.2) in \cite{IS1} or Section III.C.1 in \cite{BIV} and the references therein.
  Since this falls beyond the context of our further argument,
  we do not discuss this issue in the present paper. 
 \end{remark} 

 As an easy corollary from Theorem \ref{SREDNIEO} we conclude that
 \begin{corollary}\label{DODAANE}
  With $A \geq 1,\alpha$ and $\gamma$ as in Theorem \ref{SREDNIEO}
  and with $C_1$ in Theorem \ref{SREDNIEO} large enough we have
  uniformly
   $$ {\Bbb P}\left( \left| 
    \blk_L\left({\cal A}^{[\beta]}_{{\Bbb R}^2:\gamma} \cup \gamma \right) -
    {\Bbb E}\blk_L\left({\cal A}^{[\beta]}_{{\Bbb R}^2:\gamma} \cup \gamma \right)
    \right| >  A \right)
    \leq \exp \left( - c \left[ \frac{A^2}{L^2} \wedge \frac{A}{\alpha} \right]\right)
    \vee O(L^2 \exp(-c \alpha)) $$
 with some constant $c > 0.$
 \end{corollary}
 \paragraph{Proof}
 This is a direct conclusion of Theorem \ref{SREDNIEO} combined
 with the observation that the variational distance between the laws
 ${\cal L}({\cal A}^{[\beta]}_{{\Bbb R}^2:\gamma} \cap {\Bbb B}_2(L))$ and
 ${\cal L}({\cal A}^{[\beta];\alpha,{\Bbb B}_2(L)}_{{\Bbb R}^2:\gamma} \cap {\Bbb B}_2(L))$
 is of order $O(L^2 \exp(-c\alpha)),$ in full analogy with the argument
 leading to (\ref{PRAWIEROWNOSC}) above. $\Box$\\

 Another useful corollary relies on a straightforward observation
 that the proof of Theorem \ref{SREDNIEO} applies with only minor
 modifications for ${\cal A}^{[\beta];\alpha,{\Bbb B}_2(L)}$
 replaced by ${\cal A}^{[\beta,h];\alpha,{\Bbb B}_2(L)}_{{\Bbb B}_2(L)}$
 with $|h| \leq H \slash \alpha,\; H$ small enough. In formal terms, 
 \begin{corollary}\label{SREDNIEODLAPOL}
  With $H > 0$ small enough, for each $\beta$ large enough there
  exists a positive constant $C = C(\beta,H)$ such that, uniformly in
  $L, \alpha \geq C \log L,$ finite collection $\gamma$ of polygonal
  contours in ${\Bbb R}^2$ and $|h| \leq H \slash \alpha,$ we have for
  all $0<A \leq C^{-1} L^2 \slash \log L$
  $$ {\Bbb P}\left( \left| 
     \blk_L\left({\cal A}^{[\beta,h];\alpha,{\Bbb B}_2(L)}_{{\Bbb B}^2(L):\gamma} \cup \gamma \right) -
     {\Bbb E}\blk_L\left({\cal A}^{[\beta,h];\alpha,
     {\Bbb B}_2(L)}_{{\Bbb B}^2(L):\gamma} \cup \gamma
     \right) \right| >  A \right)
     \leq \exp \left( - c \left[ \frac{A^2}{L^2} \wedge \frac{A}{\alpha} \right]\right) $$
  with some constant $c > 0.$
 \end{corollary}
 We omit the proof of this corollary which is just a simple repetition
 of the proof of Theorem \ref{SREDNIEO}.   

 Below, we provide some further auxiliary results related to
 moderate deviation probabilities for cut-off ensembles. 
 Note first that we can establish a bound analogous to (\ref{KUM2}) for
 the third cumulant of $\mu^{\alpha,h}_{L,\gamma}:$
 \begin{equation}\label{KUM3}
  \frac{\partial^3}{\partial h^3} \log Z[h] = O(L^2)
 \end{equation}
 uniformly in $\alpha,L,$ finite contour collection $\gamma$
 and $h$ satisfying (\ref{KGDLAPOL}).  We omit the details
 of the argument, based on the relation (\ref{ZANIKGRPOL}),
 since it goes along the same lines as the proof of Lemma
 5.3 in Baryshnikov \& Yukich \cite{BY}  (valid for arbitrary order
 cumulants in fact). In particular, in view of (\ref{PRZEDTAYLOR})
 combined with (\ref{KUM3}),
 we get for $h$ within the validity range of (\ref{KGDLAPOL})
 \begin{equation}\label{TAYLOR2}
  {\Bbb E}\mu^{\alpha,h}_{L,\gamma} = {\Bbb E}\mu^{\alpha}_{L,\gamma}
   + h \Var[\mu^{\alpha}_{L,\gamma}] + O(h^2 L^2)
 \end{equation} 
 uniformly in $L,\gamma.$
 To proceed, assume that $\gamma$ is a finite contour collection
 in ${\Bbb B}_2(L)$ with $\Area(\gamma \oplus {\Bbb B}_2(\log^2 L))
   \leq L^2 \slash \log L.$ We claim that under this condition
 we have (\ref{KUM2}) strengthened to 
 \begin{equation}\label{KUM2UZUP}
  \Var[\mu^{\alpha}_{L,\gamma}] = \Theta(L^2)
 \end{equation}
 uniformly in $\gamma, L.$ Indeed, observe that by (\ref{ZANIKGR})
 the probability of the event
 $\{ \An_0(\gamma) \not\subseteq \gamma \oplus {\Bbb B}_2(\log^2 L) \}$
 is of order at most  $O(L^2 \exp(-c \log^2 L)),\; c>0,$ whence under
 the canonical coupling with probability $1-O(L^2\exp(-c \log^2 L))$
 the field ${\cal A}^{[\beta];\alpha,{\Bbb B}_2(L)}_{{\Bbb R}^2:\gamma}$ 
 coincides with ${\cal A}^{[\beta];\alpha,{\Bbb B}_2(L)}$ over the whole
 complement of $\gamma \oplus {\Bbb B}_2(\log^2 L).$ Consequently,
 $$|\Var[\mu^{\alpha}_L]-\Var[\mu^{\alpha}_{L,\gamma}]| = o(L^2).$$
 Now, mimicking the proof of (\ref{PRAWIEROWNOSC}) we check that
 $$|\Var[\mu^{\alpha}_L]-\Var[\mu_L]| = O(L^6 \exp(-c \alpha)) =
  o(L^2)$$ provided $\alpha \geq C \log L$ with $C$ large enough.  
 This will yield the required relation (\ref{KUM2UZUP}) as soon
 as we show that for the field ${\cal A}^{[\beta]}$ the variance of
 magnetisation has the required order 
 \begin{equation}\label{WARI}
  \Var[\mu_L] = \Omega(L^2).
 \end{equation}
 To this end, we fix some large $\lambda > 0,$ large $k \in {\Bbb N}$
 and small $\epsilon > 0$ and we note that
 \begin{equation}\label{WARPOM}
 \inf \left\{ \Var[\blk_D({\cal A}^{[\beta]}_D)] \;|\;
       \Area(D) \geq \epsilon \lambda^2,\;
       \card \Ver(D) \leq k,\;
       \lgth(\partial D) \in [\epsilon \lambda,\epsilon^{-1}
                \lambda] \right\}  > 0,
 \end{equation}
 with the infimum taken over all bounded domains $D$ with polygonal
 boundary, possibly chopped off by intersecting with ${\Bbb B}_2(L).$
 Indeed, this can be proven by observing first that the
 mapping $D \mapsto \phi(D) := \Var[\blk_D({\cal A}^{[\beta]}_D)]$ admits
 only strictly positive values and it is continuous with respect
 to the pseudo-metric $\rho^*_H(D_1,D_2) := \inf_{x \in {\Bbb R}^2}
 \rho_H(D_1,x + D_2).$ Thus, putting ${\cal D}[\lambda,k,\epsilon] 
 := \{ D \;|\;\Area(D) \geq \epsilon \lambda^2,\; \card \Ver(D) \leq k,\;
       \lgth(\partial D) \in [\epsilon \lambda,\epsilon^{-1}
    \lambda] \}$ and noting that ${\cal D}[\lambda,k,\epsilon]$
 is compact in $\rho^*_H,$ we see that $\phi_0 := 
 \inf_{D \in {\cal D}[\lambda,k,\epsilon]} \phi(D) > 0,$ which yields
 the required relation (\ref{WARPOM}).  
 To proceed, note that the variance $\Var[\mu_L]$ in (\ref{WARI})
 is bounded below by the expectation of the conditional variance 
 of $\mu_L$ given the ensemble of external (outermost) contours
 $\theta$ in ${\cal A}^{[\beta]}
 \cap {\Bbb B}_2(L)$ satisfying the constraints of the infimum in
 (\ref{WARPOM}) for $D := \innt \theta.$ Thus, taking into account
 that given the presence of such $\theta$ the behaviour of the process
 ${\cal A}^{[\beta]}$ inside $\theta$ is independent of that outside $\theta$
 and then using (\ref{WARPOM}) to conclude that each such $\theta$ present brings
 a contribution of at least $\phi_0$ to the considered conditional variance,
 we have $\Var[\mu_L]$ bounded below by $\phi_0$ times the expected number 
 of external (outermost) contours $\theta$ in ${\cal A}^{[\beta]} \cap {\Bbb B}_2(L)$ as in
 (\ref{WARPOM}) with $D = \innt \theta.$ Since this number is clearly of the
 area order $\Omega(L^2),$ the required relation (\ref{WARI}) has
 been established, which completes the argument for (\ref{KUM2UZUP}).  
 
% Now, observing that the mean number of contours $\theta$ in 
% ${\cal A}^{[\beta]} \cap {\Bbb B}_2(L)$ satisfying the constraints
% of the infimum in (\ref{WARPOM}) for $D := \innt \theta$ is of
% the area order $\Omega(L^2),$ noting that given the presence
% of such $\theta$ the behaviour of the process ${\cal A}^{[\beta]}$
% inside $\theta$ is independent of that outside $\theta,$ and 
% then applying (\ref{WARPOM}) yields the required relation
% (\ref{WARI}) by a standard argument, thus completing the   
% argument for (\ref{KUM2UZUP}). 

 Putting together (\ref{TAYLOR2}), (\ref{KUM2UZUP}) and the observations
 that $h = o(1)$ by (\ref{KGDLAPOL}) and
 that ${\Bbb E}\mu^{\alpha,h}_{L,\gamma}$ strictly increases with $h$ we
 come to
 \begin{corollary}\label{POLAUSTAWIONE}
  There are positive constants $K_0 = K_0(\beta)$ and 
  $C = C(\beta)$ such that for each $\alpha \geq C \log L,$
  each $\Delta$ with $|\Delta| \leq K_0 L^2 \slash \alpha$ and
  each finite contour collection $\gamma$ with
  $\Area(\gamma \oplus {\Bbb B}_2(\log^2 L)) \leq L^2 \slash \log L$ 
  there exists a unique value $h = h[\Delta,L,\gamma]$ of 
  external magnetic field such that 
  $$ {\Bbb E}\mu^{\alpha,h}_{L,\gamma} = {\Bbb E}\mu^{\alpha}_{L,\gamma} +
     \Delta $$
  and
  $$ h = \Theta(\Delta \slash L^2) $$
  uniformly in $\alpha,\Delta,L,\gamma.$
 \end{corollary}  

 Our next statement provides a lower bound for moderate
 deviation probabilities of $\mu^{\alpha}_{L,\gamma},$ 
 complementary to the upper bound of Theorem \ref{SREDNIEO}.
 \begin{lemma}\label{DOLNESREDNIEO}
  For all
  $0 \leq \Delta \ll L^2 \slash \alpha,$ with
  $\alpha$ and $\gamma$ as in Corollary \ref{POLAUSTAWIONE}
  and with $\alpha \ll L \slash \log L$ we have
  uniformly in $\Delta,\alpha,L,\gamma$
  $$ {\Bbb P}\left( \mu^{\alpha}_{L,\gamma} > 
  {\Bbb E}\mu^{\alpha}_{L,\gamma} + \Delta \right) 
  \geq
  \exp(-O([\Delta+L\log L]^2 \slash L^2)). $$
 \end{lemma}
 \paragraph{Proof}
  Write using Corollary \ref{POLAUSTAWIONE}, putting for brevity
  $h[\cdot] := h[\cdot,L,\gamma],$
  $$ {\Bbb P}(\mu^{\alpha}_{L,\gamma} > {\Bbb E}\mu^{\alpha}_{L,\gamma}
      + \Delta) \geq {\Bbb P}(|\mu^{\alpha}_{L,\gamma} -
     {\Bbb E}\mu^{\alpha}_{L,\gamma} - \Delta -L\log L| < L \log L)
    \geq $$ 
 $$ \exp\left(-h[\Delta+L\log L] [{\Bbb E}\mu^{\alpha}_{L,\gamma}+
    \Delta+2L\log L]\right)
    {\Bbb E}\exp\left(h[\Delta+L\log L] \mu^{\alpha}_{L,\gamma}\right) $$ $$
         {\Bbb P}\left(\left|\mu_{L,\gamma}^{\alpha,h[\Delta+L\log L]}-
         {\Bbb E}
         \mu_{L,\gamma}^{\alpha,h[\Delta+L\log L]}\right| < L \log L\right) $$
 and use Jensen's inequality  to bound it below by
 $$  {\Bbb P}\left(\left|\mu_{L,\gamma}^{\alpha,h[\Delta+L\log L]}-{\Bbb E}
             \mu_{L,\gamma}^{\alpha,h[\Delta+L\log L]}\right| < L\log L\right)   
     \exp(-h[\Delta+L\log L][\Delta+2L\log L]). $$    
 Thus, taking into account that $h[\Delta+L\log L] =
 \Theta([\Delta+L\log L] \slash L^2)$
 by Corollary \ref{POLAUSTAWIONE} 
 and that ${\Bbb P}(|\mu_{L,\gamma}^{\alpha,h[\Delta+L\log L]}-{\Bbb E}
             \mu_{L,\gamma}^{\alpha,h[\Delta+L\log L]}| < L\log L) = 1-o(1)$
 in view of Corollary \ref{SREDNIEODLAPOL},
 completes the proof of the lemma. $\Box$ 
 
% we conclude that 
% \begin{equation}\label{DOSROWN}
%  {\Bbb P}(\mu_L > {\Bbb E}\mu_L + \Delta) \geq 
%  \exp(-O(\Delta^2 \slash L^2))
% \end{equation}

 \section{Decoupling lemma}\label{DECRE}
  The purpose of this section is to establish Lemma \ref{MALAROZNICA} stating that
  the avoidance probabilities for the field ${\cal A}^{[\beta]}$ over disjoint 
  regions can be very well approximated by the product of the corresponding 
  avoidance probabilities for individual regions. Even though this lemma
  is a direct conclusion from the graphical construction, we state it
  in a separate section due to its importance in our further argument.  

 \begin{lemma}\label{MALAROZNICA}
  Assume that $U_1,U_2,...,U_k,\; k \geq 1$ are disjoint bounded regions in ${\Bbb R}^2$
  such that $\min_{i \neq j} \dist(U_i,U_j) > \Delta \gg \log [k \sup_{i=1}^k \diam(U_i)].$
  Then, for some $C > 0$ we have
  $$ {\Bbb P}\left({\cal A}^{[\beta]} \cap \bigcup_{j=1}^k U_j = \emptyset \right) = $$
  $$ \left(1+O\left((\log k)\e^{-C \Delta}\sum_{i=1}^k \Area(U_i \oplus {\Bbb B}_2(1)) \right)\right)
     \prod_{j=1}^k {\Bbb P}\left( {\cal A}^{[\beta]} \cap U_j = \emptyset\right). $$
 \end{lemma}

 \paragraph{Proof}
  The exponential decay of ancestor clan sizes in the graphical construction (\ref{ZANIKGR}) yields
  \begin{equation}\label{ODCIECIE0}
   {\Bbb P}({\cal E}^c_i) = O(\Area(U_i \oplus {\Bbb B}_2(1))\exp(-C \Delta)),\; i=1,\ldots,k
  \end{equation}
  with
  $$ {\cal E}_i := \{ \An_0(U_i) \subseteq U_i \oplus {\Bbb B}_2(\Delta \slash 2) \}. $$
  Write ${\cal I}_j,\; j=1,...,k$ for the event
  $$ {\cal I}_j := \{ {\cal A}^{[\beta]} \cap U_j = \emptyset \} $$
  and use the canonical coupling of the graphical construction for ${\cal A}^{[\beta]}$ with
  the conditional graphical construction for the field ${\cal A}^{[\beta]}_{{\Bbb R}^2:
  [\bigcup_{i=1}^{\lfloor k \slash 2\rfloor} U_i]}$ as provided in Section \ref{KONSGRAF}
  to conclude that 
  \begin{equation}\label{WARPORZ}
   \left| {\Bbb P}\left( \bigcap_{i=\lfloor k \slash 2\rfloor+1}^k {\cal I}_i \right| \left.
    \bigcap_{i=1}^{\lfloor k \slash 2\rfloor} {\cal I}_i \right) -
    {\Bbb P}\left( \bigcap_{i=\lfloor k \slash 2\rfloor+1}^k {\cal I}_i \right) \right|
    \leq {\Bbb P}\left(\bigcup_{i=1}^k {\cal E}^c_i\right).
  \end{equation}
  Combining (\ref{ODCIECIE0}) with (\ref{WARPORZ}) leads to
  \begin{equation}\label{DOILOCZYNU}
   {\Bbb P}\left( \bigcap_{i=1}^k {\cal I}_i \right)
   = \left(1+O\left( \sum_{i=1}^k \Area(U_i \oplus {\Bbb B}_2(1)) \e^{-C \Delta} \right)\right)
     {\Bbb P}\left( \bigcap_{i=1}^{\lfloor k \slash 2 \rfloor} {\cal I}_i \right)
     {\Bbb P}\left( \bigcap_{i=\lfloor k \slash 2\rfloor+1}^k {\cal I}_i \right).
  \end{equation}
  The assertion of the lemma follows now by recursive application of (\ref{DOILOCZYNU}). $\Box$ 

% Dotad 10 wrzesnia rano !

\section{Existence and properties of surface tension}\label{SEKCJANAPPOW}
 This section deals with the existence of the limit (\ref{NAPPOW}) defining
 the surface tension functional specific for our model. The argument below
 relies on a number of technical properties of the quantity
 $T^{[\beta]}_{(\cdot)}[\cdot \leftrightarrow \cdot]$ and is split into
 several subsections. Our main tool here is the random walk representation
 of surface tension, stated in Lemma \ref{BLLOS}, and our main effort is
 concentrated on establishing  the crucial finite volume approximation
 Lemma \ref{PRZYBLNPT}. As everywhere in this paper, the results below
 are valid for $\beta$ large enough.
  
 \subsection{Optimising and freezing initial segments}\label{SNAPPOW1}
 It will be convenient for our further purposes to switch between
 several alternative but asymptotically equivalent variants
 and representations of the surface tension. In this subsection
 we argue that modifying and freezing the directions of the initial
 segments of the polygonal path in the original definition (\ref{T1})
 of the functional $T^{[\beta]}_{(\delta)}[x \leftrightarrow y]$
 does not alter its asymptotic behaviour for large $\dist(x,y).$   
 To this end we consider a version
 $\hat{T}^{[\beta]}_{(\delta)}[x \leftrightarrow y]$ of the quantity
 $T^{[\beta]}_{(\delta)}[x \leftrightarrow y],$ which arises
 as the supremum of the integrals as in (\ref{T1}), but in which the initial
 point of the first segment is now allowed in the whole ${\Bbb B}_2(x,\delta)$
 rather than just on $\partial {\Bbb B}_2(x,\delta),$ the endpoint
 of the last segment is allowed in the whole ${\Bbb B}_2(y,\delta)$ rather than
 just on $\partial {\Bbb B}_2(y,\delta),$ and the directions of both segments
 are fixed so that the integration is carried out over the remaining segments only.
 It is easily checked that, provided the distance between $x$ and $y$
 is large enough, 
 \begin{equation}\label{ROZZIEW}
  C^{-1} T^{[\beta]}_{(\delta)}[x \leftrightarrow y] \leq
         \hat{T}^{[\beta]}_{(\delta)}[x \leftrightarrow y] \leq
  C T^{[\beta]}_{(\delta)}[x \leftrightarrow y]
 \end{equation}
 for some $C = C(\beta,\delta) >1$ independent of $D,x,y.$ Indeed, the impact of
 taking the first and last segments fixed in the optimal way rather than
 integrating over them is easily seen to be only confined to close
 neighbourhoods of the initial point and the endpoint of the path, and
 can be compensated at a constant probability cost by appropriately
 adjusting a small number of initial and final segments.
 We also consider finite volume versions $T^{[\beta]}_{(\delta;D)}$
 and $\hat{T}^{[\beta]}_{(\delta;D)}$ of  $T^{[\beta]}_{(\delta)}$
 and $\hat{T}^{[\beta]}_{(\delta)},$ putting in analogy with (\ref{T1})
 $$ T^{[\beta]}_{(\delta;D)}[x \leftrightarrow y] :=
    \int_{{\cal C}^{x \leftrightarrow y;\delta}}
    {\Bbb P}({\cal A}^{[\beta]} \cap \theta = \emptyset)
    [\Theta_D^{x \leftrightarrow y;\delta}]^{[\beta]}(d\theta) $$
 and defining $\hat{T}^{[\beta]}_{(\delta;D)}$ in the same way as
 $\hat{T}^{[\beta]}_{(\delta)}$ with the additional requirement that
 the whole path be contained in $D.$ If the domain $D$ contains
 neighbourhoods of $x$ and $y$ (say, ${\Bbb B}_2(x,2\delta)
 \subseteq D$ and ${\Bbb B}_2(y,2\delta) \subseteq D$), a relation
 analogous to (\ref{ROZZIEW}) is easily verified to hold for
 $x$ and $y$ far enough
 \begin{equation}\label{ROZZIEWOGR}
  C^{-1} T^{[\beta]}_{(\delta;D)}[x \leftrightarrow y] \leq
         \hat{T}^{[\beta]}_{(\delta;D)}[x \leftrightarrow y] \leq
  C T^{[\beta]}_{(\delta;D)}[x \leftrightarrow y]
 \end{equation}
 with some $C := C(\beta,\delta) > 1$ independent of $D,x,y.$  

 We close this subsection with one more quantity, to be of use in 
 the sequel, for which a relation analogous to (\ref{ROZZIEW})
 and (\ref{ROZZIEWOGR}) is valid. Write
 \begin{equation}\label{DODDZIWNE}
  \vartheta^{[\beta]}_{(\delta)}[x \leftrightarrow y] =
  \int_{{\cal C}^{x \leftrightarrow y;\delta}} [\Theta^{x\leftrightarrow y;\delta}]^{[\beta]}(d\theta) =
  [\Theta^{x\leftrightarrow y;\delta}]^{[\beta]}({\cal C}^{x\leftrightarrow y;\delta})
 \end{equation}
 and, as in the definition of $\hat{T}^{[\cdot]}_{(\cdot)}[\cdot \leftrightarrow
 \cdot],$ let $\hat{\vartheta}^{[\beta]}_{(\delta)}[x\leftrightarrow y]$ be the supremum
 of integrals as in (\ref{DODDZIWNE}), but with the initial point of the first segment
 now allowed in the whole ${\Bbb B}_2(x,\delta)$ rather than just on $\partial
 {\Bbb B}_2(x,\delta),$ the endpoint of the last segment allowed in the whole
 ${\Bbb B}_2(y,\delta)$ rather than just on $\partial {\Bbb B}_2(y,\delta),$
 and the directions of both segments fixed so that the integration is carried
 out over the remaining segments only. Clearly, in full analogy to (\ref{ROZZIEW}),
 we have with $\dist(x,y)$ large enough
 \begin{equation}\label{ROZZIEWDODZ}
  C^{-1} \vartheta^{[\beta]}_{(\delta)}[x \leftrightarrow y] \leq
         \hat{\vartheta}^{[\beta]}_{(\delta)}[x \leftrightarrow y] \leq
  C \vartheta^{[\beta]}_{(\delta)}[x \leftrightarrow y]
 \end{equation}
 for some $C = C(\beta,\delta) >1$ independent of $D,x,y.$

\subsection{Random walk representation}
 The quantity $T^{[\beta]}_{(\delta)}[x \leftrightarrow y]$ admits a
 particularly useful interpretation in terms of a killed continuum
 random walk in environment with random obstacles. To see it consider
 a continuous-time random walk $Z_{t;{\Bbb B}_2(x,\delta)} := Z_t$
 in ${\Bbb R}^2$ independent of ${\cal A}^{[\beta]}$ and governed by
 the following dynamics
 \begin{itemize}
  \item between critical events specified below move in a constant direction with speed $1,$
  \item with intensity given by $4$ times the covered length element update
        the movement direction, choosing the angle $\phi \in (0,2\pi)$ between
        the old and new direction according to the density $|\sin(\phi)| \slash 4.$
 \end{itemize}
 The starting point and the initial velocity direction for $Z_t$ are chosen
 by taking a straight line $l$ crossing ${\Bbb B}_2(x,\delta)$ according
 to the measure $\mu(\cdot) \slash \mu(\{ l \;|\; l \cap {\Bbb B}_2(x,\delta)
 \neq \emptyset \}).$ The starting point of $Z_t$ is now taken to be one
 of the intersection points of $l$ with $\partial {\Bbb B}_2(x,\delta),$
 each picked with probability $1 \slash 2,$ while the initial velocity
 vector lies on $l$ pointing outwards ${\Bbb B}_2(x,\delta).$
 Let $\tilde{Z}_t = \tilde{Z}_{t;{\Bbb B}_2(x,\delta)}$ be the random walk $Z_t$
 killed whenever hitting its past trajectory. A crucial observation is that
 the probability element of the walk $Z_t$ containing a given polygonal path 
 $\theta \in {\cal C}^{x\leftrightarrow y;\delta}$ as its initial subpath 
 is exactly 
 \begin{equation}\label{RWR}
 \frac{1}{2 \mu(\{ l \;|\; l \cap {\Bbb B}_2(x,\delta)
 \neq \emptyset \})} \exp(-4\lgth(\theta)) \prod_{i=1}^k d\mu(l[e_k]),
 \end{equation}
 where $e_1,\ldots,e_k$ are the segments of $\theta$ while $l[e_i]$
 stands for the straight line determined by $e_i.$ Indeed, the prefactor
 $[2 \mu(\{ l \;|\; l \cap {\Bbb B}_2(x,\delta) \neq \emptyset \})]^{-1}$
 comes from the choice of the initial segment of $Z_t$ [the line 
 on which it lies and one out of two equiprobable directions, whence the
 extra $2^{-1}$] while for the 
 remaining segments we use the fact that, for any given straight line 
 $l_0,$ we have $\mu(\{l\;|\; l \cap l_0 \in d\ell,\; \angle(l,l_0) \in d\phi\})
 = |\sin \phi| d\ell d\phi$ with $d\ell$ standing for the length element 
 on $l_0$ and with $\angle(l_0,l)$ denoting the angle between $l$ and
 $l_0,$ see Proposition 3.1 in \cite{AS1} as well as the argument
 justifying the dynamic representation of the Arak in Section 4 ibidem
 and the proof of Lemma 1 in Schreiber \cite{SC2}. Note that the direction
 update intensity for $Z_t$ was set to $4$ to cancel out with the normalising
 constant $\int_0^{2\pi} |\sin\phi| d\phi = 4$ in the density
 $|\sin \phi| \slash 4$ for the new angle choice.   
 Clearly, the formula (\ref{RWR}) is also valid for $Z_t$ replaced
 by $\tilde{Z}_t$ since the paths in ${\cal C}^{x\leftrightarrow y;\delta}$
 are by definition self-avoiding. Thus, taking into account that, by
 standard integral geometry, $\mu(\{ l \;|\; l \cap {\Bbb B}_2(x,\delta)
 \neq \emptyset \}) = 2\pi \delta$ and recalling (\ref{WOLNESCIEZKI})
 and (\ref{WOLNESCIEZKIB}) we rewrite (\ref{RWR}) as 
 $\frac{1}{4\pi\delta} [\Theta^{x \leftrightarrow y;\delta}]^{[2]}(d\theta).$
 Consequently, recalling the definition of 
 ${\cal C}^{x\leftrightarrow y;\delta}$ and using (\ref{RWR}) we
 come to   
 \begin{lemma}\label{BLLOS}
  For each $C \subseteq {\cal C}^{x \leftrightarrow y;\delta}$ the
  following representation formula is valid for the value of
  $[\Theta^{x \leftrightarrow y;\delta}]^{[2]}(C)$
  $$
   [\Theta^{x \leftrightarrow y;\delta}]^{[2]}(C) = 4\pi\delta {\Bbb E}
   \card \{ \tilde{t} > 0 \;|\; \tilde{Z}_{\tilde{t}} \in (\tilde{Z}_{t})_{t
         \geq 0} \cap_{{\rm in}} \partial {\Bbb B}_2(y,\delta),\;
         \tilde{Z}_{[0,\tilde{t}]} \in C \},
  $$
  where $(\tilde{Z}_t)_{t \geq 0} \cap_{{\rm in}} \partial {\Bbb B}_2(y,\delta)$
  stands for the collection of {\it entry} points of $\tilde{Z}_t$ into
  ${\Bbb B}_2(y,\delta),$ with {\it exit} points not taken into account.
 \end{lemma}
 %Indeed, $|\sin(\phi)| \slash 2$ is the density of the law of
 %the typical angle between  two lines of the Poisson line process $\Lambda$
 %while $2$ is the rate  at which the lines of $\Lambda$ hit a fixed straight line, see
 %Sections 3 and 4 ibidem.   
 %Note that the prefactor $2$ before the expectation in Lemma \ref{BLLOS} is
 %due to choosing one of two equiprobable initial directions on the initial line.
 A simple yet useful conclusion of Lemma \ref{BLLOS} is that, denoting by 
 $\tilde{Z}^{[\beta]}_t = \tilde{Z}^{[\beta]}_{t;{\Bbb B}_2(x,\delta)}$
 the random walk $Z_t$ killed at rate $\beta$ and, in addition, killed
 whenever hitting its past trajectory, we have for $\beta \geq 2$
 \begin{equation}\label{REPRENOWE}
   [\Theta^{x \leftrightarrow y;\delta}]^{[\beta]}(C) = 4\pi\delta {\Bbb E}
   \card \{ \tilde{t} > 0\;|\; 
         \tilde{Z}^{[\beta-2]}_{\tilde{t}} \in (\tilde{Z}^{[\beta-2]}_{t})_{t
         \geq 0} \cap_{{\rm in}} \partial {\Bbb B}_2(y,\delta),\;
         \tilde{Z}^{[\beta-2]}_{[0,\tilde{t}]} \in C \}.
 \end{equation}
 Consequently, writing now $\hat{Z}^{[\beta]}_t = \hat{Z}^{[\beta]}_{t;{\Bbb B}_2(x,\delta)}$
 for the random walk $Z_t$ killed at rate $\beta$ and, in addition, killed whenever hitting
 its past trajectory {\it or} a contour of ${\cal A}^{[\beta]},$ in view of Lemma \ref{BLLOS}
 and (\ref{REPRENOWE}) the definition (\ref{T1}) yields 
 \begin{equation}\label{REPRE2}
  T^{[\beta]}_{(\delta)}[x \leftrightarrow y] = 4\pi\delta {\Bbb E}\card [
   (\hat{Z}^{[\beta-2]}_t)_{t \geq 0}
   \cap_{{\rm in}} \partial {\Bbb B}_2(y,\delta)].
 \end{equation}
 A similar representation can be provided for $T^{[\beta]}_{(\delta;D)},$
 by additionally killing the random walk whenever it hits $\partial D.$ A corresponding
 representation for $\hat{T}^{[\beta]}_{(\delta)}$ and $\hat{T}^{[\beta]}_{(\delta;D)}$
 can also be given, yet we omit it because it is unessential for our further purposes
 and involves certain technicalities due the fixed last segment.

 \subsection{Finite volume approximations}
 The following lemma shows that $T^{[\beta]}_{(\delta)}[x \leftrightarrow y]$
 is well approximated by $T^{[\beta]}_{(\delta;D)}[x \leftrightarrow y]$ for
 sufficiently large domains $D.$ We write $\Pi(x \leftrightarrow y;\delta)$
 for the square of sidelength $2\delta + \dist(x,y)$ with one
 pair of its sides parallel and equidistant to $[x,y]$ and with
 the remaining two sides at the distance $\delta$ from $x$ and
 $y$ respectively, perpendicular to $[x,y].$ 
 \begin{lemma}\label{PRZYBLNPT}
  For each sufficiently large $\beta > 2$ there exists a constant $C = C(\beta,\delta) > 0$ such that
  $$ C^{-1} T^{[\beta]}_{(\delta)}[x \leftrightarrow y]
     \leq T^{[\beta]}_{(\delta;\Pi(x \leftrightarrow y;\delta))}[x \leftrightarrow y]
     \leq T^{[\beta]}_{(\delta)}[x \leftrightarrow y] $$
  whenever $\dist(x,y)$ is large enough.
 \end{lemma}

 \paragraph{Proof}
  The relation
  $T^{[\beta]}_{(\delta;\Pi(x \leftrightarrow y;\delta))}[x \leftrightarrow y]
   \leq T^{[\beta]}_{(\delta)}[x \leftrightarrow y]$ is obvious and only
  the remaining inequality $T^{[\beta]}_{(\delta)}[x \leftrightarrow y] \leq
  C T^{[\beta]}_{(\delta;\Pi(x \leftrightarrow y;\delta))}[x \leftrightarrow y]$
  requires verification. In view of the random walk representation (\ref{REPRE2}) 
  it will follow as soon as we show that
  \begin{equation}\label{DOPOK}
   P^{[\beta]}_{(\delta;\Pi(x\leftrightarrow y;\delta))}[x \leftrightarrow y]
   \geq C^{-1} P^{[\beta]}_{(\delta)}[x \leftrightarrow y]
  \end{equation}
  for some $C > 0,$ where
  $$ P^{[\beta]}_{(\delta)}[x \leftrightarrow y] := {\Bbb P}\left((\hat{Z}_t^{[\beta-2]})_{t \geq 0}
   \mbox{ reaches } \partial {\Bbb B}_2(y,\delta)\right) $$
  and
  $$ P^{[\beta]}_{(\delta;D)}[x \leftrightarrow y] :=
     {\Bbb P}\left((\hat{Z}_t^{[\beta-2]})_{t \geq 0} \mbox{ reaches } \partial {\Bbb B}_2(y,\delta)
              \mbox{ before hitting } \partial D \right). $$
  Indeed, it is easily argued that upon hitting $\partial {\Bbb B}_2(y,\delta)$
  once, the random walk $\hat{Z}^{[\beta-2]}_t$ is unlikely to hit it too many more times and,
  consequently, the expectation on the right-hand side of (\ref{REPRE2}) is bounded
  above and below by some constant multiplicities of the probability on the right-hand side of (\ref{DOPOK}),
  the same observation holds for the corresponding representation of the finite-volume
  quantity $T^{[\beta]}_{(\delta;\Pi(x \leftrightarrow y;\delta))}[x \leftrightarrow y].$
  We omit the tedious technical details of this conceptually simple argument.

  The proof of (\ref{DOPOK}) splits into two parts. First, denoting by
  $R_1(x \leftrightarrow y;\delta)$ the infinite strip between the lines
  determined by the sides of $\Pi(x \leftrightarrow y;\delta)$ perpendicular
  to $[x,y],$ we show that
  \begin{equation}\label{TRUDNE1}
   P^{[\beta]}_{(\delta)}[x \leftrightarrow y] \leq C
   P^{[\beta]}_{(\delta;R_1(x\leftrightarrow y;\delta))}[x \leftrightarrow y]
  \end{equation}
  for some $C > 0.$ Below it will be convenient to use the name $x$-line
  (resp. $y$-line) for the boundary line (side) of $R_1(x\leftrightarrow y;\delta)$
  at the distance $\delta$ from $x$ (resp. $y$), perpendicular to $[x,y]$.
  Next, writing $R_2(x \leftrightarrow y;\delta)$ for the infinite strip
  contained between the lines determined by the sides
  of $\Pi(x \leftrightarrow y;\delta)$ parallel to $[x,y],$ we show
  that
  \begin{equation}\label{LATWE1}
   P^{[\beta]}_{(\delta)}[x \leftrightarrow y] \leq
   P^{[\beta]}_{(\delta;R_2(x\leftrightarrow y;\delta))}[x \leftrightarrow y] (1+o(1)).
  \end{equation}
  as $\dist(x,y) \to \infty.$ Write
  $$ P^{[\beta]}_{(\delta;\Pi(x\leftrightarrow y;\delta))}[x \leftrightarrow y]
     \geq P^{[\beta]}_{(\delta)}[x \leftrightarrow y]
          - \left( P^{[\beta]}_{(\delta)}[x \leftrightarrow y]-
             P^{[\beta]}_{(\delta;R_1(x\leftrightarrow y;\delta))}[x \leftrightarrow y]\right)
          - $$ $$  \left( P^{[\beta]}_{(\delta)}[x \leftrightarrow y]-
             P^{[\beta]}_{(\delta;R_2(x\leftrightarrow y;\delta))}[x \leftrightarrow y] \right) =
     P^{[\beta]}_{(\delta;R_1(x\leftrightarrow y;\delta))}[x \leftrightarrow y] +
     P^{[\beta]}_{(\delta;R_2(x\leftrightarrow y;\delta))}[x \leftrightarrow y] - $$ $$ 
     P^{[\beta]}_{(\delta)}[x \leftrightarrow y]. $$
 Combining this with (\ref{TRUDNE1}) and (\ref{LATWE1}) yields (\ref{DOPOK}) as required for
 completing the proof of the lemma.

  To proceed with the verification of (\ref{TRUDNE1}), on the event that
  the random walk $\hat{Z}^{[\beta-2]}_t$ reaches $\partial {\Bbb B}_2(y,\delta)$
  before being killed we decompose its trajectory into three subpaths
  \begin{itemize}
   \item $\zeta_{x \leftrightarrow y; \delta} := (\hat{Z}_t^{[\beta-2]})_{[\tau_x,\tau_y]},$
         where $\tau_y$ is the first time $\hat{Z}^{[\beta-2]}_t$
         hits the $y$-line while $\tau_x$ is the last time
         $\hat{Z}^{[\beta-2]}_t$ hits the $x$-line before $\tau_y.$

   \item $\zeta_x := (\hat{Z}_t^{[\beta-2]})_{[0,\tau_x]},$
   \item $\zeta_y := (\hat{Z}_t^{[\beta-2]})_{t \geq \tau_y}, $
  \end{itemize}
  with the additional convention that $\tau_x := 0$ if $\hat{Z}^{[\beta-2]}_t$
  does not reach the $x$-line and $\tau_y := +\infty$ if $\hat{Z}^{[\beta-2]}_t$
  does not reach the $y$-line (we set respectively $\zeta_x := \emptyset$ and
  $\zeta_y := \emptyset$ in these cases).  On the $x$-line we construct
  a double sequence $(x_i)_{i \in {\Bbb Z}}$ of points with $x_{i+k}$
  lying at the distance $|k|\delta$ from $x_i$ (say above for $k >0,$
  below for $k < 0$) and with $x_0$ coinciding
  with the intersection point of the $x$-line and
  the line  extending $[x,y].$ 
  The sequence $(y_i)_{i \in {\Bbb Z}}$ on the $y$-line is constructed
  in the same way and ordered in the same direction as $(x_i).$ Let
  $\hat{x}$ denote the point in $(x_i)_{i \in {\Bbb Z}}$ which lies the
  closest to $\hat{Z}^{[\beta]}_{\tau_x}$ if $\tau_x > 0$ and
  $\hat{x} := x$ otherwise. Likewise, let $\hat{y}$ be the point
  in $(y_i)_{i \in {\Bbb Z}}$ lying the closest to $\hat{Z}^{[\beta-2]}_{\tau_y}$
  if $\tau_y < + \infty$ and $\hat{y} := y$ otherwise. With this notation
  it is easily seen that
  \begin{equation}\label{COSTAM}
   P^{[\beta]}_{(\delta)}[x \leftrightarrow y] \leq
        \sum_{\hat{x} \in \{ x \} \cup \{ x_i,\;i \in {\Bbb Z}\}}
        \sum_{\hat{y} \in \{ y \} \cup \{ y_j,\;j \in {\Bbb Z}\}}
        P^{[\beta]}_{(\delta;R_1(x \leftrightarrow y;\delta))}[\hat{x}
        \leftrightarrow \hat{y}] Q^{[\beta]}_{(\delta)}[x\leftrightarrow \hat{x};
        \hat{y}\leftrightarrow y],
  \end{equation}
  where $Q^{[\beta]}_{(\delta)}[x \leftrightarrow \hat{x};\hat{y}\leftrightarrow y]$
  stands for the supremum over the possible realisations of $\zeta_{x \leftrightarrow y}$
  connecting ${\Bbb B}_2(\hat{x},\delta)$ with ${\Bbb B}_2(\hat{y},\delta)$
  of the conditional probability, given $\zeta_{x \leftrightarrow y},$ that the
  random walk $\hat{Z}^{[\beta-2]}_t$ connects ${\Bbb B}_2(x,\delta)$ to
  ${\Bbb B}_2(\hat{x},\delta)$ and ${\Bbb B}_2(\hat{y},\delta)$ to
  ${\Bbb B}_2(y,\delta).$ Since $\hat{Z}^{[\beta-2]}_t$ is killed with the
  constant rate $\beta-2 > 0,$ for arbitrarily small $\epsilon$ we can find
  $\lambda = \lambda(\epsilon)$ such that, uniformly over $x,y$ with
  $\dist(x,y)$ large enough,
  \begin{equation}\label{OGRAE}
   \sum_{x_i,\; \dist(x_i,x) > \lambda} \sum_{y_j} Q^{[\beta]}_{(\delta)}[x \leftrightarrow x_i;
    y_j \leftrightarrow y] +
    \sum_{x_i} \sum_{y_j,\; \dist(y_j,y) > \lambda} Q^{[\beta]}_{(\delta)}[x \leftrightarrow x_i;
    y_j \leftrightarrow y] \leq \epsilon.
  \end{equation}
  Putting (\ref{COSTAM}) and (\ref{OGRAE}) together yields
  \begin{equation}\label{ICODALEJ}
    P^{[\beta]}_{(\delta)}[x \leftrightarrow y] \leq
        \sum_{x_i,\; \dist(x_i,x) \leq \lambda; }
        \sum_{y_j,\; \dist(y_j,y) \leq \lambda }
         P^{[\beta]}_{(\delta;R_1(x \leftrightarrow y;\delta))}[x_i \leftrightarrow y_j]
        + \epsilon \sup_{x_i,y_j}
        P^{[\beta]}_{(\delta;R_1(x \leftrightarrow y;\delta))}[x_i \leftrightarrow y_j].
  \end{equation}
  For $\dist(x,y)$ large enough the double sum in (\ref{ICODALEJ}) can be bounded above
  by some constant $C[\lambda]$ times $P^{[\beta]}_{(\delta;R_1(x \leftrightarrow y;\delta))}
  [x \leftrightarrow y]$ because each path of $\hat{Z}_t^{[\beta-2]}$ connecting
  $\partial {\Bbb B}_2(x_i,\delta)$ to $\partial {\Bbb B}_2(y_j,\delta)$ in
  $R_1(x\leftrightarrow y;\delta)$ with
  $\dist(x,x_i) \leq \lambda$ and $\dist(y,y_j) \leq \lambda$ can be modified into
  a path connecting $\partial {\Bbb B}_2(x,\delta)$ to $\partial {\Bbb B}_2(y,\delta)$
  in $R_1(x\leftrightarrow y;\delta)$ by an appropriate surgery between $x$ and $x_i$
  and between $y$ and $y_j$ at a probability cost depending only on $\lambda.$
  % Dotad 13 wrzesnia w dzien
  It seems natural to expect that the supremum
  $\sup_{x_i,y_j} P^{[\beta]}_{(\delta;R_1(x \leftrightarrow y;\delta))}
  [x_i \leftrightarrow y_j]$ admits an upper bound very close to $P^{[\beta]}_{(\delta)}
  [x \leftrightarrow y],$ because $\dist(x_i,y_j) > \dist(x,y)$ for all $x_i, y_j.$
  While we are not able to establish such a bound, we easily show that there exists
  a positive constant $C'$ with
  \begin{equation}\label{JAKBYMALENIE}
   \sup_{x_i,y_j} P^{[\beta]}_{(\delta;R_1(x \leftrightarrow y;\delta))}
   [x_i \leftrightarrow y_j] \leq C' P^{[\beta]}_{(\delta)}[x \leftrightarrow y]
  \end{equation}
  uniformly in $x,y$ with $\dist(x,y)$ large enough. Indeed, this is done
  much along the same lines as in the considerations leading to (\ref{COSTAM})
  and (\ref{ICODALEJ}), so we only sketch the argument omitting technical details.
  We split each path of the random walk $\hat{Z}_{t;{\Bbb B}_2(x_i,\delta)}^{[\beta-2]}$
  connecting $\partial {\Bbb B}_2(x_i,\delta)$ to $\partial {\Bbb B}_2(y_j,\delta)$
  into two subpaths: the initial subpath $\zeta_1$
  connecting $\partial {\Bbb B}_2(x_i,\delta)$ to some $\partial {\Bbb B}_2(z,\delta),\;
  z \in \delta {\Bbb Z}^2$ with $|\dist(x_i,z)-\dist(x,y)| \leq \delta$ (in fact,
  $z$ can be chosen as the $\delta {\Bbb Z}^2$-lattice point closest to the point
  where the random walk $\hat{Z}_{t;{\Bbb B}_2(x_i,\delta)}^{[\beta-2]}$ first gets
  at the distance $\dist(x,y)$ away from $x_i$) and the remaining subpath $\zeta_2.$
  Integrating over $\zeta_1$ for fixed $z$ yields a value bounded above by a constant
  multiplicity of $P^{[\beta]}_{(\delta)}[x \leftrightarrow y]$ with this prefactor
  (arbitrarily close to $1$ for $\delta$ small enough) due to the fact that
  $\dist(x_i,z)$ differs slightly from $\dist(x,y).$ Integrating over $\zeta_2$
  conditioned on $\zeta_1$ and summing over $z$ yields only a constant prefactor
  \--- the sum of integrals converges due to the constant killing rate $\beta - 2 > 0$
  along $\zeta_2.$
  This proves (\ref{JAKBYMALENIE}). Combining now (\ref{ICODALEJ}) with (\ref{JAKBYMALENIE})
  and with the discussion directly following (\ref{ICODALEJ}) we obtain
  \begin{equation}\label{ITODALEJ}
     P^{[\beta]}_{(\delta)}[x \leftrightarrow y] \leq
        C[\lambda] P^{[\beta]}_{(\delta;R_1(x \leftrightarrow y;\delta))}[x \leftrightarrow y]
      + \epsilon C' P^{[\beta]}[x \leftrightarrow y].
  \end{equation}
  Choosing $\epsilon$ small enough so that $\epsilon C' < 1$ (recall that
  $C'$ does not depend on $\lambda$) completes the proof of (\ref{TRUDNE1}).

  To establish (\ref{LATWE1}) we denote by $\vec{v_{xy}}$ the unit vector
  pointing from $x$ to $y,$ i.e. $\vec{v_{xy}} := (y-x) \slash \dist(x,y),$ and
  for small $\eta > 0$ we consider the event ${\cal E}_{(\delta;\eta)}[x\leftrightarrow y]$
  that
  \begin{itemize}
   \item The random walk $(\hat{Z}^{[\beta-2]}_{t})_{t \geq 0}$ reaches $\partial {\Bbb B}_2(y,\delta),$
   \item The scalar product of $\vec{v_{xy}}$ and the current velocity vector
         of $\hat{Z}^{[\beta-2]}_t$ is in $[1-\eta,1]$ for all time moments $t \geq 0$
         before $\partial {\Bbb B}_2(y,\delta)$ is reached.
  \end{itemize}
  Observe that on the event ${\cal E}^{[\beta]}_{(\delta;\eta)}[x \leftrightarrow y]$
  the total length of the path of $\hat{Z}^{[\beta-2]}_t$ connecting $\partial {\Bbb B}_2(x,\delta)$
  to $\partial {\Bbb B}_2(y,\delta)$ cannot exceed $[\dist(x,y) + 2\delta] \slash [1-\eta].$
  Since ${\cal A}^{[\beta]}$ is stochastically dominated by the Poisson contour process
  ${\cal P}_{\Theta^{[\beta]}},$ as follows by the graphical construction of Section 
  \ref{KONSGRAF}, we conclude that there exists $\kappa > 0$ such that for all $\beta$
  large enough
  \begin{equation}\label{OGRANAE}
   {\Bbb P}\left( {\cal E}^{[\beta]}_{(\delta;\eta)}[x \leftrightarrow y] \right)
   \geq \exp\left(-\left[\frac{\beta-2}{1-\eta} + \kappa\right] (\dist(x,y) + 2\delta) \right).
  \end{equation}
  Indeed, to see it we:
  \begin{itemize}
   \item Split the strip $R_1(x \leftrightarrow y;\delta)$ with equidistant straight lines
         perpendicular to $[x,y]$ into $\Theta(\dist(x,y))$ equal-sized strips.
   \item Construct a path of the random walk $\tilde{Z}$ connecting $\partial {\Bbb B}_2(x,\delta)$
         to $\partial {\Bbb B}_2(y,\delta)$ and such that the scalar product of
         $\vec{v_{xy}}$ and the current velocity vector of $\tilde{Z}_t$ falls
         into $[1-\eta,1]$ for all time moments before  $\partial {\Bbb B}_2(y,\delta)$
         is reached. This is done by constructing and patching together subpaths of
         $\tilde{Z}$ crossing individual strips, at a constant probability cost per strip.
   \item Use the stochastic domination of ${\cal A}^{[\beta]}$ by
         ${\cal P}_{\Theta^{[\beta]}}$ to conclude that the probability that 
         the so constructed path of $\tilde{Z}_t$ avoids ${\cal A}^{[\beta]}$
         is bounded below by $\exp(-\Theta(\dist(x,y))).$ 
   \item Check for survival of the so constructed path under $\beta-2$-killing,
         which yields a probability prefactor bounded below by
         $\exp(- \frac{\beta-2}{1-\eta} [\dist(x,y) + 2\delta]).$
  \end{itemize}
  By the definition of ${\cal E}^{[\beta]}_{(\delta;\eta)},$
  for $\dist(x,y)$ large enough this procedure allows us to bound below
  the probability of this event by
  $\exp\left(-(\frac{\beta-2}{1-\eta}+\kappa) [\dist(x,y) + 2\delta]\right)$ for some $\kappa > 0,$
  as required. Since ${\cal P}_{\Theta^{[\beta]}}$ stochastically dominates
  ${\cal P}_{\Theta^{[\beta']}}$ for $\beta' > \beta,$ this technique works
  uniformly in $\beta$ large enough. We omit tedious technical details of this
  standard argument.
  To proceed, define the event ${\cal R}^{[\beta]}_{(\delta)}[x \leftrightarrow y]$ that
  \begin{itemize}
   \item The random walk $(\hat{Z}^{[\beta-2]}_t)_{t \geq 0}$ reaches $\partial {\Bbb B}_2(y,\delta),$
   \item The random walk $(\hat{Z}^{[\beta-2]}_t)_{t \geq 0}$ hits $\partial R_2(x\leftrightarrow y;\delta)$
         before reaching $\partial {\Bbb B}_2(y,\delta),$
  \end{itemize}
  and observe that on ${\cal R}^{[\beta]}_{(\delta)}[x\leftrightarrow y]$ the length of the
  path connecting $\partial {\Bbb B}_2(x,\delta)$ to $\partial {\Bbb B}_2(y,\delta)$
  has to exceed $\sqrt{5/4} \dist(x,y) - 2\delta$ and, hence,
  \begin{equation}\label{OGRANAR}
   {\Bbb P}\left( {\cal R}^{[\beta]}_{(\delta)}[x \leftrightarrow y] \right)
   \leq \exp\left(-(\beta-2)[\sqrt{5/4} \dist(x,y)-2\delta]\right).
  \end{equation}
  Noting that ${\Bbb P}\left( {\cal E}^{[\beta]}_{(\delta;\eta)}[x \leftrightarrow y] \right)
  \leq P^{[\beta]}_{(\delta;R_2(x\leftrightarrow y;\delta))}[x \leftrightarrow y]$
  for sufficiently small $\eta$ 
  and putting (\ref{OGRANAE}) together with (\ref{OGRANAR}) we see that,
  for $\beta$ large enough, 
  $$ P^{[\beta]}_{(\delta)}[x \leftrightarrow y] -
     P^{[\beta]}_{(\delta;R_2(x\leftrightarrow y;\delta))}[x \leftrightarrow y]
     =
     {\Bbb P}\left( {\cal R}^{[\beta]}_{(\delta)}[x \leftrightarrow y] \right)
       = o\left(P^{[\beta]}_{(\delta;R_2(x\leftrightarrow y;\delta))}[x \leftrightarrow y]\right). $$
  This yields (\ref{LATWE1}) and hence the required relation (\ref{DOPOK}). The proof
  of Lemma  \ref{PRZYBLNPT} is complete. $\Box$

% Dotad 14 wrzesnia rano !
\subsection{Existence and finiteness of surface tension}
 In this subsection we use the preceding results of this section
 to show that 
 \begin{lemma}\label{NPISTN}
  The limit defining the surface tension functional $\tau^{[\beta]}$
  in (\ref{NAPPOW}) exists, is finite and strictly positive.
 \end{lemma}
 
 \paragraph{Proof}
  The main work has already been done in Lemma \ref{PRZYBLNPT}. In
  view of the relation $\sup_{\lambda > 2\delta} \tau^{[\beta]}_{\lambda}
  < \infty$ as easily deduced from (\ref{OGRANAE}), the required
  existence of the limit in (\ref{NAPPOW}) will follow by a standard
  almost-subadditivity argument once we establish the following auxiliary
  lemma.
 \begin{lemma}\label{POMOCNP}
  For $D[\lambda_1,\lambda_2] := (\log [\lambda_1+\lambda_2])^2$
  and $\lambda_1,\lambda_2$ large enough we have
  $$(\lambda_1+\lambda_2 + D[\lambda_1,\lambda_2]) \tau^{[\beta]}_{\lambda_1 + \lambda_2 + D[\lambda_1,\lambda_2]}
  \leq \lambda_1 \tau^{[\beta]}_{\lambda_1} + \lambda_2 \tau^{[\beta]}_{\lambda_2} +
  O(D[\lambda_1,\lambda_2])).$$
 \end{lemma}

 \paragraph{Proof of Lemma \ref{POMOCNP}}
  For fixed $\delta > 0$ consider disjoint squares $\Pi_1 := \Pi(0 \leftrightarrow
  \lambda_1 {\bf e}_x;\delta)$ and $\Pi_2 := \Pi((\lambda_1 + D[\lambda_1,\lambda_2])
  {\bf e}_x \leftrightarrow (\lambda_1+\lambda_2 + D[\lambda_1,\lambda_2]) {\bf e}_x;\delta),$
  separated by a moat of width $D[\lambda_1,\lambda_2],$ and observe that, since
  $D[\lambda_1,\lambda_2] \gg \log [\lambda_1+\lambda_2],$ by the decoupling Lemma
  \ref{MALAROZNICA} applied to $U_1 := \Pi_1 \cap \theta,\; U_2 := \Pi_2 \cap \theta,$
  with $\theta$ standing for the integrand polygonal path in the definition (\ref{T1})
  of $T^{[\beta]}_{(\cdot)}[\cdot \leftrightarrow \cdot],$ it follows that 
  $$ 
    T^{[\beta]}_{(\delta)}[0 \leftrightarrow (\lambda_1+\lambda_2 + D[\lambda_1,\lambda_2]) {\bf e}_x]
    \geq $$
  \begin{equation}\label{ROZBICIE}
    \e^{-O(D[\lambda_1,\lambda_2])} T^{[\beta]}_{(\delta;\Pi_1)}[0 \leftrightarrow \lambda_1 {\bf e}_x]
    T^{[\beta]}_{(\delta;\Pi_2)}[(\lambda_1 + D[\lambda_1,\lambda_2]){\bf e}_x \leftrightarrow (\lambda_1
    +\lambda_2 + D[\lambda_1,\lambda_2]){\bf e}_x],
  \end{equation}
  with the prefactor $\e^{-O(D[\lambda_1,\lambda_2])}$ due to patching together pairs of
  paths $\theta_1$ in ${\cal C}^{0\leftrightarrow \lambda_1 {\bf e}_x;\delta}$ and
  $\theta_2$ in ${\cal C}^{(\lambda_1+D[\lambda_1,\lambda_2]) {\bf e}_x \leftrightarrow 
  (\lambda_1+\lambda_2+D[\lambda_1,\lambda_2])
  {\bf e}_x;\delta},$ both disjoint with ${\cal A}^{[\beta]},$ into paths
  $\theta$ falling into ${\cal C}^{0 \leftrightarrow (\lambda_1+\lambda_2+D[\lambda_1,\lambda_2])
  {\bf e}_x;\delta}$ disjoint 
  with ${\cal A}^{[\beta]},$ by constructing a path connecting $\theta_1$ and $\theta_2$
  across the moat of width $D[\lambda_1,\lambda_2]$ separating $\Pi_1$ and $\Pi_2,$
  according to a procedure completely analogous to that used
  in the argument leading to (\ref{OGRANAE}). Note that the fact that the patching procedure
  involves here conditioning on ${\cal A}^{[\beta]}$ being disjoint with $\theta_1$ and $\theta_2$
  does not affect this argument because the conditional graphical construction of 
  the process ${\cal A}^{[\beta]}_{{\Bbb R}^2:\theta_1 \cup \theta_2}$ guarantees
  that it is stochastically bounded by ${\cal P}_{\Theta^{[\beta]}:\theta_1 \cup \theta_2}$
  and hence by ${\cal P}_{\Theta^{[\beta]}}$ as used in the proof of (\ref{OGRANAE}).   
  To proceed, apply Lemma \ref{PRZYBLNPT} to conclude that the quantities
  $T^{[\beta]}_{(\delta)}[0 \leftrightarrow \lambda_1 {\bf e}_x]$ and  $T^{[\beta]}_{(\delta)}
  [(\lambda_1 + D[\lambda_1,\lambda_2]) {\bf e}_x \leftrightarrow
  (\lambda_1+\lambda_2+D[\lambda_1,\lambda_2]){\bf e}_x]$
  are bounded above by constant multiplicities of their respective finite volume counterparts
  $T^{[\beta]}_{(\delta;\Pi_1)}[0 \leftrightarrow \lambda_1 {\bf e}_x]$ and
  $T^{[\beta]}_{(\delta;\Pi_2)}[(\lambda_1 + D[\lambda_1,\lambda_2])
   {\bf e}_x \leftrightarrow (\lambda_1+\lambda_2+ D[\lambda_1,\lambda_2]){\bf e}_x].$
  Combining this conclusion with (\ref{ROZBICIE}) shows that
  $T^{[\beta]}_{(\delta)}[0 \leftrightarrow (\lambda_1+\lambda_2+D[\lambda_1,\lambda_2]) {\bf e}_x] \geq 
  \exp(-O(D[\lambda_1,\lambda_2])) T^{[\beta]}_{(\delta)}[0 \leftrightarrow \lambda_1 {\bf e}_x]
  T^{[\beta]}_{(\delta)}[(\lambda_1+D[\lambda_1,\lambda_2]) {\bf e}_x
  \leftrightarrow (\lambda_1+\lambda_2+D[\lambda_1,\lambda_2]){\bf e}_x]$
  for some $C > 0,$ which completes the proof in view of the definition (\ref{PRZYBLNAP})
  of $\tau^{[\beta]}_{\lambda_i},\; i=1,2.$ 
 $\Box$
 
 \paragraph{Completing the proof of Lemma \ref{NPISTN}}
  With the existence of the limit in (\ref{NAPPOW}) established
  we now easily conclude its strict positivity from the positivity
  of killing rate in the  random walk representation (\ref{REPRE2})
  while the 
  finiteness of $\tau^{[\beta]}$ follows by the probability
  lower bound (\ref{OGRANAE}). $\Box$

\section{Skeleton estimates}\label{SZKIEE}
 The purpose of this section is to provide coarse-graining estimates 
 based on skeleton calculus. 
 For $\alpha,\delta > 0,$ always assumed to satisfy $L \gg \alpha \gg \delta$
 and to tend to $\infty$ as $L \to \infty,$ 
 by an $(\alpha,\delta)$-skeleton in ${\Bbb B}_2(L)$ we
 shall understand a collection $(I_1,E_1,I_2,E_2,$ $\ldots,I_m,E_m)$
 of pairwise different points ({\it skeleton vertices}) in
 ${\Bbb B}_2(L) \cap {\Bbb Z}^2,$ with $I_1,I_2,\ldots$ referred to
 as the {\it initial points}, $E_1,E_2,\ldots$ as the corresponding
 {\it endpoints} and $[I_1,E_1],[I_2,E_2],\ldots$ as the
 {\it skeleton segments}, where the following is satisfied for all $i=1,\ldots,m$
 \begin{description}
  \item{\bf (S1)} $\alpha - \sqrt{2} \leq \dist(I_i,E_i) \leq \alpha + \sqrt{2}.$
 \end{description}
 We say that a collection $\gamma$ of $\alpha$-large polygonal contours is compatible 
 with an $(\alpha,\delta)$-skeleton $\Sigma = (I_1,E_1,\ldots,I_m,E_m),$
 write $\gamma \sim \Sigma,$ if the following holds for all $i=1,\ldots,m$
 \begin{description}
  \item{\bf (S2)}  There exists a contour $\theta_i \in \gamma$ and points
                   $I_i^{\gamma}, E_i^{\gamma} \in \theta_i$
                   such that $\dist(I_i,I_i^{\gamma}) \leq \frac{1}{\sqrt{2}},
                   \dist(E_i,E_i^{\gamma}) \leq \frac{1}{\sqrt{2}}$ and
                   $\dist(I_i,x) \leq \alpha + \frac{1}{\sqrt{2}}$ for all
                   $x \in \theta_i[I_i^{\gamma},E_i^{\gamma}]$ with
                   $\theta_i[I_i^{\gamma},E_i^{\gamma}]$ standing for
                   the polygonal path from $I_i^{\gamma}$ to $E_i^{\gamma}$
                   along $\theta_i$ (note that we do not require that
                   $\theta_i \neq \theta_j$ for $i \neq j$),
 \item{\bf (S3)}   Either we have
                   $\dist(I_i,\{ I_1,\ldots,I_{i-1} \})$ $\leq \alpha + \delta + \sqrt{2}$
                   or $i$ is the smallest index with $I^{\gamma}_i \in \theta_i$
                   for some $\theta_i \in \gamma,$
 \item{\bf (S4)}   For each $x \in \gamma$ we have $\dist(x,\{I_1,\ldots,I_m\}) \leq
                   2 \alpha + \delta + \sqrt{2},$
 \item{\bf (S5)}   The polygonal paths $\theta_i[I_i^{\gamma},E_i^{\gamma}]$
                   are in a distance at least $\delta$ away from each other.
 \end{description}
 Roughly speaking, the motivation underlying this definition is the following. For two distant
 points $x,y$ with $\dist(x,y) = \Omega(\alpha)$ connected by a polygonal subpath of a contour we want
 to find a collection of approximately equal-sized segments $[I_i,E_i]$ of length $\alpha (1+o(1)),$
 lying on this path and such that their overall length is at least $\dist(x,y) (1+o(1)).$ Being
 only concerned with this total length condition, as ensured by {\bf (S3)} stating that the
 distance between initial points is close to the single segment length,  we do not require that
 these segments form
 themselves a connected polygonal path or that their ordering agree with the orientation of
 the path. On the other hand, we do impose an explicit lower bound {\bf (S5)} for distance
 between polygonal subpaths crossing different segments, thus ensuring the applicability
 of the decoupling Lemma \ref{MALAROZNICA} in our further argument. It should be emphasised that
 this approach, considerably simplifying our argument in the sequel, can only work in an isometry
 invariant setting, as ours, where it is justified to look only at the total length of phase
 interfaces while ignoring their local directions, ordering etc. 

 We say that a collection $\gamma$ of $\alpha$-large contours dominates an
 $(\alpha,\delta)$-skeleton $\Sigma,$
 write $\gamma \succeq \Sigma,$ iff  $\gamma$ contains a sub-family of
 contours $\gamma'$ with $\gamma' \sim \Sigma.$ 
 Further, by the length of a skeleton $\Sigma = (I_1,E_1,\ldots,I_m,E_m),$
 denoted $\lgth(\Sigma),$ we understand the total length of skeleton
 segments $\sum_{i=1}^m \dist(I_i,E_i).$ We write also $N(\Sigma)$
 for the total number of initial and endpoints in $\Sigma.$ We say
 that a collection $\gamma$ of $\alpha$-large polygonal contours
 is well covered by an $(\alpha,\delta)$-skeleton $\Sigma,$ write
 $\gamma \propto \Sigma,$ if the following holds
 \begin{itemize}
  \item $\gamma \sim \Sigma,$
  \item $\Sigma$ maximises $\lgth(\Sigma)$ among skeletons compatible with $\gamma.$ 
 \end{itemize} 
 
 For an $(\alpha,\delta)$-skeleton $\Sigma$ we consider
 the corresponding {\it black phase area}, denoted in the sequel by
 $\Area(\Sigma),$ and given by
 $$ \Area(\Sigma) := \sup_{\gamma \propto \Sigma}
    \Area\left({\rm black}[\bigcup_{\theta \in \gamma} \theta]\right). $$
 In other words, $\Area(\Sigma)$ is the supremum value of possible
 black phase area which can be enclosed by a collection $\gamma$
 of $\alpha$-large contours well covered by $\Sigma.$ We note that
 for some $\Sigma$ there may be no such $\gamma$ in which
 case we put by convention $\Area(\Sigma) := 0.$  
   
 In the sequel when no ambiguity occurs we will often use the
 $\sim, \succeq, \propto$ notation for contour collection containing also $\alpha$-small
 contours, in which case we shall always mean that the approriate relation holds for
 the corresponding sub-ensemble of $\alpha$-large contours.

 \begin{lemma}\label{SZKIELETY}
  For a collection $\gamma$ of $\alpha$-large contours there
  exists a compatible $(\alpha,\delta)$-skeleton $\Sigma.$
 \end{lemma}

 \paragraph{Proof}
  Choose an initial point $I_1 \in {\Bbb Z}^2 \cap {\Bbb B}_2(L)$
  at a distance less that $1 \slash \sqrt{2}$ from some $\theta_1
  \in \gamma,$ set $I^{\gamma}_1$ to be the point of $\theta_1$ minimising
  the distance to $I_1$ and let $E_1^{\gamma}$ be the first point (say in
  clockwise order) on $\theta_1$ at the distance $\alpha$ from $I^{\gamma}_1$
  (note that the distance considered here and below is the usual 
  Euclidean distance and not the distance along the contour $\theta_1$ !).
  Set $E_1$ to be the point of ${\Bbb B}_2(L) \cap {\Bbb Z}^d$
  which lies the closest to $E^{\gamma}_1.$ The conditions 
  {\bf (S1),(S2)} for $i=1$ is now easily verified.  
  Further, if existing,
  choose $I_2^{\gamma}$ to be the point minimising the distance
  to $\theta_1[I^{\gamma}_1,E^{\gamma}_1]$ (with ties broken
  in an arbitrary way) among the points $I_2^{\gamma}$ in $\gamma$ 
  with the property that there exists $E^{\gamma}_2 \in \theta_2$
  with $\dist(I^{\gamma}_2,E^{\gamma}_2) = \alpha,$
  $\dist(I^{\gamma}_2,x) \leq \alpha$ for all $x \in \theta_2[I^{\gamma}_2,E^{\gamma}_2]$
  and
  $\dist(\theta_1[I^{\gamma}_1,E^{\gamma}_1],\theta_2[I^{\gamma}_2,E^{\gamma}_2])
  \geq \delta,$ where $\theta_2$ is the contour of $\gamma$ containing
  $I^{\gamma}_2.$ Note that if $\theta_1 = \theta_2$ then
  $\dist(I_2^{\gamma},\theta_1[I^{\gamma}_1,E^{\gamma}_1]) = \delta$
  and hence  $\dist(I_1^{\gamma},I_2^{\gamma}) \leq \alpha + \delta.$
 % Note that the pair $I^{\theta}_2,
 % E^{\theta}_2$ occurs on $\theta$ before $\theta$ reaches
 % distance larger than $\alpha + \delta$ from $\theta[I_1^{\theta},
 % E^{\theta}_1]$ \--- in fact whenever this distance is reached by $\theta$,
 % $I_2^{\theta}$ exists
 % and lies at the distance $\delta$ from $\theta[I^{\theta}_1,E^{\theta}_1],$
 % hence at a distance at most $\alpha + \delta$ from $I^{\theta}_1.$
 % Moreover then for each $x \in \theta[E^{\gamma}_1,I^{\theta}_2]$
 % we have $\dist(x,\theta[I^{\theta}_1,E^{\theta}_1]) < \alpha + \delta,$
 % hence $\dist(x,I^{\theta}_1) < 2\alpha + \delta.$
  In case such $I^{\gamma}_2$ and $E^{\gamma}_2$ exist, we
  define $I_2$ and $E_2$ as the best approximations in
  ${\Bbb Z}^2 \cap {\Bbb B}_2(L)$ of $I^{\gamma}_2$ and $E^{\gamma}_2$
  respectively, getting the required relations {\bf (S1),(S2),(S3),(S5)} for $i=2.$
  On the other hand, if such a pair $(I^{\gamma}_2,E^{\gamma}_2)$ fails
  to exist, we conclude that no point of $\gamma$ lies further than
  $\alpha+\delta$ away from $\theta_1[I^{\gamma}_1,E^{\gamma}_1],$
  for otherwise we could find $I^{\gamma}_2$ and $E^{\gamma}_2$ with
  desired properties. In this case by {\bf (S2)} we have $\dist(x,I^{\gamma}_1) 
  \leq 2\alpha+\delta$ for all $x$ in $\gamma$ which yields {\bf (S4)},
  and {\bf (S3),(S5)} are obvious, whence $(I_1,E_1)$ is already
  an $(\alpha,\delta)$-skeleton compatible with $\gamma.$

  We proceed inductively with this construction, adding new pairs
  $(I_{i+1},E_{i+1})$ obtained as the best lattice approximations of
  $(I^{\gamma}_{i+1},E^{\gamma}_{i+1})$ with $I^{\gamma}_{i+1}$ arising
  as the point minimising the distance to $\bigcup_{j \leq i} \theta_j[I^{\gamma}_j,E^{\gamma}_j]$
  among the points $I^{\gamma}_{i+1} \in \theta_{i+1} \in \gamma$ for which there exists
  $E^{\gamma}_{i+1} \in \theta_{i+1}$ with $\dist(I^{\gamma}_{i+1},E^{\gamma}_{i+1}) = \alpha,$ 
  $\dist(I^{\gamma}_{i+1},x) \leq \alpha$ for all $x \in \theta_{i+1}[I^{\gamma}_{i+1},E^{\gamma}_{i+1}],$
  and such that $\dist(\theta_{i+1}[I^{\gamma}_{i+1},E^{\gamma}_{i+1}],\bigcup_{j \leq i}
  \theta_j[I^{\gamma}_j,E^{\gamma}_j]) \geq \delta.$ We note that if $\theta_{i+1} = \theta_j$
  for some $j \leq i$ then $\dist(I^{\gamma}_{i+1},\bigcup_{j \leq i} \theta_j[I^{\gamma}_j,E^{\gamma}_j])
  = \delta$ and hence $\dist(I^{\gamma}_{i+1},\{ I^{\gamma}_1,\ldots,I^{\gamma}_i \}) \leq \alpha + \delta.$ 
  The construction terminates when no further pair can be found, and it is easily
  verified as in the argument above that the resulting collection
  $(I_1,E_1,I_2,E_2,...)$ is an $(\alpha,\delta)$-skeleton
  compatible with $\gamma.$ The proof is complete. $\Box$\\

   Recalling that, by the definition, skeletons have their vertices pairwise different
   and belonging to ${\Bbb B}_2(L) \cap {\Bbb Z}^2$ and hence their number is finite,
   we obtain the following corollary as an immediate conclusion of Lemma \ref{SZKIELETY}.

  \begin{corollary}\label{WYPELNIENIE}
   Each finite collection $\gamma$ of $\alpha$-large contours
   can be well covered by some $(\alpha,\delta)$-skeleton
   $\Sigma.$
  \end{corollary}

  A particular feature of the notion of skeleton as introduced in 
  this section is that if two polygonal subpaths of some contours go
  very close to each other, it may happen that only one
  of these subpaths will contribute to the total length of a
  well-covering skeleton because of the requirement that 
  subpaths going along the segments of the skeleton keep
  distance at least $\delta$ from each other as imposed
  in {\bf (S5)} above. However, this does not lead to problems
  in our  further argument, since we are mainly concerned with
  minimising the skeleton length given the enclosed area, where
  collections consisting of multiple contours are outperformed
  by  singleton ones. This is made formal in the isoperimetric
  lemma below.  
 \begin{lemma}\label{IZOP}
  Assume that $A \ll \alpha^6.$ Then for each $(\alpha,\delta)$-skeleton
  $\Sigma$ in ${\Bbb B}_2(L)$ with $\Area(\Sigma) = A,\; A \in [0,\pi L^2],$
  we have
  $$ \lgth(\Sigma) \geq 2 \sqrt{\pi A}[1-O(\delta \slash \alpha)] - O(\alpha).$$ 
 \end{lemma}

 \paragraph{Proof}
  Below, we restrict our attention to skeletons $\Sigma$ with
  $\lgth(\Sigma) \leq 2\pi L,$ since otherwise our assertion
  is obvious. 

  Pick some collection of $\alpha$-large contours $\gamma^* \propto \Sigma$
  with $\Area({\rm black}[\bigcup_{\theta^* \in \gamma^*} \theta^*]) 
  = \Area(\Sigma) - o(1) = A - o(1),$
 % assumed with no loss of generality
 % to satisfy $\sum_{\theta^*\in \gamma^*} \lgth(\theta^*) = O(L),$
  and observe that, by the definition of the relation $\propto,$
  to prove the lemma it is enough to construct $\Sigma^*$ with
  $\gamma^* \sim \Sigma^*$ and such that, for $A \gg \alpha^2,$ 
  \begin{equation}\label{INDD1}
   \lgth(\Sigma^*) \geq A \psi(A)
  \end{equation}
  for a non-increasing function $A \mapsto \psi(A) = \psi(A;\alpha,\delta)$
  with $A \mapsto A \psi(A)$ non-decreasing, satisfying
  \begin{equation}\label{INDD2}
   \psi(A) = 2 \sqrt{\pi \slash A}[1-O(\delta\slash\alpha)] - O(\alpha \slash A) 
  \end{equation}
  (note that the statement of the lemma trivialises for $A = O(\alpha^2)$). 
  Without loss of generality we can and do assume that $\gamma^*$ contains
  no nested contours, for otherwise we could  simply remove the internal contours
  increasing the area enclosed by $\gamma^*,$ proceed with the construction below
  for the so reduced $\gamma^*$ obtaining $\Sigma^*$ of required length, and then
  construct some additional skeleton segments for the internal contours and add 
  them to $\Sigma^*$ thus increasing its length even further. We also assume
  that $\gamma^*$ contains only contours for which
  $\dist(\theta^*,\bigcup_{\delta^* \in \gamma^* \setminus \{ \theta^* \}} \delta^*) > 64 \alpha.$
  This does not result in loss of generality because finding a sub-collection $\hat{\gamma}^*$
  of contours satisfying this condition and such that all other contours of $\gamma^*$ are
  contained in $64\alpha$-neighbourhood of $\bigcup \hat{\gamma}^*,$ and then constructing
  $\hat{\Sigma}^*$ for $\hat{\gamma}^*,$ we see that the total area enclosed by the
  contours in $\gamma^* \setminus \hat{\gamma}^*$ is of order $O(\alpha \lgth(\hat{\Sigma}^*)),$
  whence by (\ref{INDD1}) for $\hat{\Sigma}^*$ we get $\lgth(\hat{\Sigma}^*) \geq
  [A-O(\alpha \lgth(\hat{\Sigma}^*))] \psi(A-O(\alpha \lgth(\hat\Sigma^*))$ and
  consequently, by (\ref{INDD2}), $\lgth(\Sigma^*) \geq \lgth(\hat{\Sigma}^*) \geq 
  A \psi(A) - O(\alpha)$ provided $\alpha \lgth(\hat\Sigma^*) = o(A).$ 
  The remaining case $\lgth(\hat\Sigma^*) = \Omega(A \slash \alpha)$ is easily
  handled directly, by considering subcases $A = O(\alpha^2)$ and
  $A \gg \alpha^2.$   

  The proof of existence of $\Sigma^*$ satisfying (\ref{INDD1}) goes by induction
  with respect to the number $n^*$ of contours in $\gamma^* = \{ \theta^*_1,\ldots,
  \theta^*_{n^*}\},$ assumed to be ordered by decreasing enclosed area. 
  For $n^*=1$ the assertion follows immediately by standard
  isoperimetric argument: note that the correcting term $A O(\alpha \slash A) =
  O(\alpha)$ coming to the RHS of (\ref{INDD1}) when substituting (\ref{INDD2})
  is due to the admissible distance $\Theta(\alpha)$ between a skeleton and a compatible
  polygonal path [see {\bf (S4)}], while the prefactor $1-O(\delta \slash \alpha)$
  there comes from the fact that, in the single contour case, the distance between
  the initial point of a given skeleton segment and the set of preceding initial
  points may exceed the length of the segment by at most $\delta + 2\sqrt{2}$ 
  [see {\bf (S1),(S3)}], which is fraction $O(\delta \slash \alpha)$ of the
  segment length. 

  To proceed, take $n^* > 1.$ We split our argument into
  three possible cases.
 \begin{description}
 \item{\bf Case 1:} Say that a point $x \in \theta^*_{n^*}$
  is $\alpha$-seen from a contour $\theta_i^* \in \gamma^*$
  iff $\dist(x,\theta^*_i) \leq 4 \alpha.$ Assume that the total
  length of the set $\seen(\theta^*_{n^*},\theta^*_i;\alpha)$
  of all such points
  exceeds $16\alpha$ for some $i < n^*$ and recall that, as
  assumed above, there exists $x \in  \theta^*_{n^*}$ with
  $\dist(x,\bigcup_{j<n^*} \theta_j^*) > 64 \alpha.$ Patching
  $\theta^*_{n^*}$ and $\theta^*_i$ together with additional
  polygonal paths at two extreme points $x_1,x_2$ of
  $\seen(\theta^*_{n^*},\theta^*_i;\alpha)$ and removing
  the internal parts of both contours between $x_1$ and $x_2,$
  we replace $\theta_{n^*}^*$ and $\theta_i^*$ by  a single
  contour $\theta_+^*,$ which can be made disjoint with all
  remaining contours $\theta^*_j,\; j \neq i, j \neq n^*.$
  Denote by $\gamma_+^*$ the contour collection resulting from
  $\gamma^*$ by replacing $\theta^*_{n^*}$ and $\theta_i^*$ by
  $\theta_+^*$ and possibly removing some further contours which 
  would become nested due to this replacement. It is easily seen
  that, by our assumptions above, any skeleton $\Sigma^*_+ \sim \gamma^*_+$
  can be modified into $\Sigma^* \sim \gamma^*$ with $\lgth(\Sigma^*)
  \geq \lgth(\Sigma^*_+).$ Thus, the assertion (\ref{INDD1}) for
  $\gamma^*$ will follow if we are able to find such $\Sigma^*_+$ with
  $\lgth(\Sigma^*_+) \geq  A \psi(A).$ However, this is ensured by the
  inductive hypothesis in view of the relation
  $\Area({\rm black}[\bigcup_{\theta^* \in \gamma^*_+} \theta^*])
   \geq \Area({\rm black}[\bigcup_{\theta^* \in \gamma^*} \theta^*]).$  
 
 \item{\bf Case 2:} Next, suppose that
  $\lgth(\seen(\theta^*_{n^*},\theta^*_i;\alpha)) \leq 16 \alpha$
  for all $i < n^*$ and that $A_{n^*} \gg \alpha^2,$ where $A_{n^*}$
  stands for the area enclosed by $\theta^*_{n^*}.$ Recall 
  in addition that there exists $x \in \theta^*_{n^*}$
  with $\dist(x,\bigcup_{j<n^*} \theta_j^*) > 64 \alpha.$
  We construct an
  $(\alpha,\delta)$-skeleton $\Sigma^*$ as follows.
  Put $\gamma^*_- := \gamma^* \setminus \{ \theta^*_{n^*} \}$ and observe that
  the area enclosed by $\gamma^*_-$ is $A - A_{n^*} - o(1),$ which is
  due to the fact that there is no contour nesting in $\gamma^*$ as
  assumed above. We let $\Sigma_-^*$ be an $(\alpha,\delta)$-skeleton
  such that 
  \begin{equation}\label{NOO1}
   \lgth(\Sigma_-^*) \geq [A-A_{n^*}] \psi(A-A_{n^*}) \geq [A-A_{n^*}] \psi(A), 
  \end{equation} 
  with its existence guaranteed by the inductive hypothesis [note that
  $A-A_{n^*} \gg \alpha^2$ since the contours are ordered by
  decreasing enclosed area]. The skeleton $\Sigma^*_-$ can be extended to a
  skeleton $\Sigma^*$ compatible with $\gamma^*$ by the procedure
  described in the proof of Lemma \ref{SZKIELETY}. Denoting by
  ${\cal S}^* := \Sigma^* \setminus \Sigma^*_-$ the collection of
  newly added segments we see by our assumptions for {\bf Case 2}
  that ${\cal S}^*$ can be in its turn extended to an $(\alpha,\delta)$-skeleton
  $\hat{\cal S}^*$ compatible with $\{ \theta_{n^*}^* \}$ by adding at
  most $O(n^*)$ new segments covering $\seen(\theta^*_{n^*},\theta^*_i;\alpha).$
  Thus, using isoperimetric argument, as applied for the case $n^* = 1$ above, we are led to
  \begin{equation}\label{NOO2}
   \lgth({\cal S}^*) \geq A_{n^*} \psi(A_{n^*}) - O(n^*).
  \end{equation}
  Recall now that the contours $\theta^*_1,\theta^*_2,\ldots$ 
  are ordered by decreasing enclosed area, whence
  \begin{equation}\label{OA}
   A_{n^*} \leq A \slash n^*.
  \end{equation}
  Using (\ref{OA}) to rewrite (\ref{NOO2}) as 
  $\lgth({\cal S}^*) \geq A_{n^*} [\psi(A_{n^*}) - O(n^* \slash A_{n^*})]
  \geq A_{n^*} [\psi(A_{n^*}) $ $- O(A \slash A^2_{n^*})]$ and then applying
  (\ref{INDD2}), noting that $A_{n^*} \leq A \slash 2$ by (\ref{OA}) and
  resorting to standard calculus in order to check that, for $\alpha$
  large enough, we have $\psi(A_{n^*}) - O(A \slash A^2_{n^*}) \geq \psi(A)$
  for $A_{n^*} \gg A^{2 \slash 3},$ we conclude from (\ref{NOO2}) that 
  $\lgth({\cal S}^*) \geq A_{n^*} \psi(A)$ for $A_{n^*} \gg A^{2 \slash 3}.$
  On the other hand, the trivial bound $\lgth({\cal S}^*) \geq \alpha$
  is easily seen to yield $\lgth({\cal S}^*) \geq A_{n^*} \psi(A)$ 
  whenever $A_{n^*} \ll \alpha \sqrt{A}.$ Since we assumed that 
  $A \ll \alpha^6$ in the statement of the lemma, we get 
  $\alpha \sqrt{A} \gg A^{2 \slash 3}$ which leads to
  \begin{equation}\label{NOO3}
   \lgth({\cal S}^*) \geq A_{n^*} \psi(A)
  \end{equation}
  for all $A_{n^*}$ within range of (\ref{OA}). 
  Combining (\ref{NOO3}) with (\ref{NOO1}) and recalling that 
  $\lgth(\Sigma^*) = \lgth(\Sigma^*_-) + \lgth({\cal S}^*)$ 
  yields the required relation (\ref{INDD1}) for {\bf Case 2}.

 \item{\bf Case 3:}
  Assume now that $A_{n^*} = O(\alpha^2).$ Then the required
  relation (\ref{INDD1}) can be obtained along the same lines as in
  {\bf Case 2} by recalling that $\dist(\theta^*_{n^*},\bigcup_{j<n^*} \theta^*_j)
  >  64 \alpha$ and noting that putting such $\theta^*_{n^*}$ into $\gamma^*$
  results in large added length to added area ratio, exceeding $\psi(A).$ The
  only reason for discussing this case separately is the technical fact that
  $\psi(A_{n^*})$ is formally not defined for $A_{n^*} \leq C \alpha^2$
  unless $C$ is large enough.     
\end{description} 
            
 The proof is now complete by induction.
 $\Box$ 

  With the concept of an $(\alpha,\delta)$-skeleton discussed above
  we are now in a position to proceed to the main result of this section. 
  \begin{lemma}\label{SZKIELETOSZ}
   With $\alpha \to \infty, \delta \to \infty$ and $L \gg \alpha \gg \delta \gg \log L,$
   we eventually have for each $(\alpha,\delta)$-skeleton $\Sigma$ in ${\Bbb B}_2(L)$
   $$ {\Bbb P}\left({\cal A}^{[\beta]} \succeq \Sigma\right) \leq
      \exp\left(-\tau^{[\beta]}_{\alpha} \lgth(\Sigma) \right). $$
   % (1+ O(N(\Sigma) \exp(-C \delta))) $$
   % for some positive constant $C.$
  \end{lemma}

 \paragraph{Proof}
  For a contour collection $\gamma \succeq \Sigma,\; \gamma = \{ \theta_1,\ldots,\theta_k\}$
  we consider the partition $\Sigma[\gamma] = \{ {\cal S}[\theta_1],\ldots,{\cal S}[\theta_k] \}$
  of $\Sigma$ into disjoint sub-skeletons ${\cal S}[\theta_j]$ composed of segments $[I,E]$ with
  the corresponding points $I^{\gamma},E^{\gamma},$ as given by {\bf (S2)}, lying on $\theta_j.$
  Note that some ${\cal S}[\theta_j]$ may be empty.  Moreover, for a non-empty sub-skeleton
  ${\cal S} \subseteq \Sigma$ we write $[{\cal S}]$ to denote the family of all contours
  $\theta$ such that $\dist(I,\theta) \leq 1\slash \sqrt{2}$ and $\dist(E,\theta) \leq
  1 \slash \sqrt{2}$ for all segments $[I,E] \in {\cal S}.$ In particular, we always
  have $\theta_j \in [{\cal S}[\theta_j]]$ provided ${\cal S}[\theta_j] \neq \emptyset.$   
  In view of (\ref{WARUNKOWKT}) or, equivalently, by the graphical construction
  of Section \ref{KONSGRAF}, we see that
  \begin{equation}\label{PRZYR1}
     {\Bbb P}\left({\cal A}^{[\beta]} \succeq \Sigma\right) \leq
     \sum_{\{{\cal S}_1,\ldots,{\cal S}_k \}} \int_{[{\cal S}_1] \times ... \times [{\cal S}_k]} 
     {\Bbb P}\left(\bigcup_{j=1}^k \theta_j \cap {\cal A}^{[\beta]} = \emptyset \right)
     {\bf 1}_{\{\{ \theta_1,\ldots,\theta_k \} \sim \Sigma \}} 
     \prod_{j=1}^k d \Theta^{[\beta]}(\theta_j),
  \end{equation}
  where the sum ranges over all possible partitions $\{ {\cal S}_1,\ldots,{\cal S}_k \}$ of 
  $\Sigma$ and with the inequality rather than equality above due to the fact that we do not
  restrict the domain of integration to non-intersecting contours $\theta_j$ and that
  more than one contour of ${\cal A}^{[\beta]}$ might occur in $[{\cal S}_j],$ moreover
  it is not guaranteed that ${\cal S}_j = {\cal S}[\theta_j].$  
  % Dotad 11 wrzesnia wieczor
  We fix a partition $\Sigma = {\cal S}_1 \cup \ldots \cup {\cal S}_k$ and,
  to distinguish between vertices coming from different sub-skeletons
  ${\cal S}_j,\; j=1,\ldots,k,$ we subscript skeleton vertices with the
  corresponding sub-skeleton names, writing $I_{i;{\cal S}_j}$ and
  $E_{i;{\cal S}_j}.$ We also put $\gamma := \{ \theta_1,\ldots,\theta_k \}.$ 
  Denote by $P_{i;j}$ the polygonal subpath 
  $\theta_j[\hat{I}^{\theta_j}_i,\hat{E}^{\theta_j}_i]$
  of the contour $\theta_j \in [{\cal S}_j]$ in the above integral, with
  $\hat{I}^{\theta_j}_i$ and $\hat{E}^{\theta_j}_i$ standing for the points
  of $\theta_j$ closest to $I_{i;{\cal S}_j}$ and $E_{i;{\cal S}_j}$ respectively. 
  Note that
  % even if ${\cal S}_j = {\cal S}[\theta_j],$
  the points
  $\hat{I}^{\theta_j}_i$ and $\hat{E}^{\theta_j}_i$ do not have
  to coincide with $I^{\gamma}_{i;{\cal S}_j}$ and $E^{\gamma}_{i;{\cal S}_j}$
  as specified by the correspondence {\bf (S2)} implied by $\gamma \succeq \Sigma;$ yet we clearly have
  $\dist(I^{\gamma}_{i;{\cal S}_j},\hat{I}^{\theta_j}_i) \leq \sqrt{2}$ and
  $\dist(E^{\gamma}_{i;{\cal S}_j},\hat{E}^{\theta_j}_i) \leq \sqrt{2}.$
  The reason for introducing $\hat{I}^{\theta_j}_i$ and $\hat{E}^{\theta_j}_i$
  rather than simply using $I^{\gamma}_{i;{\cal S}_j}$
  and $E^{\gamma}_{i;{\cal S}_j}$ in their stead is to ensure measurable
  dependence of $P_{i,j}$ on $\theta_j.$   

  Observe that by condition {\bf (S5)} the distance between
  different $P_{i;j}$ does not fall below $\delta.$ Given the collections
  $(\theta) := (\theta_j)$ and $(P):=(P_{i;j})_{i,j}$ we consider the events
  $$ {\cal I}_{i;j}[\theta_j] := \{ {\cal A}^{[\beta]} \cap P_{i;j} = \emptyset \}. $$
  Taking into account that $\Area(P_{i,j} \oplus {\Bbb B}_2(1)) % = O(\lgth(P_{i,j}))
  = O(\alpha^2)$ (see {\bf (S2)}) and using the decoupling Lemma \ref{MALAROZNICA} yields uniformly in
  $(\theta)$
  \begin{equation}\label{NIEZAAL}
   {\Bbb P}\left( \bigcap_{i,j} {\cal I}_{i;j}[\theta_j] \right) =
   \prod_{i,j}^k {\Bbb P}\left( {\cal I}_{i;j}[\theta_j] \right)
   (1+O(\exp(-C \delta) \alpha \lgth(\Sigma) \log N(\Sigma))).
  \end{equation}
  To proceed, note that, by (\ref{PRZYR1}),
  $$
    {\Bbb P}\left({\cal A}^{[\beta]} \succeq \Sigma\right) \leq
    \sum_{\{ {\cal S}_1,\ldots,{\cal S}_k \}} 
    \int_{[{\cal S}_1] \times ... \times [{\cal S}_k]}
     {\Bbb P}\left(\bigcap_{i,j} {\cal I}_{i;j}[\theta_j] \right)
     {\bf 1}_{\{\{ \theta_1,\ldots,\theta_k \} \sim \Sigma \}}     
     \prod_{j=1}^k d \Theta^{[\beta]}(\theta_j)
  $$
  and hence, in view of (\ref{NIEZAAL}), applying the rough bounds
  $N(\Sigma) = O(L^2),\; \alpha = O(L)$ and $\lgth(\Sigma) = O(L^3)$ we obtain
  $$
   {\Bbb P}\left({\cal A}^{[\beta]} \succeq \Sigma\right) \leq
     \sum_{\{ {\cal S}_1,\ldots,{\cal S}_k \}}  
     \int_{[{\cal S}_1] \times ... \times [{\cal S}_k]}
     \prod_{i,j} {\Bbb P}\left({\cal I}_{i;j}[\theta_j] \right)
     \prod_{j=1}^k d \Theta^{[\beta]}(\theta_j)
  $$
  $$ 
   (1+O(\exp(-C \delta) L^4 \log L)).
  $$
  For an endpoint $E_{i;{\cal S}_j}$ we write $\varsigma(E_{i;{\cal S}_j})$
  to denote the skeleton vertex $I_{i';{\cal S}_j}$ or $I_{i';{\cal S}_j}$
  directly succeeding $E_{i;{\cal S}_j}$ in clockwise order on $\theta_j.$  
  Then, by the formulae (\ref{T1}) and (\ref{DODDZIWNE}) for
  $T^{[\beta]}_{(\cdot)}[\cdot \leftrightarrow \cdot]$
  and $\vartheta^{[\beta]}_{(\cdot)}[\cdot \leftrightarrow \cdot]$ respectively,
  in view of the definitions of $\hat{T}^{[\beta]}_{(\cdot)}[\cdot \leftrightarrow \cdot]$
  and $\hat{\vartheta}^{[\beta]}_{(\cdot)}[\cdot \leftrightarrow \cdot]$  as provided
  in Subsection \ref{SNAPPOW1}, and by the definition (\ref{WOLNEKONTURY}) of the
  free contour measure and (\ref{WOLNESCIEZKI}) of the free path measure we are led to
  \begin{equation}\label{PRZYR2} 
    {\Bbb P}\left({\cal A}^{[\beta]} \succeq \Sigma\right) \leq
     \sum_{\{ {\cal S}_1,\ldots,{\cal S}_k\}} \sum_{\varsigma}
     \prod_{i,j} \left( \hat{T}^{[\beta]}_{(\frac{1}{\sqrt{2}})} [I_{i;{\cal S}_j}
     \leftrightarrow E_{i;{\cal S}_j}] \;
     \hat{\vartheta}^{[\beta]}_{(\frac{1}{\sqrt{2}})} [E_{i;{\cal S}_j} \leftrightarrow \varsigma(E_{i;{\cal S}_j})]
    \right) 
   \end{equation}
   $$ 
   (1+O(\exp(-C \delta) L^4 \log L)) C_1^{N(\Sigma)},\; C_1 > 0, 
   $$ 
   where the inner sum ranges over all possible successor assignments $\varsigma$
   and where the extra factor $C_1^{N(\Sigma)}$
   comes from integrating out the configuration of contours $\theta_j$ within
   $1 \slash \sqrt{2}$-neighbourhoods of $I_{i;{\cal S}_j}$ and $E_{i;{\cal S}_j},\; i=1,\ldots,$
   which are subject to optimisation rather than integration in definitions
   of $\hat{T}^{[\beta]}_{(\cdot)}[\cdot
   \leftrightarrow \cdot]$ and $\hat{\vartheta}^{[\beta]}_{(\cdot)}[\cdot \leftrightarrow \cdot].$   
   Recall that $\dist(E_{i;{\cal S}_j},\varsigma(E_{i;{\cal S}_j})) > \delta - \sqrt{2}$ in view of
   {\bf (S5)} and then use the random walk representation of Lemma \ref{BLLOS}
   and (\ref{REPRENOWE}) combined with (\ref{ROZZIEWDODZ}) to conclude that
   $\hat{\vartheta}^{[\beta]}_{(\frac{1}{\sqrt{2}})}[x \leftrightarrow y] = \exp(-\Omega(\delta)).$  
   Thus, taking into account that both the total number of possible partitions $\{ {\cal S}_1,\ldots,
   {\cal S}_k \}$ and the total number of possible successor assignments $\varsigma$ are of order
   $\exp(O(N(\Sigma) \log N(\Sigma))),$ in view of (\ref{PRZYBLNAP}) the relation (\ref{PRZYR2})
   combined with (\ref{ROZZIEW}) gives us
   $$ {\Bbb P}\left({\cal A}^{[\beta]} \succeq \Sigma\right) \leq
      \prod_{i,j} \exp\left(-\tau^{[\beta]}_{\alpha} \dist(I_{i;{\cal S}_j},E_{i;{\cal S}_j})
                  + O(N(\Sigma) \log N(\Sigma) - \Omega(\delta N(\Sigma))) \right) $$ 
   $$ (1+O(\exp(-C \delta) L^4 \log L)). $$
  Since, by definition, $\lgth(\Sigma) = \sum_{i,j} \dist(I_{i;{\cal S}_j},E_{i;{\cal S}_j})$
  and, moreover, $\exp(-C \delta) L^4 \log L = o(1)$ and 
  $\delta N(\Sigma) \gg N(\Sigma) \log N(\Sigma)$ by the assumptions of the lemma,
  we conclude that
  $$ {\Bbb P}\left({\cal A}^{[\beta]} \succeq \Sigma\right) \leq
     \exp\left(-\tau^{[\beta]}_{\alpha} \lgth(\Sigma) \right) $$
  for $\alpha,\delta,L$ large enough, as required. $\Box$

  % Cala powyzsza sekcja sprawdzona 12 IX wieczorem.

\section{Lower bound}\label{LOBO}
 Below, we provide a lower bound for the occurrence probabilities of 
 large contours in ${\cal A}^{[\beta]}.$ This is complementary to the
 upper bounds obtained in the preceding Section \ref{SZKIEE}.  
 For $\alpha, \delta > 0$ and for a piecewise smooth closed curve $\sigma$
 in ${\Bbb R}^2$ we consider the event ${\cal U}[\sigma;\alpha]$ that there
 exists a contour $\theta \in {\cal A}^{[\beta]}$ such that
 $\rho_H(\sigma,\theta) \leq 2\alpha$ with $\rho_H(\cdot,\cdot)$ 
 standing for the usual Hausdorff distance. The following lemma
 gives a lower bound for the probability of such event for 
 $\sigma := {\Bbb S}_1(R) = \partial {\Bbb B}_2(R).$ 
 \begin{lemma}\label{DOLNE}
  With $\alpha \to \infty, \delta \to \infty, R \to \infty$ such that
  $\log R \ll \delta \ll \alpha \ll R$ we have
  $$ {\Bbb P}({\cal U}[{\Bbb S}_1(R);\alpha]) \geq
     \exp\left(-2\pi R \tau^{[\beta]}_{\alpha} - O(\delta R \slash \alpha) \right). $$
 \end{lemma}

 \paragraph{Proof}
  Note that
  \begin{equation}\label{POCZATEK1}
    {\Bbb P}({\cal U}[{\Bbb S}_1(R);\alpha]) \geq
    \int_{\{ \theta \in {\cal C}\;|\;\rho_H(\theta,{\Bbb S}_1(R)) \leq 2 \alpha \}}
    {\Bbb P}\left(\theta \cap {\cal A}^{[\beta]} = \emptyset \right)
    \Theta^{[\beta]}(d\theta) - {\Bbb P}({\cal U}^{(>1)}[{\Bbb S}_1(R);
    \alpha]),
  \end{equation}
  where ${\cal U}^{(>1)}[{\Bbb S}_1(R);\alpha]$ is the event that 
  there exist at least two contours $\theta_1,\theta_2,\ldots$
  in ${\cal A}^{[\beta]}$ such that $\rho_H({\Bbb S}_1(R),\theta_i)
  \leq 2\alpha,\; i=1,2,\ldots.$ Using the conditional graphical
  construction with forbidden regions we easily see that
  ${\Bbb P}({\cal U}^{(>1)}[{\Bbb S}_1(R);\alpha] |
            {\cal U}[{\Bbb S}_1(R);\alpha]) = o(1),$ whence
  (\ref{POCZATEK1}) becomes 
  \begin{equation}\label{POCZATEK2}
    {\Bbb P}({\cal U}[{\Bbb S}_1(R);\alpha]) \geq
    \int_{\{ \theta \in {\cal C}\;|\;\rho_H(\theta,{\Bbb S}_1(R)) \leq 2 \alpha \}}
    {\Bbb P}\left(\theta \cap {\cal A}^{[\beta]} = \emptyset \right)
    \Theta^{[\beta]}(d\theta) (1-o(1)).
  \end{equation} 
  To proceed, 
  we partition the circle ${\Bbb S}_1(R)$ into disjoint segments
  $[I_i,E_i],\; i=1,...,N(R;\alpha,\delta) $ $ = \Theta(R \slash \alpha)$
  separated by spacings of length $\delta$ and such that $\dist(I_i,E_i)
  = \alpha,\; i=1,\ldots,$ $N(R;\alpha,\delta).$ Denote by $\Pi_i$ the square
  $\Pi(I_i \leftrightarrow E_i;1\slash \sqrt{2})$ as defined in the lines preceding
  Lemma \ref{PRZYBLNPT}. Clearly, $\Pi_i$ are disjoint and $\dist(\Pi_i,\Pi_j)
  = \Theta(\delta)$ for $i \neq j.$
  The integral in (\ref{POCZATEK2}) can be bounded below by restricting
  the domain of integration to the family ${\cal C}[I_1,E_1,...]$ of
  paths $\theta$ such that, for all $i=1,\ldots,N(R;\alpha,\delta),$
  $\theta$ contains a subpath $\theta_i$ connecting $\partial {\Bbb B}_2(I_i,1 \slash \sqrt{2})$
  to $\partial {\Bbb B}_2(E_i,1\slash \sqrt{2})$ within $\Pi_i.$
  Using the decoupling Lemma \ref{MALAROZNICA}, with $U_i := \theta \cap
  \Pi_i$ there, we can factorize the integral
  $$ \int_{\{ \theta \in {\cal C}[I_1,E_1,\ldots]\;|\;
    \rho_H(\theta,{\Bbb S}_1(0,R)) \leq 2 \alpha \}}
    {\Bbb P}\left(\theta \cap {\cal A}^{[\beta]} = \emptyset \right)
    \Theta^{[\beta]}(d\theta) $$
  into the product of $T^{[\beta]}_{(1\slash \sqrt{2};\Pi_i)}[I_i \leftrightarrow E_i],\;
  i=1,\ldots,N(R;\alpha,\delta)$ with a prefactor
  $$ (1+O(R \exp(-C_1 \delta) \log N(R;\alpha,\delta))) \exp(O( \delta N(R;\alpha,\delta))),
     \; C_1 > 0,$$
  where $(1+O(R \exp(-C_1 \delta) \log N(R;\alpha,\delta)))$ is the factorization correction
  from Lemma \ref{MALAROZNICA} while $\exp(O(\delta N(R;\alpha,\delta)))$
  comes from {\it patching} the contour $\theta$ by joining together the
  subpaths $\theta_i$ passing through adjacent $\delta$-distant squares
  $\Pi_i$ so as to keep the resulting path 
  within distance $2\alpha$ from ${\Bbb S}_1(R),$ see the discussion of
  (\ref{ROZBICIE}) and (\ref{OGRANAE}) above.
  Since $R \exp(-C_1 \delta) \log N(R;\alpha,\delta) = o(1),$ we obtain
  $$  {\Bbb P}({\cal U}[{\Bbb S}_1(R);\alpha]) \geq
      \exp(O(\delta N(R;\alpha,\delta)))
      \prod_{i=1}^{N(R;\alpha,\delta)}
      T^{[\beta]}_{(1 \slash \sqrt{2};\Pi_i)}[I_i \leftrightarrow E_i]. $$
  Applying Lemma \ref{PRZYBLNPT} we conclude that
  $$  {\Bbb P}({\cal U}[{\Bbb S}_1(R);\alpha]) \geq
       O(C_2^{N(R;\alpha,\delta)})
       \exp(O(\delta N(R;\alpha,\delta)))
     \prod_{i=1}^{N(R;\alpha,\delta)}
      T^{[\beta]}_{(1 \slash \sqrt{2})}[I_i \leftrightarrow E_i],\; C_2 > 0. $$
  Observing that $N(R;\alpha,\delta) = \Theta(R \slash \alpha)$
  completes the proof in view of the definition (\ref{PRZYBLNAP})
  of $\tau^{[\beta]}_{\alpha}.$ $\Box$

\section{Proof of the main theorem}\label{GLOTWI}
 Throughout this proof we shall put
 \begin{equation}\label{AADD}
  \alpha = \alpha[L] := \sqrt{L} \log L \mbox{ and } \delta = \delta[L] := (\log L)^2.
 \end{equation}
 As in the classical DKS theory, our argument below uses
 the decomposition of the contour ensemble
 ${\cal A}^{[\beta]} \cap {\Bbb B}_2(L)$ into the collection
 ${\Bbb L}_{\alpha;L} := {\Bbb L}_{\alpha;L}\left[{\cal A}^{[\beta]}\right]$ of
 $\alpha$-large contours and the remaining family of $\alpha$-small contours,
 and it relies on an application of the skeleton bounds of Section
 \ref{SZKIEE} and complementary estimates of Section \ref{LOBO},
 combined with the use of moderate deviation results of Section \ref{MDCE}. 

 \subsection{Lower bound for (\ref{PSTWAA})}
  In order to prove (\ref{PSTWAA}) we establish first the lower bound
  \begin{equation}\label{DOLNOOG}
   {\Bbb P}\left( \blk_L({\cal A}^{[\beta]}) \geq \blk[\beta] \pi L^2 + a L^2,\;
    {\cal N}[\alpha;L] \; {\rm holds} \right)
    \geq \exp\left( -  \sqrt{\frac{2\pi a}{|\blk[\beta]|}} L
                       \tau^{[\beta]}_{\alpha} +O(\alpha) \right).
 \end{equation}
 To show it, put
 \begin{equation}\label{DOR1}
   R := L \sqrt{\frac{a}{2\pi|\blk[\beta]|}} + C \alpha
 \end{equation}
 for some constant $C$ large enough so that
 \begin{equation}\label{DOR2}
    {\Bbb P}\left( \blk_L({\cal A}^{[\beta]}) \geq \blk[\beta] \pi L^2 + a L^2
    \;|\; {\cal U}[{\Bbb S}_1(R);\alpha] \right) > 1 \slash 2,
 \end{equation}
 with the event ${\cal U}[{\Bbb S}_1(R);\alpha],$ indicating the existence 
 of a contour $\theta$ of ${\cal A}^{[\beta]}$ with $\rho_H(\theta,{\Bbb S}_1(R))
 $ $\leq 2 \alpha,$ defined as in Section \ref{LOBO}. Clearly, $R < L$ for 
 $L$ large enough because $a < 2 \pi |\blk[\beta]|.$ To see that the required
 choice of $C$ in (\ref{DOR1}) is indeed possible note first that 
 $$ {\Bbb P}\left( \blk_L({\cal A}^{[\beta]}) \leq \blk[\beta] \pi L^2 + a L^2
    \;|\; {\cal U}[{\Bbb S}_1(R);\alpha] \right) \leq $$ 
 $$ \frac{1}{{\Bbb P}({\cal U}[{\Bbb S}_1(R);\alpha])}
    \int_{\{ \theta \in {\cal C},\; \rho_H(\theta,{\Bbb S}_1(R)) \leq 2 \alpha \}}
         {\Bbb P}\left(\blk_L({\cal A}^{[\beta]}_{{\Bbb R}^2;\theta} \cup \theta) \leq
         \blk[\beta] \pi L^2 + a L^2\right) \Theta^{[\beta]}(d\theta). $$
 Then use (\ref{RATUJACE}) to conclude that, for $\rho_H(\theta,{\Bbb S}_1(R)) \leq 2\alpha,$  
 $$ {\Bbb E}\blk_L\left({\cal A}^{[\beta]}_{{\Bbb R}^2;\theta} \cup \theta \right) 
    = \blk[\beta] \pi L^2 + a L^2 + 4 \pi |\blk[\beta]| C R \alpha + O(L \alpha), $$
 which can be made larger than $\blk[\beta] \pi L^2 + a L^2$ by a term of order
 $\Theta(L \alpha)$ under appropriate choice of $C.$ In view of Corollary
 \ref{DODAANE} this makes the integrand probability 
 ${\Bbb P}(\blk_L({\cal A}^{[\beta]}_{{\Bbb R}^2;\theta}
 \cup \theta) \leq $ $  \blk[\beta] \pi L^2 + a L^2 )$ arbitrarily close to $0,$
 uniformly in $\theta$ with $\rho_H({\Bbb S}_1(R),\theta) \leq 2\alpha.$ In 
 particular, (\ref{DOR2}) is seen to hold under such choice of $C,$ as required.    
 To proceed, observe that the conditional probability
 ${\Bbb P}({\cal N}^c[\alpha;L] | {\cal U}[{\Bbb S}_1(R);\alpha]),$
 being bounded above by $[{\Bbb P}({\cal U}[{\Bbb S}_1(R);\alpha])]^{-1}
    \int_{\{ \theta \in {\cal C},\; \rho_H(\theta,{\Bbb S}_1(R)) \leq 2 \alpha \}}
         {\Bbb P}( {\cal N}^c[\alpha;L] \mbox{ holds for } {\cal A}^{[\beta]}_{{\Bbb R}^2;\theta} )
         \Theta^{[\beta]}(d\theta),$ 
 tends to $0$ as $L \to \infty$ by the results of Lemma
 \ref{DLG} in Section \ref{EXPTI} specialised for
 ${\cal A}^{[\beta]}_{{\Bbb R}^2;\theta}.$ 
 Thus, we conclude from (\ref{DOR2}) that
 for sufficiently large $L$
 $$ {\Bbb P}\left( \blk_L({\cal A}^{[\beta]}) \geq \blk[\beta] \pi L^2 + a L^2,\;
    {\cal N}[\alpha;L] \; {\rm holds} \;|\; {\cal U}[{\cal S}_1(R);\alpha] \right)
    > 1 \slash 4. $$
  The required relation (\ref{DOLNOOG}) follows now by Lemma \ref{DOLNE} in view of
  (\ref{DOR1}).
 
 \subsection{Upper bound for (\ref{PSTWAA})}
 To complete the proof of (\ref{PSTWAA}) we shall establish the following upper bound,
 complementary to (\ref{DOLNOOG}),
 \begin{equation}\label{GOROG}
   {\Bbb P}\left( \blk_L({\cal A}^{[\beta]}) \geq \blk[\beta] \pi L^2 + a L^2,\;
    {\cal N}[\alpha;L] \; {\rm holds} \right)
    \leq \exp\left( -  \sqrt{\frac{2\pi a}{|\blk[\beta]|}} L
                       \tau^{[\beta]}_{\alpha} +O(\alpha) \right).
 \end{equation}
 To this end use the exponential tightness bound in Lemma \ref{DLG} to get for
 some $C_1 = C_1(a)$ and $C_2 > \sqrt{\frac{2\pi a}{|\blk[\beta]|}} \tau^{[\beta]}$
 $$
  {\Bbb P}\left( \blk_L( {\cal A}^{[\beta]}) \geq \blk[\beta] \pi L^2 + a L^2,\;
  {\cal N}[\alpha;L] \; {\rm holds} \right) \leq $$
 $$
  {\Bbb P}\left( \blk_L\left( {\cal A}^{[\beta]} \right) \geq \blk[\beta] \pi L^2 + a L^2,\;
                 \lgth ( {\Bbb L}_{\alpha;L}) \leq C_1 L,\; {\cal N}[\alpha;L]
                 \; {\rm holds} \right) + O(\exp(-C_2 L)). $$
 Applying Lemma \ref{SZKIELETY} together with Corollary \ref{WYPELNIENIE}
 we see that this probability is bounded above by
 \begin{equation}\label{ROZKLADPSTWA}
   \sum_{\Sigma}
    {\Bbb P}\left( \blk_L({\cal A}^{[\beta]}) \geq \blk[\beta] \pi L^2 + a L^2,\;
    \lgth ( {\Bbb L}_{\alpha;L}) \leq C_1 L,\; {\Bbb L}_{\alpha;L} \propto
  \Sigma \right) + O(\exp(-C_2 L)),
 \end{equation}
 where the sum above is taken over all $(\alpha,\delta)$-skeletons $\Sigma$ contained
 in ${\Bbb B}_2(L-[4\alpha-\delta-\sqrt{2}]).$ Note that we could restrict our
 attention to $\Sigma \subseteq {\Bbb B}_2(L-[4\alpha-\delta-\sqrt{2}])$ because of working
 on the event ${\cal N}[\alpha,L],$ see {\bf (S4)}. It should also be noted that any contour
 collection well covered by such $\Sigma$ is completely contained in ${\Bbb B}_2(L)$ 
 for $L$ and $\alpha[L]$ large enough. 
 % with the
 % additional condition that $\Sigma$ keeps distance more than $2\alpha - \delta - \sqrt{2}$
 % from ${\Bbb S}_1(L).$
 Under the imposed requirement that $\lgth ( {\Bbb L}_{\alpha;L} ) \leq C_1 L,$ the total
 length of skeleton segments in any $\Sigma$ with ${\Bbb P}({\Bbb L}_{\alpha;L}
 \propto \Sigma) > 0$ is also of order $O(L),$ whence
 the sum in (\ref{ROZKLADPSTWA}) can be restricted only to skeletons of such
 length order. Observe now that the number of such skeletons is of
 order $\exp(O(\alpha^{-1} L \log L ))$ which, by (\ref{AADD}), is
 $\exp(O(\sqrt{L})).$ Indeed, constructing the skeleton {\it segment
 after segment} at each step we have at most $O(L^2)$ possibilities of choosing
 new initial$\slash$end point. However, the total number of such steps, coinciding
 with twice the number of segments, is at most of order $O(L \slash \alpha),$
 because, as stated above, we only consider skeletons $\Sigma$
 with $\lgth(\Sigma) = O(L)$ and the length of a single segment is close
 to $\alpha,$ see {\bf (S1)}. We put this statement as a remark for
 further reference
 \begin{remark}\label{ILKONT}
  The number of $(\alpha,\delta)$-skeletons $\Sigma$
  in ${\Bbb B}_2(L)$ with $\lgth(\Sigma) = O(L)$ is
  of order  $\exp(O(\alpha^{-1} L \log L)).$ 
 \end{remark}
 Consequently, by (\ref{ROZKLADPSTWA}), in order
 to establish (\ref{GOROG}) it is enough to show that
 $$
  \max_{\Sigma} {\Bbb P}\left( \blk_L({\cal A}^{[\beta]}) \geq \blk[\beta] \pi L^2 + a L^2,\;
    \lgth ({\Bbb L}_{\alpha;L}) \leq C_1 L,\; {\Bbb L}_{\alpha;L} \propto \Sigma
   \right)
 $$
 \begin{equation}\label{DOPOOK}
   \leq \exp\left( - \sqrt{\frac{2\pi a}{|\blk[\beta]|}} L \tau^{[\beta]}_{\alpha} +
    O(\alpha) \right),
 \end{equation}
 with the maximum taken over all $(\alpha,\delta)$-skeletons $\Sigma$
 satisfying the conditions specified above (i.e. contained in ${\Bbb B}_2(L-[4\alpha-\delta-\sqrt{2}])$
 and with total length of order $O(L)$).
% and distance more than $2\alpha - \delta - \sqrt{2}$
% from ${\Bbb S}_1(R)$). 
 To proceed with the verification of (\ref{DOPOOK}) choose a skeleton $\Sigma_0$
 which achieves the above maximum. Putting $\nu := \Area(\Sigma_0)$ and
 $\lambda := \lgth(\Sigma_0)$ we conclude by the isoperimetric Lemma \ref{IZOP}
 that
 $$ \lambda \geq 2 \sqrt{\pi \nu} [1-O(\delta \slash \alpha)] - O(\alpha). $$ 
 Using that $\nu = O(L^2)$ and that $\delta L \slash \alpha = \alpha$  we obtain
 \begin{equation}\label{ZASTRZ}
  \lambda \geq 2\sqrt{\pi \nu} - O(\alpha). 
 \end{equation}
 % Dotad 16 wrzesnia rano
 To proceed, recall that $\blk[\beta] \in (-1,0)$ and observe that on the event
 ${\Bbb L}_{\alpha;L} \propto \Sigma_0$ we get
 by (\ref{RATUJACE}) and {\bf (S4)}
 $$ {\Bbb E}\blk_L({\cal A}^{[\beta];\alpha,{\Bbb B}_2(L)}_{{\Bbb R}^2:\gamma} \cup \gamma)
    \leq \blk[\beta] (\pi L^2 - \nu) - \nu \blk[\beta] + O(L \alpha). $$
 Thus, noting that the field ${\cal A}^{[\beta]}$
 conditioned on ${\Bbb L}_{\alpha;L} = \gamma$ coincides in law with
 ${\cal A}^{[\beta];\alpha,{\Bbb B}_2(L)}_{{\Bbb R}^2:\gamma} \cup \gamma$
 and recalling that $N(\Sigma) = O(\lambda \slash \alpha),$ 
 we conclude from Lemma \ref{SZKIELETOSZ} and Theorem \ref{SREDNIEO}
 applied conditionally on ${\Bbb L}_{\alpha;L}$ that the probability
 maximised in (\ref{DOPOOK}) is bounded above by
 \begin{equation}\label{DOMAKS}
  \exp\left( -\tau^{[\beta]}_{\alpha} \lambda \right)
  \exp\left( - c \left[ \frac{\Delta^2}{L^2} \wedge \frac{\Delta}{\alpha} \right] \right),
 \end{equation}
 where $\Delta := [\blk[\beta] (\pi L^2 - \nu) - \nu \blk[\beta] + O(L\alpha)] - [\blk[\beta]\pi L^2 + a L^2]
 = 2\nu |\blk[\beta]| - aL^2 + O(L\alpha)$ is the difference between the expected and actual
 (required) magnetisation on the event ${\Bbb L}_{\alpha;L} \propto \Sigma_0,$
 and with $\nu$ and $\lambda$ related by (\ref{ZASTRZ}). Recalling that $\lambda = O(L),
 \alpha = \sqrt{L} \log L,$ $\delta = (\log L)^2$ and applying the lower 
 bound (\ref{DOLNOOG}) we see that the maximum in (\ref{DOMAKS}) has to be reached
 with $\Delta = O(L^{3\slash 2}\log L) = O(L\alpha)$ and, consequently,
 \begin{equation}\label{DLANU}
  \nu = \frac{a L^2}{2 |\blk[\beta]|} + O(L \alpha),
 \end{equation} 
 whence, by (\ref{ZASTRZ}), 
 \begin{equation}\label{DLALAMBDY}
  \lambda = L\sqrt{\frac{2\pi a}{|\blk[\beta]|}} + O(\alpha),
 \end{equation}
 with the equality rather than inequality in the last formula
 due to (\ref{DOLNOOG}). 
 By Lemma \ref{SZKIELETOSZ} this yields the required relation
 (\ref{DOPOOK}) and hence completes the proof of (\ref{GOROG}).

 \subsection{Existence of a large contour}\label{ISTNKONT}
  In view of the lower bound (\ref{DOLNOOG}), the argument leading to (\ref{DOMAKS})
  with the optimal skeleton $\Sigma_0$ replaced by a generic skeleton $\Sigma$ 
  shows that, conditionally on the event $\{ \blk_L\left( {\cal A}^{[\beta]}
  \right) > \blk[\beta] \pi L^2 + a L^2,\; {\cal N}[\alpha,L] \mbox{ holds}
  \},$ with probability tending to $1$ we can have ${\Bbb L}_{\alpha,L} 
  \propto \Sigma$ only for those $(\alpha,\delta)$-skeletons
  $\Sigma$ which satisfy (\ref{DLANU}) and (\ref{DLALAMBDY})
  with $\nu = \Area(\Sigma)$ and $\lambda = \lgth(\Sigma).$ By the
  definition of the relation $\propto$ and by the proof of the
  isoperimetric Lemma \ref{IZOP} this means that with conditional probability
  tending to $1$ on the event  $\{ \blk_L\left( {\cal A}^{[\beta]}
  \right) > \blk[\beta] \pi L^2 + a L^2,\; {\cal N}[\alpha,L] \mbox{ holds}
  \}$ there exists at least one contour $\theta_{\rm large}$ of length
  $L\sqrt{\frac{2\pi a}{|\blk[\beta]|}} + O(\alpha)$ and enclosing 
  area $\frac{a L^2}{2 |\blk[\beta]|} + O(L \alpha).$ In fact, we claim that 
  for $K$ large enough, conditionally on $\{ \blk_L\left( {\cal A}^{[\beta]}
  \right) > \blk[\beta] \pi L^2 + a L^2,\; {\cal N}[\alpha,L] \mbox{ holds}
  \},$ with probability arbitrarily close to $1$ the contour $\theta_{\rm
  large}$ is the only $K \alpha$-large contour of ${\cal A}^{[\beta]}$
  in ${\Bbb B}_2(L).$ Indeed, for each $\Sigma$ as above, i.e. satisfying
  (\ref{DLANU}) and (\ref{DLALAMBDY}), we have 
  $$ {\Bbb P}\left( {\Bbb L}_{\alpha,L} \propto
      \Sigma,\; {\cal A}^{[\beta]} \mbox{ contains more than one
      $K \alpha$-large contour in } {\Bbb B}_2(L),\; {\cal N}[\alpha,L]
      \mbox{ holds } \right) $$
  $$ \leq \int_{\theta_{\rm large}} {\Bbb P}\left(
     \lgth {\Bbb L}_{\alpha,L}
     \left({\cal A}^{[\beta]}_{{\Bbb R}^2:\theta_{\rm large}}\right) 
     \geq K \alpha \right) \Theta^{[\beta]}(d\theta_{\rm large}), $$
  where the integral ranges over $\theta_{\rm large}$ in ${\Bbb B}_2(L)$
  of length $L\sqrt{\frac{2\pi a}{|\blk[\beta]|}} + O(\alpha)$ and
  enclosing area $\frac{a L^2}{2 |\blk[\beta]|} + O(L \alpha).$ 
  Now, Remark \ref{ILKONT} and Lemma \ref{SZKIELETOSZ} imply that the
  total mass $\Theta^{[\beta]}(\cdot)$ of such $\theta_{\rm large}$'s
  is of order $\exp\left( -  \sqrt{\frac{2\pi a}{|\blk[\beta]|}} L
  \tau^{[\beta]}_{\alpha} +O(\alpha) \right).$ Moreover, by Lemma
  \ref{DLG} applied to ${\cal A}^{[\beta]}_{{\Bbb R}^2:\theta_{\rm large}}$
  the integrand probability is uniformly of order 
  $O(\exp(-K\alpha)).$ We now conclude our claim for $K$ 
  large enough in view of the lower bound (\ref{DOLNOOG}).
 % that, conditionally on $\{ \blk_L\left( {\cal A}^{[\beta]}
 % \right) > \blk[\beta] \pi L^2 + a L^2,\; {\cal N}[\alpha,L] \mbox{ holds}
 % \},$ with probability arbitrarily close to $1$ the contour $\theta_{\rm
 % large}$ is the only $\alpha \log L$-large contour of ${\cal A}^{[\beta]}$
 % in ${\Bbb B}_2(L).$      

\subsection{Uniqueness of the large contour, excluding intermediate contours}
 It follows by the previous Subsection \ref{ISTNKONT} that, conditionally on the event
 $\{ \blk_L\left( {\cal A}^{[\beta]} \right) > \blk[\beta] \pi L^2 +
      a L^2,\; {\cal N}[\alpha,L] $ $ \mbox{ holds} \},$ with overwhelming
 probability there exists one large contour $\theta_{\rm large}$ of length 
 $L\sqrt{\frac{2\pi a}{|\blk[\beta]|}} + O(\alpha),$ enclosing phase
 area $\frac{a L^2}{2 |\blk[\beta]|} + O(L \alpha),$ and this is the only
 $K \alpha$-large contour of ${\cal A}^{[\beta]}$ hitting ${\Bbb B}_2(L),$
 with $K$ large enough. Below, we argue that for sufficiently large
 $C_{\rm large},$ 
 with overwhelming conditional probability, $\theta_{\rm large}$ is in
 fact the unique $C_{\rm large} \log L$-large contour of
 ${\cal A}^{[\beta]}$ hitting
 ${\Bbb B}_2(L).$ The first step in this direction is showing in
 Lemma \ref{ODCHYLKI}, similar to Lemma 4.2.4 in \cite{IS1}, that the phase of $K \alpha$-large contours adjusts
 very tightly to the micro-canonical constraint $\blk_L({\cal A}^{[\beta]}) >
 [\blk[\beta] \pi+a] L^2$ and, roughly speaking, 'not much work is left
 for small contours'. Next, in Lemma \ref{TYLKOJEDEN} we use this 
 knowledge to deduce the uniqueness of the large contour $\theta_{\rm large}$
 and to exclude the presence of any other $C_{\rm large} \log L$-large contours with
 overwhelming probability under the micro-canonical constraint.      

 To proceed with the first of the afore-mentioned steps, we claim
 first that 
\begin{lemma}\label{ODCHYLKI}
 With $K$ as specified above we have
 $$ {\Bbb P}\left( \left. [\blk[\beta] \pi +a] L^2  - {\Bbb E}\left(
    \left. \blk_L({\cal A}^{[\beta]}) \right| {\Bbb L}_{K\alpha;L} \right)
     > L^{4 \slash 3} \right| \right. $$ 
    $$\left. \blk_L({\cal A}^{[\beta]}) > [\blk[\beta] \pi+a] L^2,
    {\cal N}[\alpha,L] \; {\rm holds} \right) = o(1). $$
\end{lemma} 

\paragraph{Proof}
 We set
 $$ \rho = \rho[L] := L^{7 \slash 12}. $$
 Applying Lemma \ref{SZKIELETY} and Corollary \ref{WYPELNIENIE} we get 
 $${\Bbb P}\left( [\blk[\beta] \pi + a] L^2 - 
   {\Bbb E}\left( \left. \blk_L({\cal A}^{[\beta]}) \right| {\Bbb L}_{K\alpha;L} \right)
                    > L^{4 \slash 3},\; \blk_L({\cal A}^{[\beta]}) > [\blk[\beta] \pi+a] L^2 \right) \leq $$
 $$ \sum_{\Sigma} {\Bbb P}\left(  \blk_L({\cal A}^{[\beta]}) > [\blk[\beta] \pi+a] L^2,\;
                  [\blk[\beta] \pi + a] L^2 -  
 {\Bbb E}\left( \left. \blk_L({\cal A}^{[\beta]}) \right| {\Bbb L}_{K\alpha;L} \right) 
                    >
                  L^{4 \slash 3}, \right. $$ 
 $$ \left. {\Bbb L}_{K\alpha;L} \stackrel{(K\alpha,\delta)}
                      {\propto} \Sigma \right) $$
 with the sum ranging over all $(K\alpha,\delta)$-skeletons $\Sigma$
 contained in ${\Bbb B}_2(L)$ and with $\stackrel{(K\alpha,\delta)}{\propto}$
 used as an indexed version of $\propto$ to denote the well-covering relation of
 $(K\alpha,\delta)$-contours by $(K\alpha,\delta)$-skeletons.  
 Use the exponential tightness results of
 Lemma \ref{DLG} in Section \ref{EXPTI}
 to conclude that, with arbitrarily large $C_1$ and with  $C_2$ large enough,
 this sum can be bounded above by 
 $$ \sum_{\Sigma,\; \lgth(\Sigma) \in \left[L
    \sqrt{\frac{2\pi a}{|\blk[\beta]|}} - \rho,
    C_2 L \right]} {\Bbb P}\left(  [\blk[\beta] \pi + a] L^2 -
     {\Bbb E}\left( \left. \blk_L({\cal A}^{[\beta]}) \right| {\Bbb L}_{K\alpha;L} \right) 
                    > L^{4 \slash 3},\;{\Bbb L}_{K\alpha;L}
                      \stackrel{(K\alpha,\delta)}
                      {\propto} \Sigma \right) $$ 
 \begin{equation}\label{NIEISTOTNE}
  + \sum_{\Sigma,\; \lgth(\Sigma) < L \sqrt{\frac{2\pi a}{|\blk[\beta]|}} - \rho}
                {\Bbb P}\left(\blk_L({\cal A}^{[\beta]}) > [\blk[\beta] \pi+a] L^2,\;
                      {\Bbb L}_{K\alpha;L} \stackrel{(K\alpha,\delta)}
                      {\propto} \Sigma \right)
                    + \exp(-C_1 L).
 \end{equation}
 We proceed by showing that all consecutive terms in (\ref{NIEISTOTNE}),
 for brevity denoted below by $P_1,P_2$ and $P_3$ respectively, are
 negligibly small compared to the probability $P_4 :=
 {\Bbb P}( \blk_L({\cal A}^{[\beta]}) > [\blk[\beta] \pi+a] L^2,
 {\cal N}[\alpha,L] \; {\rm holds}).$ 
 To begin with the first term $P_1,$ use Remark \ref{ILKONT} to conclude that the
 number of summands in this sum is of order $\exp(O(\alpha^{-1} L \log L)).$
 Moreover, applying Lemma \ref{SZKIELETOSZ}, noting that conditionally 
 on ${\Bbb L}_{K\alpha;L} = \gamma$ the field ${\cal A}^{[\beta]}$ coincides
 in law with ${\cal A}^{[\beta];K\alpha,{\Bbb B}_2(L)}_{{\Bbb R}^2:\gamma}
 \cup \gamma$ and using Theorem \ref{SREDNIEO} conditionally on 
 ${\Bbb L}_{K\alpha;L},$ we uniformly bound above each summand of $P_1$ by
 $$ \exp\left( - \left[\sqrt{\frac{2\pi a}{|\blk[\beta]|}} L - \rho \right]
     \tau^{[\beta]}_{K \alpha} \right)
     \exp\left( - c \left[ \frac{L^{8 \slash 3}}{L^2}
             \wedge \frac{L^{4 \slash 3}}{K\alpha}\right] \right) =
 $$ 
 $$ \exp\left( - \sqrt{\frac{2\pi a}{|\blk[\beta]|}} L \tau^{[\beta]}_{K\alpha}
    + O(\rho) \right) \exp\left( - c L^{2 \slash 3} \right). $$
 Recalling the definition of $\alpha = \alpha[L] = \sqrt{L} \log L,\;
 \rho=L^{7 \slash 12}$ and
 using  the lower bound (\ref{DOLNOOG}) of Theorem \ref{GLOWNE}
 with $\alpha$ replaced there by $K\alpha,$ we conclude that 
 \begin{equation}\label{DLAP1}
  P_1 = o(P_4).
 \end{equation}
 To show that 
 \begin{equation}\label{DLAP2}
  P_2 = o(P_4)
 \end{equation} 
 observe that, by isoperimetric Lemma \ref{IZOP}, $\lgth(\Sigma) < 
 L \sqrt{\frac{2\pi a}{|\blk[\beta]|}} - \rho$ implies that 
 $\Area(\Sigma) < \frac{a L^2}{2 |\blk[\beta]|} - \Omega(L \rho),$
 whence, by (\ref{RATUJACE}),  
 ${\Bbb E}\left( \left. \blk_L({\cal A}^{[\beta]}) \right|
  {\Bbb L}_{K\alpha;L} \right) \leq [\blk[\beta]\pi + a] L^2
  - \Omega(L \rho)$ almost surely on the event
 $\{ {\Bbb L}_{K\alpha;L} \stackrel{(K\alpha,\delta)}{\propto} \Sigma \}.$
 Consequently, by Theorem \ref{SREDNIEO} applied conditionally 
 on ${\Bbb L}_{K\alpha;L},$ each summand in $P_2$ is bounded above by  
 $\exp\left( - c \left[ \frac{L^2 \rho^2}{L^2} \wedge \frac{L
                 \rho}{\alpha} \right] \right) = \exp\left(-c \frac{L\rho
  }{\alpha} \right).$ Using the lower bound (\ref{DOLNOOG}) of
  Theorem \ref{GLOWNE} (with $\alpha$ replaced there by $K\alpha$) 
  and recalling that $\frac{\rho L}{\alpha} \gg L$ we obtain (\ref{DLAP2}).
  Observing that, by the same lower bound (\ref{DOLNOOG}),
  $P_3 = o(P_4)$ provided $C_1$ is chosen large enough, we complete
  the proof of the Lemma by combining (\ref{DLAP1}) and (\ref{DLAP2}).
  $\Box$ \\

 As announced above, our next statement will allow us to exclude
 with overwhelming conditional probability under the micro-canonical constraint
 the presence of $C_{\rm large} \log L$-large contours different
 than $\theta_{\rm large}.$ 
 
\begin{lemma}\label{TYLKOJEDEN}
 There exists a constant $C_{\rm large} > 0$ such that uniformly in
 collections $\gamma$ of $K \alpha$-large contours in ${\Bbb B}_2(L)$
 with $\lgth(\gamma) \leq L \log L$ and in $\Delta \leq L^{4 \slash 3}$ we have
 $$ {\Bbb P}\left( \left.
    {\cal A}^{[\beta];K\alpha,{\Bbb B}_2(L)}_{{\Bbb R}^2:\gamma} \;
    {\rm contains \; a \;} C_{\rm large} \log L{\rm-large\; contour} \; \right| 
    \blk_L\left({\cal A}^{[\beta];K\alpha,{\Bbb B}_2(L)}_{{\Bbb R}^2:\gamma}
    \cup \gamma \right) > \right. $$
 $$ \left. {\Bbb E}\blk_L\left({\cal A}^{[\beta];K\alpha,
    {\Bbb B}_2(L)}_{{\Bbb R}^2:\gamma} \cup \gamma\right) + \Delta \right) = o(1). $$
\end{lemma}   

\paragraph{Proof}
 For brevity write
 $$ \mu^{K\alpha}_{L,\gamma} := \blk_L\left({\cal A}^{[\beta];\alpha,
    {\Bbb B}_2(L)}_{{\Bbb R}^2:\gamma} \cup \gamma\right),\;
   \mu_{L,\gamma}^{K\alpha,h} :=
   \blk_L\left({\cal A}^{[\beta,h];\alpha,{\Bbb B}_2(L)}_{{\Bbb B}_2(L):\gamma}
  \cup \gamma \right) $$
 and let ${\cal E}[C_{\rm large},L]$ be the event that
 ${\cal A}^{[\beta];\alpha,{\Bbb B}_2(L)}_{{\Bbb R}^2:\gamma}$ contains
 no $C_{\rm large} \log L$-large contours hitting ${\Bbb B}_2(L)$ 
 and ${\cal E}^h[C_{\rm large},L]$
 the event that 
 ${\cal A}^{[\beta,h];K\alpha,{\Bbb B}_2(L)}_{{\Bbb B}_2(L):\gamma}$
 contains no $C_{\rm large} \log L$-large contours hitting ${\Bbb B}_2(L).$ 
 From Corollary \ref{POLAUSTAWIONE} it follows in particular that
 for each $\eta \in \left[\Delta,L^{4 \slash 3} \log L\right]$ 
 there exists a unique value of the external magnetic field 
 $h[\eta,L] := h[\eta,L,\gamma] = \Theta(\eta \slash L^2)$
 such that
 \begin{equation}\label{DOPASPOLE}
  {\Bbb E} \mu_{L,\gamma}^{K\alpha,h[\eta,L]} = {\Bbb E}
  \mu_{L,\gamma}^{K\alpha} + \eta
 \end{equation}
 and, moreover, $h[\eta,L]$ increases with $\eta$ given $L,$ whence
 $h[\eta,L] \in [h^-[L],h^+[L]]$ with
 $$ h^-[L] := h[\Delta, L] = \Theta(\Delta \slash L^2)$$
 and 
 $$ h^+[L] := h[L^{4 \slash 3} \log L, L] = \Theta(L^{-2 \slash 3} \log L).$$
 For each $L > 0$ we split the interval
 $[\Delta,L^{4 \slash 3} \log L]$ into $\Theta(h^+[L] L^{4 \slash 3} \log L) =
  \Theta(L^{2 \slash 3} \log^2(L))$ equal-sized subintervals
 $[\Delta=\eta_0,\eta_1),[\eta_1,\eta_2),\ldots$ of length
 $\Theta(1 \slash h^+[L]) = \Theta(L^{2 \slash 3} \slash \log L)$ each
 and we put $h_{k,L} := h\left[\frac{\eta_k+\eta_{k+1}}{2},L\right].$ 
 For each of the subintervals $[\eta_k,\eta_{k+1})$ write
 $$ {\Bbb P}({\cal E}[C_{\rm large},L] |
    \mu_{L,\gamma}^{K\alpha} \in {\Bbb E}\mu_{L,\gamma}^{K\alpha}
    + [\eta_k,\eta_{k+1}) ) \leq $$ 
  $$ \frac{\exp(-h_{k,L} [{\Bbb E}\mu_{L,\gamma}^{K\alpha} + \eta_k])
    {\Bbb P}\left({\cal E}^{h_{k,L}}[C_{\rm large},L] \mbox{ holds, }
    \mu_{L,\gamma}^{K\alpha,h_{k,L}} \in {\Bbb E}\mu_{L,\gamma}^{K\alpha}
    + [\eta_k,\eta_{k+1})
    \right)}{\exp(-h_{k,L} [{\Bbb E}\mu^{K\alpha}_{L,\gamma} + \eta_{k+1}])
    {\Bbb P}\left( \mu_{L,\gamma}^{K\alpha,h_{k,L}} \in
    {\Bbb E}\mu_{L,\gamma}^{K\alpha} + [\eta_k,\eta_{k+1})\right)} \leq
 $$
 \begin{equation}\label{NIERWAR}
   \exp(h_{k,L}[\eta_{k+1}-\eta_k]) {\Bbb P}\left( 
   {\cal E}^{h_{k,L}}[C_{\rm large},L] \left| |\mu_{L,\gamma}^{K\alpha,h_{k,L}} - 
   {\Bbb E}\mu_{L,\gamma}^{K\alpha,h_{k,L}}| \leq \frac{\eta_{k+1}-\eta_k}{2} \right. \right).
 \end{equation}
 At this point we claim that 
 \begin{equation}\label{LCLT}
  {\Bbb P}\left( \left| \mu_{L,\gamma}^{K\alpha,h_{k,L}} - 
   {\Bbb E}\mu_{L,\gamma}^{K\alpha,h_{k,L}} \right|
   \leq \frac{\eta_{k+1}-\eta_k}{2} \right) = \Omega(L^{-1 \slash
   3}\log^{-1} L)
 \end{equation}
  uniformly in $L,\gamma,\Delta,k.$ Note that this is in fact a rather
  weak statement in the spirit of local central limit theorem (LCLT)
  and an LCLT could in principle be established
  for the polygonal Markov fields in its full strength much along the same
  lines as Lemma 2.4.1 in \cite{IS1}, with standard modifications
  due to the non-lattice nature of our setting. However, since we only need the
  weaker relation (\ref{LCLT}), we provide a much shorter argument
  specialised for this case. To this end, we subdivide the disk
  ${\Bbb B}_2(L)$ into $\Theta(L)$ equal-sized squares
  $Q_{1,L},Q_{2,L},\ldots$ of side length $\Theta(\sqrt{L}),$
  separated by moats of width $\log^2(L).$ Now, in view of
  (\ref{ZANIKGRPOL}), the family of identically
  distributed random variables 
  $$ X_{i,L} := \blk_{Q_{i,L}}\left({\cal A}^{[\beta,h_{k,L}];
  K\alpha,{\Bbb B}_2(L)}_{{\Bbb B}_2(L):\gamma} \cup \gamma \right), $$
  can be coupled with a sequence of i.i.d. copies $\hat{X}_{i,L}$
  of $X_{i,L}$ in the way that ${\Bbb P}(\exists_{i} X_{i,L}
  \neq \hat{X}_{i,L}) = O(L^2 \exp(-c \log^2(L))),\; c>0.$
  Indeed, $O(L^2\exp(-c \log^2(L)))$ is the order of the 
  probability that the ancestor clans arising for different
  $Q_{i,L}$ in the graphical construction of Subsubsection
  \ref{MODGK} are not all pairwise disjoint. Write 
  $$ Y_L := \blk_{{\Bbb B}_2(L) \setminus \bigcup_{i} Q_{i,L}}
     \left({\cal A}^{[\beta,h_{k,L}];
     K\alpha,{\Bbb B}_2(L)}_{{\Bbb B}_2(L):\gamma} \cup \gamma \right) $$
  and note that
  \begin{equation}\label{PRZEDSTAA}
    \mu_{L,\gamma}^{h_{k,L},K\alpha} =
    \sum_{i} X_{i,L} + Y_L.
  \end{equation}
  Further, observe that, in complete analogy with Theorem \ref{SREDNIEO},
  \begin{equation}\label{PSTWOINNE}
   {\Bbb P}(|Y_L - {\Bbb E}Y_L| > \sqrt{L} \log^3(L)) \leq
    \exp\left( - c \left[
     \frac{[\sqrt{L} \log^{3}(L)]^2}{L \log^2(L)} \wedge
     \frac{\sqrt{L} \log^{3}(L)}{K\alpha} \right] \right) = \e^{-c \log^2(L)}.
  \end{equation}  
  Using the coupling of $X_{i,L}$ and $\hat{X}_{i,L}$ as discussed above,
  taking into account (\ref{PRZEDSTAA}) and (\ref{PSTWOINNE}) and
  recalling that $\eta_{k+1}-\eta_k = \Theta(L^{2\slash 3} \slash \log L)
  \gg \sqrt{L} \log^3(L)$ we can now deduce the required relation
  (\ref{LCLT}) by the classical local central limit theorem applied
  for $\sum_{i} \hat{X}_{i,L},$ use e.g. Theorem 1 in Wey \cite{WE} 
  with $\lambda_L := L^{2\slash 3} \slash \log(L)$ and $M_L :=
  L^{2\slash 3} \slash \log^2(L)$ there, with the assumption
  (H1) [central limit theorem for $\mu^{K\alpha,h_{k,L}}_{L,\gamma}$]
  there following by Theorem 2.10.5 of \cite{FFG2} or Theorem 2.4.R5
  and Section 5.3 in \cite{FFG1} with obvious modifications due to
  the continuum
  rather than lattice nature of our setting, and with the assumption
  (H2) in \cite{WE} satisfied in view of (\ref{KUM2}), (\ref{KUM3})
  and by the relation $\Var[\mu_{L,\gamma}^{K\alpha,h_{k,L}}] = 
  \Theta(L^2)$ which can be established along the same lines as
  (\ref{KUM2UZUP}).  
   
 Consequently, since $h_{k,L}[\eta_{k+1}-\eta_k]=O(1)$ in (\ref{NIERWAR}),
 combining the relation (\ref{NIERWAR}) with (\ref{LCLT}) and taking into
 account that ${\Bbb P}({\cal E}^{h_{k,L}}[C_{\rm large},L]) = 
 \exp(-\Omega(C_{\rm large} \log L))$ uniformly in $\Delta,L,k,\gamma$
 for $C_{\rm large}$ large enough in view of Lemma \ref{DLG} in
 Section \ref{EXPTI}, we conclude that, uniformly in $\Delta,L,k,\gamma$
 \begin{equation}\label{WAAR}
   {\Bbb P}({\cal E}[C_{\rm large},L] | \mu_{L,\gamma}^{K\alpha}
    \in {\Bbb E}\mu_{L,\gamma}^{K\alpha} + [\eta_k,\eta_{k+1}) ) = o(1).
 \end{equation}
 In view of (\ref{WAAR}) the assertion of our lemma will follow as
 soon as we show that 
 \begin{equation}\label{WAAR2}
  {\Bbb P}(\mu^{K\alpha}_{L,\gamma} > {\Bbb E} \mu^{K\alpha}_{L,\gamma}
  + L^{4\slash 3} \log L | \mu_{L,\gamma}^{K\alpha} > {\Bbb E}
   \mu_{L,\gamma}^{K\alpha} + \Delta) = o(1)
 \end{equation}
 uniformly in $\Delta,L,\gamma.$
 To this end, use Theorem \ref{SREDNIEO} to conclude that 
 \begin{equation}\label{SROWN}
   {\Bbb P}(\mu_{L,\gamma}^{K\alpha} >
   {\Bbb E} \mu_{L,\gamma}^{K\alpha} + L^{4\slash 3} \log L) \leq
  % \exp\left(-c \left[ \frac{L^{2(4 \slash 3)}\log^2 L}{L^2} \wedge
  % \frac{L^{4 \slash 3}\log L}{K\alpha} \right]\right) = 
   \exp(-c L^{2\slash 3}\log^2(L)).   
 \end{equation}
 Next, apply Lemma \ref{DOLNESREDNIEO} to get
 $$ {\Bbb P}(\mu_{L,\gamma}^{K\alpha} > {\Bbb E}\mu_{L,\gamma}^{K\alpha}
    + \Delta) \geq \exp(-O([\Delta + L \log L]^2 \slash L^2))
    \geq \exp(-O(L^{2\slash 3})). $$ 
 which yields the required relation (\ref{WAAR2}) when 
 combined with (\ref{SROWN}). The proof of the lemma
 is hence complete. $\Box$ \\

 Recalling that conditionally on 
 ${\Bbb L}_{K\alpha;L} = \gamma$ the field ${\cal A}^{[\beta]}$ coincides
 in distribution with
 ${\cal A}^{[\beta];K\alpha,{\Bbb B}_2(L)}_{{\Bbb R}^2:\gamma}$ $
 \cup \gamma$ and that, by the discussion in Subsection \ref{ISTNKONT},
 ${\Bbb L}_{K\alpha,L}
 = \{ \theta_{\rm large} \}$ with overwhelming probability under the
 micro-canonical constraint, and then combining Lemma \ref{ODCHYLKI} with
 Lemma \ref{TYLKOJEDEN} applied conditionally on
 $\gamma = {\Bbb L}_{K\alpha,L},$ shows that, conditionally on the
 event $\{ \blk_L\left( {\cal A}^{[\beta]} \right) > \blk[\beta] \pi L^2
 + a L^2,\; {\cal N}[\alpha,L] $ $ \mbox{ holds} \},$ with overwhelming
 probability $\theta_{\rm large}$ is the only $C_{\rm max} \log L$-large
 contour of ${\cal A}^{[\beta]}$ hitting ${\Bbb B}_2(L).$ This completes
 the present subsection of the proof.  

\subsection{Localising the large contour}
 It remains to show that the large contour $\theta_{\rm large}$ satisfies
 $$ \min_x \rho_H\left( \theta_{\rm large},
            {\Bbb S}_1\left(x,L\sqrt{\frac{a}{2\pi |\blk[\beta]|}}\right)
            \right) = O\left(L^{3\slash 4} \sqrt{\log L}\right). $$
 But this follows immediately by specialising to our setting for
 $\theta_{\rm large}$ the inequality (2.4.1) in Section 2.4 of
 Dobrushin, Koteck\'y \& Shlosman \cite{DKS} and combining it with 
 (\ref{DLANU}) and (\ref{DLALAMBDY}). This completes the 
 proof of Theorem \ref{GLOWNE}. $\Box$  

\section{Appendix}
 Below, we discuss the dynamic representation and some further properties of
 the basic Arak process, see Arak \& Surgailis \cite{AS1}, Section 4 for the
 dynamic representation. For a fixed bounded open convex domain $D$ we shall
 construct the basic Arak process ${\cal A}^*_D$ with free boundary conditions
 (unlike in (\ref{GREPR1}) where empty boundary conditions are imposed).   

 \subsection{Dynamic construction of the basic Arak process}\label{REPRDYN}
 We interpret the domain $D$ as a set of {\it time-space} points $(t,y) \in D,$
 with $t$ referred to as the {\it time} coordinate and with $y$ standing for
 the {\it spatial} coordinate of a particle at the time $t.$ In this language,
 a straight line segment in $D$ stands for a piece of the time-space trajectory
 of a freely moving particle. For a straight line $l$ non-parallel to the spatial
 axis and crossing the domain $D$ we define in the obvious way its entry point
 to $D,\;\iin(l,D) \in \partial D$ and its exit point $\oout(l,D) \in \partial D.$

 We choose the time-space birth coordinates for the new particles according
 to a homogeneous Poisson point process of intensity $\pi$ in $D$ (interior
 birth sites) superposed with a Poisson point process on the boundary
 (boundary birth sites) with the intensity measure
 \begin{equation}\label{KAPPAWPRO}
  \kappa(B) = {\Bbb E}\card\{ l \in \Lambda,\; \iin(l,D) \in B \},\; B \subseteq \partial D.
 \end{equation}
 Each interior birth site emits two particles, moving with initial velocities
 $v'$ and $v''$ chosen according to the joint distribution
 \begin{equation}\label{THETAWPRO}
   \theta(dv',dv'') := \pi^{-1} |v'-v''| (1+{v'}^{2})^{-3/2} (1+{v''}^{2})^{-3/2} dv' dv''.
 \end{equation}
 This can be shown to be equivalent to choosing the directions of the straight lines
 representing the space-time trajectories of the emitted particles according to the
 distribution of the {\it typical angle} between two lines of $\Lambda,$
 see Sections 3 and 4 in \cite{AS1} and the references therein.
 It is also easily seen that the value of angle $\phi \in (0,\pi)$ between
 these lines is distributed according to the density $\sin(\phi) \slash 2.$ 
 Each boundary birth site $x \in \partial D$ yields one particle
 with initial speed $v$ determined according to the distribution
 $\theta_x(dv)$ identified by requiring that the direction of the line
 entering $D$ at $x$ and representing the time-space trajectory of the
 emitted particle be chosen according to the distribution of a straight
 line $l \in \Lambda$ conditioned on the event $\{ x = \iin(l,D) \}.$

 All the  particles evolve independently in time according to the following rules.
 \begin{description}
  \item{\bf (E1)} Between the critical moments listed below each particle
                  moves freely with constant velocity so that $dy = v dt,$
  \item{\bf (E2)} When a particle touches the boundary $\partial D,$ it dies,
  \item{\bf (E3)} In case of a collision of two particles (equal spatial coordinates $y$
        at some moment $t$ with $(t,y) \in D$), both of them die,
  \item{\bf (E4)} The time evolution of the velocity $v_t$ of an individual particle
        is given by a pure-jump Markov process so that
        $$ {\Bbb P}(v_{t+dt} \in du \;|\; v_t = v) = q(v,du) dt $$
        for the transition kernel
        $$ q(v,du) := |u-v| (1+u^2)^{-3/2} du dt. $$
 \end{description}
 It is worth noting that, in full analogy with the discussion following 
 (\ref{THETAWPRO}), the (sharp) angle between the straight lines representing
 the space-time trajectories of the particle before and after the velocity 
 update is distributed according to the typical angle between two lines
 of $\Lambda.$ 

 It has been proven (see Lemma 4.1 in \cite{AS1}) that
 with the above construction of the interacting particle system, the time-space
 trajectories traced by the evolving particles coincide in distribution with
 the Arak process ${\cal A}^*_D$ defined as in (\ref{GREPR1}) with the 
 family $\Gamma_D$ of admissible polygonal configurations extended to
 $\Gamma^*_D$ allowing also for partial contours chopped off by the boundary,
 which amounts to admitting not only internal vertices of degree 2, as in
 {\bf (P2)}, but also boundary vertices of degree 1.

 \subsection{Properties of the basic Arak process}
  As already mentioned in the introductory section, and as shown
  in Arak \& Surgailis \cite{AS1}, the basic Arak process ${\cal A}^*_D$ 
  enjoys a number of striking properties. The two-dimensional germ
  Markov property, stating that the conditional distribution of the
  field inside a bounded region with piecewise smooth boundary given
  the outside configuration only depends on the trace of this
  configuration on the boundary (intersection points and intersection
  directions) is an immediate consequence of the Gibbsian definition. 
  Next important property is the consistency: for bounded open and convex
  $D_1$ and $D_2$ with $D_1 \subseteq D_2$ the restriction of
  ${\cal A}^*_{D_2}$ to $D_1$ coincides in distribution with
  ${\cal A}^*_{D_1},$ see Theorem 4.1 ibidem. This immediately 
  allows us to define the infinite volume Arak process ${\cal A},$
  which inherits the isometry invariance of the finite volume 
  Gibbsian definition and which is a thermodynamic limit for 
  ${\cal A}^{[0]}.$ By the results of Schreiber \cite{SC1},
  this corresponds to the unique infinite-volume bounded-density
  stationary evolution of the particle system discussed in
  Subsection \ref{REPRDYN} above. Interestingly, the intersection
  of the Arak process ${\cal A}$ with any fixed straight line is
  a Poisson point process of intensity $2,$ see \cite{AS1}, which
  gives us direct access to two-point correlation functions of ${\cal A}$ under
  the {\it colouring} as in Subsection \ref{LIPMF}. Moreover, the partition
  function for the Arak process can be explicitly evaluated: it is known that 
  $$ {\Bbb E} \sum_{\delta \in \Gamma^*_D(\Lambda_D)} \exp(-2\lgth(\delta)) = \exp(\pi \Area(D)), $$
  see Theorem 4.1 in \cite{AS1} [note that the
  prefactor $2\exp(\lgth(\partial D)),$ present in the quoted
  theorem, is absent here because we take the law of $\Lambda$
  rather than the unnormalised measure $\mu^*$ as the reference
  measure and, moreover, we do not sum over two different admissible
  black$\slash$white colourings of each polygonal configuration]. It
  should be emphasised that these exact results are only available
  for ${\cal A}$ and not for ${\cal A}^{[\beta]},\; \beta > 0.$   

 Interestingly, there exists a much broader class of consistent polygonal Markov fields
 admitting analogous dynamic representations, possibly enhanced to allow for vertices
 of higher degrees ($3$ and $4$), see ibidem. The question of characterising the class
 of all polygonal Markov fields admitting dynamic representations is far from being trivial
 and falls beyond the scope of this article. A conjectured description of this class has
 been provided in Arak, Clifford \& Surgailis \cite{ACS}, where a very nice alternative
 point- rather than line-based representation of polygonal fields is also discussed.   

\paragraph{Acknowledgements}
 The author gratefully acknowledges the support of the {\it Foundation for
 Polish Science} (FNP), the Polish Minister of Scientific Research and
 Information Technology grant 1 P03A 018 28 (2005-2007) and from the
 EC 6th Framework Programme Priority 2 Information Society Technology
 Network of Excellence MUSCLE (Multimedia
 Understanding through Semantics, Computation and Learning; FP6-507752).
 He also wishes to express his gratitude for hospitality of the {\it Centrum
 voor Wiskunde en Informatica} (CWI), Amsterdam, The Netherlands, where a part
 of this research was completed. Special thanks are due to anonymous referees
 whose remarks have been helpful in improving this paper.

\end{document}